\pdfoutput=1
\documentclass[11pt]{article}

\usepackage{arxiv}

\usepackage[utf8]{inputenc}
\usepackage[T1]{fontenc}
\usepackage{hyperref}
\usepackage{url}
\usepackage{booktabs}
\usepackage{tabularx}
\usepackage{amsfonts}
\usepackage{microtype}
\usepackage{graphicx}
\usepackage{doi}
\usepackage{amsmath,amssymb,amsthm}
\usepackage{xparse}
\usepackage{xcolor}
\usepackage{enumitem}
\usepackage{algorithm}
\usepackage{algpseudocode}
\usepackage{placeins}
\usepackage{subcaption}
\algrenewcommand\algorithmicrequire{\textbf{Input:}}
\algrenewcommand\algorithmicensure{\textbf{Output:}}
\algnewcommand{\IOitem}[1]{\Statex\hspace*{\algorithmicindent}#1}
\usepackage{tikz}
\usetikzlibrary{arrows.meta,positioning,decorations.pathreplacing,decorations.pathmorphing,calc}

\setlist{itemsep=1pt,topsep=3pt}
\newcolumntype{L}[1]{>{\raggedright\arraybackslash}p{#1}}
\newcolumntype{Y}{>{\raggedright\arraybackslash}X}

\newtheorem{theorem}{Theorem}[section]
\newtheorem{lemma}[theorem]{Lemma}

\newtheorem{corollary}[theorem]{Corollary}
\theoremstyle{definition}
\newtheorem{definition}[theorem]{Definition}

\makeatletter
\newcommand{\rs@heading}{}
\theoremstyle{plain}
\newtheorem*{rs@plain}{\rs@heading}
\theoremstyle{definition}
\newtheorem*{rs@definition}{\rs@heading}
\newcommand{\rs@name@theorem}{Theorem}
\newcommand{\rs@name@lemma}{Lemma}
\newcommand{\rs@name@proposition}{Proposition}
\newcommand{\rs@name@corollary}{Corollary}
\newcommand{\rs@name@definition}{Definition}
\newcommand{\rs@name@remark}{Remark}
\newcommand{\rs@style@theorem}{rs@plain}
\newcommand{\rs@style@lemma}{rs@plain}
\newcommand{\rs@style@proposition}{rs@plain}
\newcommand{\rs@style@corollary}{rs@plain}
\newcommand{\rs@style@definition}{rs@definition}
\newcommand{\rs@style@remark}{rs@definition}
\NewDocumentEnvironment{restatable}{O{} m m +b}{%
  \ifx\relax#1\relax\begin{#2}#4\end{#2}\else\begin{#2}[{#1}]#4\end{#2}\fi
  \expandafter\xdef\csname rs@num@#3\endcsname{\csname the#2\endcsname}%
  \expandafter\gdef\csname rs@env@#3\endcsname{#2}%
  \expandafter\gdef\csname rs@note@#3\endcsname{#1}%
  \expandafter\gdef\csname rs@body@#3\endcsname{#4}%
  \expandafter\gdef\csname #3\endcsname{\@ifstar{\rs@restate{#3}}{\rs@restate{#3}}}%
}{}
\newcommand{\rs@restate}[1]{%
  \begingroup
  \def\rs@heading{\csname rs@name@\csname rs@env@#1\endcsname\endcsname~\csname rs@num@#1\endcsname}%
  \edef\rs@envname{\csname rs@style@\csname rs@env@#1\endcsname\endcsname}%
  \let\label\@gobble
  \expandafter\ifx\csname rs@note@#1\endcsname\@empty
    \begin{\rs@envname}\csname rs@body@#1\endcsname\end{\rs@envname}%
  \else
    \begin{\rs@envname}[{\csname rs@note@#1\endcsname}]\csname rs@body@#1\endcsname\end{\rs@envname}%
  \fi
  \endgroup}
\makeatother

\newcommand{\JumpGreedy}{\textup{\textsc{JumpGreedy}}}
\newcommand{\Greedy}{\textup{\textsc{Greedy}}}
\newcommand{\Schedule}{\textup{\textsc{Schedule}}}

\newcommand{\LPSep}{\textup{\textsc{LPSep}}}
\newcommand{\LevelSep}{\textup{\textsc{LevelSep}}}
\newcommand{\Sep}{\textup{\textsc{Sep}}}

\newcommand{\MulticastDBDST}{\textup{\textsc{MulticastDBDST}}}
\newcommand{\Convert}{\textup{\textsc{Convert}}}
\newcommand{\DBDST}{\textup{\textsc{DBDST}}}

\newcommand{\DTM}{\textsf{DTM}}
\newcommand{\DTB}{\textsf{DTB}}

\newcommand{\OPT}{\mathrm{OPT}}

\newcommand{\dist}{\mathrm{dist}}

\newcommand{\LP}{\mathrm{LP}}
\newcommand{\TLP}{\theta_{\mathrm{LP}}}

\title{Approximation Algorithms for \\ Directed Telephone Multicast}

\author{Shahin Kamali \\
  York University, Canada \\
  \texttt{kamalis@yorku.ca} \\
  \And
  Seyed-Mohammad Seyed-Javadi \\
  York University, Canada \\
  \texttt{mohammad.sj2@gmail.com}
}

\date{}

\renewcommand{\undertitle}{}
\makeatletter
\renewcommand{\@toptitlebar}{\vskip 0.1in}
\renewcommand{\@bottomtitlebar}{\vskip 0.15in}
\makeatother
\hypersetup{
  hypertexnames=false,
  pdftitle={Approximation Algorithms for Directed Telephone Multicast},
  pdfauthor={Shahin Kamali, Seyed-Mohammad Seyed-Javadi},
  hidelinks,
  colorlinks=true,
  allcolors=black,
  pdfstartview=Fit,
  breaklinks=true
}

\begin{document}
\pagenumbering{roman}

\maketitle

\begin{abstract}

Telephone multicast is a classical problem in information dissemination. In the directed setting, a message starts at a source in an $n$-vertex digraph, and each informed vertex may call one out-neighbour per round. The goal is to inform a given set of $k$ terminals in the fewest rounds. Despite previous work on telephone multicast, a substantial approximation gap remained in the directed setting: approximating the problem within a factor of $o(\log k)$ is NP-hard, while the best existing polynomial-time approximation guarantee depending only on $k$ was $\widetilde O(\sqrt{k})$.

For every integer $\ell\ge2$, we give a deterministic combinatorial algorithm that runs in $n^{O(\ell)}$ time and returns a schedule of length $O(\ell^3\log k)(\OPT+k^{1/\ell})$, where $\OPT$ is the optimum number of rounds. This yields an $O(k^\epsilon)$ approximation in time $n^{O(1/\epsilon)}$ for every fixed $\epsilon>0$, and an $O(\log^4 k)$ approximation in quasi-polynomial time. The same algorithm gives an $O(\log k)$ approximation whenever $\OPT\ge k^\epsilon$ for a fixed $\epsilon\in(0,1)$, which is optimal up to a constant factor unless $\mathrm{P}=\mathrm{NP}$.

On planar digraphs, we obtain a deterministic polynomial-time $O(\log^2 k)$ approximation. More generally, we give an $O(p\log^2 k)$ approximation for minor-closed graph classes with the following separator property. Given any rooted spanning tree, a balanced separator can be computed in polynomial time as the union of all vertices on at most $p$ tree paths from the root to chosen vertices. Classes of bounded treewidth satisfy this property.

We also strengthen the known logarithmic approximation hardness for directed telephone multicast, extending it from instances with a polynomially large optimum to instances with $\OPT=O(\log k)$.

\end{abstract}

\clearpage
\pagenumbering{arabic}

\section{Introduction}
\label{sec:intro}

Telephone broadcast and multicast are classical problems of information
dissemination~\cite{hedetniemi1988survey}. A message starts at a source
vertex and spreads through a network in synchronous rounds. In each round,
every vertex that knows the message may call at most one neighbour,
which receives the message at the end of the round, and no vertex takes
part in more than one call per round. The goal is to reach
all vertices, or a specified set of vertices, in as few rounds as possible.

We study \emph{directed telephone multicast} (\DTM). The input is a
directed graph $G=(V,E)$ with $n=|V|$ vertices, a source $s\in V$, and a
set $K\subseteq V\setminus\{s\}$ of $k$ \emph{terminals}, all reachable
from $s$. Initially, only $s$ is \emph{informed}, and calls must follow
the direction of the edges. A \emph{schedule} specifies the calls made
in each round; we seek one that informs every terminal in the fewest
rounds. We denote this minimum by $\OPT$. \emph{Directed telephone
broadcast} (\DTB) is the special case $K=V\setminus\{s\}$. The undirected
problems correspond to digraphs in which each edge is present
in both directions.

Directed telephone multicast is hard to approximate: unless
$\mathrm{P}=\mathrm{NP}$, no polynomial-time algorithm can achieve an
approximation ratio sublogarithmic in $k$. This follows from a reduction
from Set Cover~\cite{fraigniaud2001approximation} and its known
hardness~\cite{moshkovitz2015projection,dinur2014analytical}.
Despite previous work, the best polynomial-time approximation guarantee
depending only on $k$ was $\widetilde O(\sqrt{k})$, due to Elkin and
Kortsarz~\cite{elkin2006directed} and Hathcock, Kortsarz and
Ravi~\cite{hathcock2026telephone}. This leaves a substantial gap between
the logarithmic hardness bound and the roughly $\sqrt{k}$ approximation
guarantee. By comparison, undirected multicast admits an
$O(\log k/\log\log k)$ approximation~\cite{elkin2006sublogarithmic}.
Whether directed multicast admits a polylogarithmic approximation is
an open question, stated by Elkin and
Kortsarz~\cite{elkin2005combinatorial,elkin2006directed}, who attribute it
to Ravi~\cite{ravi1994rapid}.

This paper narrows this gap. We obtain a polynomial-time
$O(k^\epsilon)$ approximation for every fixed $\epsilon>0$ and a
polylogarithmic approximation in quasi-polynomial time. For planar
digraphs, we obtain a polylogarithmic approximation in polynomial time;
this approach also extends to other graph families admitting suitable
separators. We complement these algorithms by extending logarithmic
approximation hardness to the regime %
with
$\OPT=\Theta(\log k)$.

\subsection{Our results}
\label{sec:results}

Our main results improve approximation guarantees for directed telephone multicast on general digraphs and on planar digraphs. We also extend the guarantees to other graph families, strengthen the known hardness results, and establish a connection to degree-bounded directed Steiner tree. Table~\ref{tab:results} summarizes these results alongside the previous bounds.

\paragraph{General digraphs.} 
Our main result is a deterministic combinatorial algorithm, named \JumpGreedy, that improves the approximation guarantee for directed telephone multicast from \(\widetilde O(\sqrt{k})\) to \(O(k^\epsilon)\) for every fixed \(\epsilon>0\), while retaining polynomial running time. The result also applies to the more general \emph{\(k\)-multicast problem}, studied by Hathcock, Kortsarz and Ravi~\cite{hathcock2026telephone}, in which the input contains \(t\ge k\) terminals and the goal is to inform \emph{any} \(k\) of them in the fewest rounds.

\begin{restatable}{theorem}{thmdtmcomb}\label{thm:dtmcomb}
For every integer \(\ell\ge2\), the deterministic algorithm \JumpGreedy\ with parameter \(\ell\) runs in \(n^{O(\ell)}\) time and, given a directed telephone multicast instance with \(k\ge2\) terminals, returns a schedule of length \(O(\ell^3\log k){(\mathrm{OPT}+k^{1/\ell})}\). The same guarantee holds for directed \(k\)-multicast, where \(k\ge2\) is the number of terminals to be informed. In each case, \(\mathrm{OPT}\) denotes the minimum number of rounds required by a feasible schedule.
\end{restatable}
Section~\ref{sec:techovdtm} gives an overview of \JumpGreedy, Section~\ref{sec:ovdtm} describes it in detail, and Appendix~\ref{sec:dtm} contains the proof.

The additive term \(k^{1/\ell}\) matters most when \(\mathrm{OPT}\) is small. When \(\mathrm{OPT}\ge k^{1/\ell}\), the schedule length is \(O(\ell^3\log k)\,\mathrm{OPT}\), giving an \(O(\log k)\) approximation for every fixed \(\ell\). Increasing \(\ell\) lowers this threshold, at the cost of a larger running time and a larger multiplicative factor \(\ell^3\). 
Since the number of informed vertices can at most double in each round, \(\mathrm{OPT}\ge\lceil\log_2(k+1)\rceil\). Combining this lower bound with Theorem~\ref{thm:dtmcomb} gives the following approximation guarantees.

\begin{table}[t]
\centering
\small
\resizebox{\linewidth}{!}{\begin{tabular}{@{}>{\raggedright\arraybackslash}p{3.75cm}>{\raggedright\arraybackslash}p{7cm}>{\raggedright\arraybackslash}p{5.5cm}@{}}
\toprule
Setting & This paper & Previously \\
\midrule
Polynomial time
& $O(k^\epsilon)$ in time $n^{O(1/\epsilon)}$, for every fixed
  $\epsilon>0$ (Cor.~\ref{cor:dtm})
& $\tilde O(\sqrt k)$~\cite{elkin2006directed,hathcock2026telephone} \vspace{1.5mm} \\

$\OPT\ge k^\epsilon$
& $O(\log k)$ in time $n^{O(1/\epsilon)}$, for every fixed
  $\epsilon\in(0,1)$;\newline best possible up to a constant factor
  (Cor.~\ref{cor:dtm}, Prop.~\ref{prop:largeopt})
& $O(\log k+k^{1/2-\epsilon})$;\newline $\Omega(\log k)$-hard for
  $\epsilon<1/2$~\cite{elkin2006directed} \vspace{1.5mm} \\

Quasi-polynomial time
& $O(\log^4k)$ in time $n^{O(\log k)}$ (Cor.~\ref{cor:dtm});\newline
  $O(\log^2n)$ in time $n^{O(\log n)}$, randomized
  (Cor.~\ref{cor:te})\vspace{1.5mm} 
& --- \ \vspace{1.5mm} \\

Planar digraphs
& $O(\log^2k)$ in polynomial time (Thm.~\ref{cor:plplanar})
& $\tilde O(\sqrt k)$, as for general digraphs \vspace{1.5mm} \\

$p$-root-path separable classes
& $O(p\log^2k)$ in polynomial time (Thm.~\ref{thm:plclass})
& $\tilde O(\sqrt k)$, as for general digraphs \vspace{1.5mm} \\

Treewidth $w$
& $O(w\log^2k)$ in polynomial time for fixed $w$
  (Cor.~\ref{cor:pltw})
& $\tilde O(\sqrt k)$, as for general digraphs \vspace{1.5mm} \\

Pathwidth $w$
& $O(w^{7/2}\log k)$ in time $n^{O(w)}$ (Cor.~\ref{cor:dtm})
& $\tilde O(\sqrt k)$, as for general digraphs \vspace{1.5mm} \\

$k$-multicast
& $O(k^\epsilon)$ in time $n^{O(1/\epsilon)}$;
  $O(\log^4k)$ in time $n^{O(\log k)}$;
  both directed and undirected (Cor.~\ref{cor:dtm})
& $\tilde O(t^{1/3})$, undirected;\newline
  $\tilde O(\sqrt k)$, directed~\cite{hathcock2026telephone} \vspace{1.5mm} \\

Hardness
& $(1-\epsilon)\ln k$ when $\OPT=O(\log k)$ 
  (Thm.~\ref{thm:lowopt});
  $\Omega(\log k)$ when $\OPT\ge k^\epsilon$
  (Prop.~\ref{prop:largeopt})
& $(1-\epsilon)\ln k$~\cite{fraigniaud2001approximation,moshkovitz2015projection,dinur2014analytical};
  $\Omega(\log k)$ when
  $\OPT=\Omega(\sqrt k)$~\cite{elkin2006directed} \vspace{1.5mm} \\

\bottomrule
\end{tabular}}
\caption{Summary of results. All rows concern directed telephone
multicast except the $k$-multicast row. Here $k$ is the number of
terminals, or, for $k$-multicast, the number to be informed out of
$t$ terminals. All our upper bounds except that of
Corollary~\ref{cor:te} are achieved by deterministic algorithms
and have approximation ratios independent of $n$.
Graph-class conditions refer to the underlying undirected graph;
$p$-root-path separable classes are defined in
Definition~\ref{def:plclass}.}
\label{tab:results}
\end{table}

\begin{restatable}{corollary}{cordtm}\label{cor:dtm}
For every integer $\ell\ge2$, \JumpGreedy\ has approximation ratio
$O(\ell^3(\log k+k^{1/\ell}))$ on \DTM\ instances with $k\ge2$ terminals.
In particular, suitable choices of $\ell$ give the following
deterministic guarantees:

\begin{enumerate}[label=(\alph*),ref=(\alph*)]
  \item\label{cor:dtm:poly} For every fixed $\epsilon>0$, an $O(k^\epsilon)$ approximation
    in time $n^{O(1/\epsilon)}$.

  \item\label{cor:dtm:quasi} An $O(\log^4 k)$ approximation in time $n^{O(\log k)}$.

  \item\label{cor:dtm:largeopt} For every fixed $\epsilon\in(0,1]$, an
    $O(\epsilon^{-3}\log k)$ approximation in time
    $n^{O(1/\epsilon)}$ on instances satisfying $\OPT\ge k^\epsilon$.
    For $\epsilon<1$, this is optimal up to a constant factor unless
    $\mathrm{P}=\mathrm{NP}$, by Proposition~\ref{prop:largeopt}.

  \item\label{cor:dtm:pw} If the underlying undirected graph has pathwidth at most
    $w\ge1$, an $O(w^{7/2}\log k)$ approximation in time $n^{O(w)}$.
    Thus, for every fixed $w$, the approximation ratio is $O(\log k)$
    and the running time is polynomial.
\end{enumerate}
The guarantees in parts~\ref{cor:dtm:poly} and~\ref{cor:dtm:quasi} also hold for directed
$k$-multicast, and hence for undirected $k$-multicast.
\end{restatable}

Appendix~\ref{sec:comb} derives Corollary~\ref{cor:dtm}. Part~\ref{cor:dtm:quasi} gives a polylogarithmic approximation for directed multicast in quasi-polynomial time. This addresses the open question on directed multicast stated by Elkin and Kortsarz~\cite{elkin2006directed} %
with a relaxed running-time requirement. Obtaining a \(\operatorname{polylog}(k)\)-approximation in polynomial time remains open.
Part~\ref{cor:dtm:largeopt} gives an optimal \(O(\log k)\) approximation whenever \(\mathrm{OPT}\ge k^\epsilon\), for every fixed \(\epsilon\in(0,1)\). For \(\epsilon\ge1/2\), this approximation guarantee already follows from the schedule of length \(O(\log k)\,\mathrm{OPT}+O(\sqrt k)\) by Elkin and Kortsarz~\cite{elkin2006directed}. They also observed that it is optimal when \(k=n^{\Omega(1)}\) and \(\mathrm{OPT}=\Omega(\sqrt k)\), which implies optimality for every fixed \(\epsilon<1/2\). Our results extend the guarantee to every fixed \(\epsilon\in(0,1/2)\), and Proposition~\ref{prop:largeopt} extends its optimality to every fixed \(\epsilon\in[1/2,1)\). Obtaining a \(\operatorname{polylog}(k)\)-approximation in polynomial time
without assuming a polynomial lower bound on \(\mathrm{OPT}\) remains open. Part~\ref{cor:dtm:pw} exploits a lower bound on broadcasting time in graphs of bounded pathwidth~\cite{egami2025broadcasting}. %
This bound makes the additive term small enough to obtain an \(O(\log k)\) approximation for fixed pathwidth. Appendix~\ref{sec:comb} gives the derivation.

\paragraph{Planar digraphs and related graph classes.}  On digraphs whose underlying undirected graphs are planar, we obtain a polynomial-time algorithm, named \LPSep, which has an approximation factor of \(O(\log^2 k)\):

\begin{restatable}{theorem}{corplplanar}\label{cor:plplanar}
Given a planar digraph, a source $s$, and a set of $k\ge2$ terminals
reachable from $s$, \LPSep, run with the separator algorithm of
Thorup~\cite[Lemma~2.3]{thorup2004compact}, deterministically computes a
\DTM\ schedule of length $O(\log^2 k)\,\OPT$ in polynomial time.
\end{restatable}

The guarantee extends beyond planar digraphs. The key requirement is that a balanced separator can be formed from a bounded number of paths in a rooted spanning tree. More precisely, consider a minor-closed graph class in which, given any rooted spanning tree, a balanced separator consisting of the union of at most \(p\) tree paths starting at the root can be computed in polynomial time. On digraphs whose underlying undirected graphs belong to such a class, \LPSep\ achieves an \(O(p\log^2 k)\) approximation (Theorem~\ref{thm:plclass}). Planar graphs satisfy this property with \(p=3\), yielding the \(O(\log^2 k)\) approximation above. Graphs of treewidth at most \(w\) satisfy it with \(p=w+1\), giving an \(O(w\log^2 k)\) approximation in polynomial time for every fixed \(w\) (Corollary~\ref{cor:pltw}). 
To our knowledge, these are the first polynomial-time \(\operatorname{polylog}(k)\)-approximations for directed telephone multicast on planar digraphs and on digraphs whose underlying undirected graphs have treewidth at most \(w\), for every fixed \(w\ge2\).

Section~\ref{sec:techovplanar} gives a technical overview of \LPSep. Section~\ref{sec:ovplanar} defines the separator property formally and describes the algorithm, and Appendix~\ref{sec:planar} contains the proofs.

\paragraph{A connection to degree-bounded directed Steiner tree.} In the degree-bounded directed Steiner tree problem (DB-DST), the input consists of a digraph with edge costs, a root, a set of terminals, and a degree bound for each vertex. The goal is to find a minimum-cost out-arborescence rooted at the given root that reaches every terminal and respects each vertex’s bound on its number of children. Unlike ordinary directed Steiner tree, DB-DST explicitly constrains these degrees.
We reduce directed telephone multicast to DB-DST by constructing a graph with a separate copy of each vertex for each round of a candidate schedule. In this graph, degree bounds limit how many calls a vertex can make in a round (Theorem~\ref{thm:te}). To obtain a short multicast schedule, we therefore need a DB-DST algorithm that keeps these degrees small.

The DB-DST algorithm of Guo, Kortsarz, Laekhanukit, Li, Vaz and Xian~\cite{guo2022degree} provides this degree control. Through our reduction, it gives a randomized \(O(\log^2 n)\)-approximation for directed telephone multicast in time \(n^{O(\log n)}\) (Corollary~\ref{cor:te}). The framework of Bamas, Li and Rohwedder~\cite{bamas2026randomized} gives a different tradeoff: a randomized \(O(n^\epsilon)\)-approximation in time \(n^{O(1/\epsilon)}\) for every fixed \(\epsilon>0\). Appendix~\ref{sec:timeexp} gives the reduction and derives these guarantees.

These guarantees depend on the total number of vertices \(n\). When \(n\) is polynomial in \(k\), the consequence of the algorithm of~\cite{guo2022degree} gives an \(O(\log^2 k)\) ratio in quasi-polynomial time, improving on the \(O(\log^4 k)\) quasi-polynomial-time guarantee of \JumpGreedy. However, \JumpGreedy\ is combinatorial and deterministic, and its approximation ratio depends only on the number of terminals \(k\), which is particularly useful when \(k\) is small compared with the total number of vertices \(n\). Its additive schedule-length bound also gives an \(O(\log k)\) approximation whenever \(\mathrm{OPT}\ge k^\epsilon\) for a fixed \(\epsilon\in(0,1)\).

\paragraph{Hardness results.} The guarantee of Theorem~\ref{thm:dtmcomb} depends on the size of the optimum. This raises the question of whether instances with a small optimum might admit a sublogarithmic approximation. We rule out this possibility by %
showing that logarithmic approximation hardness persists even when %
\(\mathrm{OPT}=\Theta(\log k)\), its minimum possible order:

\begin{restatable}{theorem}{thmlowopt}\label{thm:lowopt}
For every $\epsilon\in(0,1)$ there is a constant $C_\epsilon$ such that it is
NP-hard to distinguish acyclic \DTM\ instances with $\OPT\le
C_\epsilon\log_2k$ from those with $\OPT\ge(1-\epsilon)\ln k\cdot
C_\epsilon\log_2k$.
\end{restatable}

For instances with large optimum, Elkin and Kortsarz observed that an \(O(\log k)\) approximation is optimal up to a constant factor unless \(\mathrm{P}=\mathrm{NP}\), assuming \(k=n^{\Omega(1)}\) and \(\mathrm{OPT}=\Omega(\sqrt{k})\)~\cite[Section~1.2]{elkin2006directed}. We extend this hardness to the regime \(\mathrm{OPT}\ge k^\epsilon\) for every fixed \(\epsilon\in(0,1)\). This matches the approximation guarantee in Corollary~\ref{cor:dtm}\ref{cor:dtm:largeopt} up to a constant factor:

\begin{restatable}{proposition}{proplargeopt}\label{prop:largeopt}
For all fixed $\epsilon\in(0,1)$ and $\delta\in(0,1)$, it is NP-hard to
distinguish, for a given acyclic \DTM\ instance with $k$ terminals and a
given integer $\theta$, the case $\OPT\le \theta$ from the case
$\OPT>(1-\delta)(1-\epsilon)\ln k\cdot \theta$, even when the given instances
satisfy $\OPT\ge k^\epsilon$. In particular, unless $\mathrm{P}=\mathrm{NP}$, no
polynomial-time algorithm has ratio $(1-\delta)(1-\epsilon)\ln k$ on the
\DTM\ instances with $\OPT\ge k^\epsilon$.
\end{restatable}

Both hardness results hold even on acyclic digraphs. Section~\ref{sec:techovother} gives an overview of the reductions, and Appendix~\ref{sec:hard} contains the full proofs of Theorem~\ref{thm:lowopt} and Proposition~\ref{prop:largeopt}.

\paragraph{Limitations.} For general digraphs, \JumpGreedy\ gives a \(\operatorname{polylog}(k)\)-approximation in quasi-polynomial time, but its worst-case guarantee remains polynomial in \(k\) for every fixed \(\ell\). Obtaining a \(\operatorname{polylog}(k)\)-approximation in polynomial time remains open, and our hardness results do not rule it out. For directed broadcast, where \(k=n-1\), our guarantees are weaker than the \(O(\log n)\) approximation of Elkin and Kortsarz~\cite{elkin2005combinatorial}. On planar digraphs, \LPSep\ is not combinatorial: it requires solving a linear program. It also does not handle \(k\)-multicast.

\paragraph{Organization.} Section~\ref{sec:techov} gives a technical overview of our results. Section~\ref{sec:ovsched} introduces notation and preliminaries. Sections~\ref{sec:ovdtm} and~\ref{sec:ovplanar} describe \JumpGreedy\ and \LPSep, respectively, and Section~\ref{sec:conclusion} discusses open problems. The full analyses of both algorithms, together with their pseudocode and proofs, are deferred to the appendices. We also defer the detailed constructions and proofs for the reduction from \DTM\ to DB-DST and the hardness results; Section~\ref{sec:techovother} outlines their main ideas. Appendix~\ref{sec:prelim} proves the preliminary lemmas. Appendix~\ref{sec:dtm} contains the analysis of \JumpGreedy\ (Appendix~\ref{sec:comb}) and the DB-DST reduction (Appendix~\ref{sec:timeexp}). Appendix~\ref{sec:planar} contains the analysis of \LPSep\ (Appendix~\ref{sec:planalysis}), and Appendix~\ref{sec:hard} gives the hardness constructions and their proofs.

\subsection{Related work}
\label{sec:related}

\paragraph{Telephone broadcast and multicast.} 
On undirected graphs, telephone broadcast is NP-hard but linear-time solvable on trees~\cite{slater1981nptree}. Hardness persists even on cactus graphs of pathwidth \(2\), although constant-factor approximations exist for cactus graphs and graphs of bounded pathwidth~\cite{aminian2025complexity}. For pathwidth \(w\), the approximation ratio is \(O(w)\)~\cite{egami2025broadcasting}. Kortsarz and Peleg~\cite{kortsarz1995approximation} gave a broadcast schedule of length \(2\,\mathrm{OPT}+O(\sqrt n)\) on general undirected graphs. They also obtained an \(O(\log n)\) approximation on chordal graphs and on graphs with constant-size separators, including series-parallel graphs.
Ravi~\cite{ravi1994rapid} obtained an \(O(\log^2 n/\log\log n)\) approximation using the connection between broadcast schedules and spanning trees of small maximum degree and diameter. Bar-Noy, Guha, Naor and Schieber~\cite{bar2000message} obtained an \(O(\log k)\) approximation for multicast with \(k\) terminals. Elkin and Kortsarz~\cite{elkin2006sublogarithmic} subsequently improved the ratio to \(O(\log k/\log\mathrm{OPT})\), implying an \(O(\log k/\log\log k)\) approximation.

For directed multicast, Fraigniaud’s Set Cover reduction was originally stated for a related communication model~\cite{fraigniaud2001approximation}. Elkin and Kortsarz~\cite{elkin2006directed} observed that a variant using many disjoint copies of a hard instance rules out sublogarithmic approximation ratios unless \(\mathrm{P}=\mathrm{NP}\), even when \(k=n^{\Omega(1)}\) and \(\mathrm{OPT}=\Omega(\sqrt k)\). Proposition~\ref{prop:largeopt} extends this hardness to the regime \(\mathrm{OPT}\ge k^\epsilon\) for every fixed \(\epsilon\in(0,1)\). %

For undirected \(k\)-multicast, Hathcock, Kortsarz and Ravi~\cite{hathcock2026telephone} obtained an \(\widetilde O(t^{1/3})\) approximation, where \(t\) is the number of candidate terminals. They also observed that a \(\rho\)-approximation for directed multicast yields a \(\rho\cdot\operatorname{polylog}(k)\)-approximation for undirected \(k\)-multicast, using a reduction for degree-bounded trees due to Kortsarz and Nutov~\cite{kortsarz2022minimum}.

\paragraph{Directed Steiner trees.} A multicast schedule informs the terminals along an out-arborescence, connecting telephone multicast to the directed Steiner tree problem. The latter seeks a minimum-cost out-arborescence rooted at a given source that reaches all terminals. For every fixed \(\epsilon>0\), an \(O(k^\epsilon)\) approximation is achievable in polynomial time: Zelikovsky~\cite{zelikovsky1997series} obtained this guarantee for acyclic digraphs, and Charikar, Chekuri, Cheung, Dai, Goel, Guha and Li~\cite{charikar1999dst} extended it to general digraphs.
Their recursive greedy approach, combined with the height-reduction bound of Helvig, Robins and Zelikovsky~\cite{helvig2001improved}, gives an \(O(\log^3 k)\) approximation in quasi-polynomial time. %
Grandoni, Laekhanukit and Li~\cite{grandoni2019dst} improved the ratio to \(O(\log^2 k/\log\log k)\) with a randomized algorithm, and Ghuge and Nagarajan~\cite{ghuge2022quasi} obtained the same guarantee deterministically. Grandoni, Laekhanukit and Li also established a matching lower bound under the Projection Games Conjecture and an additional randomized complexity assumption~\cite{grandoni2019dst}. Earlier, Halperin and Krauthgamer~\cite[Theorem~1.1]{halperin2003polylogarithmic} ruled out a \(\log^{2-\epsilon}k\) approximation for every fixed \(\epsilon>0\), unless \(\mathrm{NP}\subseteq\mathrm{ZTIME}(n^{\operatorname{polylog}(n)})\). Their hardness holds already for group Steiner tree on trees, a special case of directed Steiner tree.

Our general-digraph algorithm draws on height reduction and algorithms for group Steiner tree on trees~\cite{garg2000polylogarithmic,chekuri2006greedy}. The objectives, however, require different forms of control. Steiner-tree cost is a sum over selected edges, whereas telephone multicast requires controlling the number of calls made by each vertex as well as the lengths of the paths carrying the message. Our analysis therefore tracks individual vertex loads rather than only an aggregate cost.
Degree-constrained versions of Steiner problems control these local degrees, for group Steiner tree~\cite{kortsarz2022minimum} and for directed Steiner tree~\cite{guo2022degree}. More recently, Bamas, Li and Rohwedder~\cite{bamas2026randomized} developed a general framework for rounding solutions described by dynamic-programming recurrences while approximately satisfying additional packing constraints. They mention degree-bounded directed Steiner tree as an application, in a remark without details. We discussed the multicast guarantees obtained from these degree-bounded algorithms in Section~\ref{sec:results}.

\paragraph{Planar digraphs.} Friggstad and Mousavi~\cite[Theorem~1]{friggstad2023planar} gave a polynomial-time \(O(\log k)\)-approximation for directed Steiner tree on planar digraphs. Our planar algorithm builds on their separator recursion. Chekuri, Jain, Kulkarni, Zheng and Zhu~\cite[Theorem~2 and Lemma~13]{chekuri2024polymatroid} used a similar recursion, together with pruning of fractional solutions, to show that the cut relaxation of directed Steiner tree has integrality gap \(O(\log^2 k)\) on planar digraphs. Chekuri and Jain~\cite{chekuri2026length} recently gave polylogarithmic bicriteria approximations for length-constrained directed Steiner tree and directed Steiner forest on planar digraphs, violating the length bound by a factor \(O(\log k)\). Length bounds correspond to our bound on the height, but they do not control out-degrees.

Separator methods have also been used for telephone communication problems on undirected graphs. Iglesias, Rajaraman, Ravi and Sundaram~\cite{iglesias2018plane} obtained a polylogarithmic approximation for multicommodity multicast, in which each source must send its message to its own destination, on undirected planar graphs and extended it to graphs excluding a fixed minor~\cite{iglesias2017plane}. Their algorithm uses shortest-path separators and a linear programming relaxation based on \emph{poise}, which for a subgraph is the sum of its maximum degree and the largest distance in it between a source and its destination. 
Its approximation guarantee includes a factor depending on \(n\), whereas our guarantee applies to directed graphs and depends only on the number of terminals \(k\).

\section{Technical overview}
\label{sec:techov}

This section explains the main ideas behind our results. Later sections describe the algorithms in detail, and the appendices contain the full proofs.

\paragraph{Schedules as trees.} From any multicast schedule, retain the call that first informs each informed vertex other than the source. These calls form an out-arborescence: a directed tree rooted at the source, with all arcs pointing away from it. For an optimal schedule, both the height and the maximum out-degree of this tree are at most \(\mathrm{OPT}\). Conversely, Ravi’s scheduling lemma~\cite{ravi1994rapid} converts an out-arborescence of height \(h\) and maximum out-degree \(d\ge2\) that reaches all \(k\) terminals and whose leaves are terminals into a schedule of length
$O(h+d\log k/\log d)$.
Thus, finding a tree with small height and small maximum out-degree gives a short multicast schedule.
This connects multicast to directed Steiner tree, but changes what the algorithm must control. Instead of minimizing a sum of edge costs, we need to bound the tree’s height and the number of children at every vertex. Controlling these individual degrees is the main difficulty: greedy covering arguments, the standard tool for
directed Steiner tree, bound a total cost, which can remain small even when too much of it is concentrated at one vertex, resulting in a long multicast schedule.

\subsection{General digraphs}
\label{sec:techovdtm}

We sketch \JumpGreedy, the algorithm of Theorem~\ref{thm:dtmcomb}.
Fix an integer \(\ell\ge2\), which controls the tradeoff between running time and approximation quality. \JumpGreedy\ searches for a \emph{jump arborescence}: a tree rooted at $s$ whose nodes are labelled by vertices of \(G\), and whose edges represent directed walks in \(G\), called \emph{jumps}. Every terminal in a jump arborescence must label at least one node. Different jumps may overlap, and a vertex may label several tree nodes. Each jump has at most \(\theta\) arcs, where \(\theta\) is a guess for \(\mathrm{OPT}\), and the tree has depth at most \(\eta=2\ell-1\). We analyze the correct guess \(\theta=\mathrm{OPT}\).
The load \(L(v)\) counts how many times the selected jumps traverse an arc leaving \(v\). These contributions add up as jumps are added. The union of the jumps has out-degree at most \(L(v)\) at each vertex, and every terminal is reachable within \(\eta\theta\) arcs of the source. We can therefore extract an out-arborescence from a suitable jump arborescence and apply Ravi’s scheduling lemma. As a result, finding a jump arborescence with small loads gives a short multicast schedule.

We first prove that a jump arborescence with the required depth and load bounds exists. Starting from the tree of an optimal schedule, our \emph{height-reduction} construction groups representatives at ancestor vertices called hubs. Initially, the representatives are terminals. At each stage, a vertex becomes a hub when it gathers at least \(\rho=\lceil k^{1/\ell}\rceil\) representatives, and the hubs become representatives for the next stage.
Each stage reduces their number by roughly a factor of \(\rho\), and after \(\ell-1\) stages the source connects to the remaining representatives. Crucially, connections pass through the original children of each hub, rather than requiring a separate jump from the hub to every representative. This controls every vertex’s load, giving depth at most \(2\ell-1\) and maximum load
$\Lambda=O\!\left(\ell(\mathrm{OPT}+k^{1/\ell})\right).
$
This construction is a comparison solution for the analysis; the algorithm does not compute it.

To search for a suitable jump arborescence, we unfold \(G\) into 
a tree recording all possible sequences of at most \(\eta\) jumps from the source, represented by their endpoints.
This tree has \(n^{O(\ell)}\) nodes. Finding a subtree that contains an occurrence of every terminal resembles the group Steiner tree problem on a tree. We adapt the recursive greedy algorithms of~\cite{charikar1999dst} and of Chekuri, Even and Kortsarz~\cite{chekuri2006greedy}, replacing their cost objective with a penalty for vertex loads.
Specifically, we use the exponential potential
$\Phi(L)=\sum_{v\in V}\left(e^{L(v)/\Gamma}-1\right),$
as in online load balancing~\cite{aspnes1997online}. The scale \(\Gamma\) determines how quickly the penalty grows. Reusing a heavily loaded vertex becomes expensive, encouraging the algorithm to distribute the load. At each step, the recursive procedure generates candidate partial solutions and chooses one minimizing the potential increase divided by the number of newly covered terminals. The search takes \(n^{O(\ell)}\) time. 

The main difficulty in the analysis is that the vertex penalties change as the algorithm adds jumps. We compare each greedy choice with the height-reduced optimal tree, priced using the current loads, and control how those prices change during recursive calls. Choosing \(\Gamma=\Theta(\ell\Lambda)\) allows this comparison to work across all recursion levels. The analysis bounds the potential by a polynomial in \(k\), implying that every final load is \(O(\Gamma\log k)\).

Converting the resulting structure into a schedule gives length
$O(\ell^3\log k)(\mathrm{OPT}+k^{1/\ell}).$
The three factors of \(\ell\) come from height reduction, controlling the changing penalties, and scheduling; the logarithmic factor comes from greedy covering. Section~\ref{sec:ovdtm} describes \JumpGreedy\ in detail, and Appendix~\ref{sec:dtm} contains the full proof.

\subsection{Planar digraphs}
\label{sec:techovplanar}

For planar digraphs, \LPSep, the algorithm of Theorem~\ref{cor:plplanar}, uses balanced separators to divide the problem into smaller regions. Thorup’s separator lemma~\cite[Lemma~2.3]{thorup2004compact} guarantees that at most three paths from the root of any spanning tree form a separator whose removal leaves at most half the terminals in each component. Following Friggstad and Mousavi~\cite{friggstad2023planar}, we choose these paths in a breadth-first arborescence, contract their union into the root, and recurse on the remaining components. This gives \(O(\log k)\) recursion levels.

Even though the regions are disjoint, several may require calls from the same previously informed vertex. If many regions require calls from the same vertex, its call load becomes a bottleneck: each call requires a separate round.
We refer to a call from a previously informed vertex into one of the remaining regions as an \emph{entry call}. The number of entry calls assigned to a vertex at a given recursion level is its \emph{per-level load}.
Controlling the per-level load is the main difficulty in constructing a short multicast schedule.

We solve a linear program that assigns edge capacities and finds a fractional unit flow from the source to each terminal along paths of at most \(\theta\) arcs. Each terminal’s flow separately respects the same capacities, and the capacities leaving any vertex sum to at most \(\theta\). These constraints provide a fractional counterpart to the routing and call limits of a \(\theta\)-round schedule. An optimal schedule makes the LP feasible at \(\theta=\mathrm{OPT}\), so the smallest feasible value satisfies \(\theta\le\mathrm{OPT}\).

To choose entry calls, we assign each possible caller an exponential weight and allow each region to use only sufficiently cheap callers. Specifically, a caller’s weight must be at most \(1/\epsilon\) times the total capacity-weighted price of the region’s possible entries. By Markov’s inequality, the excluded entry arcs have total capacity at most \(\epsilon\). Removing them therefore loses at most \(\epsilon\) of each terminal’s flow. Rescaling restores unit flows, and choosing \(\epsilon=\Theta(1/\log k)\) keeps the accumulated rescaling bounded over the recursion.
The weights account for both calls already assigned at the current level and entry capacity in regions still to be processed. The remaining-capacity term, scaled by \(O(1/\epsilon)\), accounts for future calls. Processing a region removes its contribution to the remaining-capacity term while adding its chosen entry calls to the current loads. Our choice of cheap entries ensures that the decrease pays for the increase: the sum of the weights never increases during a level. This sum acts as a pessimistic estimator~\cite{raghavan1988probabilistic}, yielding per-level load \(O(\theta\log k)\).

The schedule informs the separators level by level. At each level, making the entry calls takes at most the maximum per-level load in rounds; completing the short separator paths takes another \(O(\theta)\) rounds. Over \(O(\log k)\) levels, this gives \(O(\theta\log^2 k)\) rounds and hence an \(O(\log^2 k)\) approximation. One logarithmic factor comes from controlling entry congestion while preserving fractional feasibility, and the other from the recursion depth.

Planarity enters only through the separators. The approach gives an \(O(p\log^2 k)\) approximation on the minor-closed classes admitting separators formed by at most \(p\) such tree paths. Section~\ref{sec:ovplanar} defines these classes and describes \LPSep\ in detail; Appendix~\ref{sec:planar} contains the proofs.

\subsection{The DB-DST reduction and hardness}
\label{sec:techovother}

We next outline the reduction to degree-bounded directed Steiner tree and explain how Set Cover yields hardness results for both small and large optima.

To reduce \DTM\ to DB-DST, we encode a candidate schedule of length \(\theta\) in a \emph{time-expanded digraph}. It contains a copy of each vertex for each round, together with nodes representing calls. Degree bounds ensure that an informed vertex can make at most one call per round while retaining the message for later rounds. A \(\theta\)-round multicast schedule therefore gives a feasible DB-DST solution with zero edge costs. If a DB-DST algorithm exceeds the degree bounds by a factor \(\beta\), we spread the extra calls over consecutive rounds. The resulting schedule has length at most \(\beta\theta\). Appendix~\ref{sec:timeexp} gives the construction and proves the conversion guarantee (Theorem~\ref{thm:te}).

Both hardness reductions start from Set Cover and adapt a gadget of Fraigniaud~\cite{fraigniaud2001approximation}, with a vertex for each set and a terminal for each element. In the basic gadget, the source calls the vertices corresponding to chosen sets. Each informed set vertex then spreads the message through a binary tree of auxiliary vertices to the terminals corresponding to its elements. Informing every terminal requires the chosen sets to form a cover. Since the source makes only one call per round, the schedule length is at least the minimum cover size. For a small optimum, the source delegates these calls to many hubs, so that a cover is called within \(O(\log k)\) rounds (Theorem~\ref{thm:lowopt}). For a large optimum, the gadget is applied to many disjoint copies of the Set Cover instance, which multiplies the optimum without changing the gap (Proposition~\ref{prop:largeopt}). Appendix~\ref{sec:hard} gives both reductions and their proofs.

\section{Preliminaries}
\label{sec:ovsched}

We introduce our notation, explain the connection between schedules and out-arborescences, and state two scheduling bounds used by our algorithms. Appendix~\ref{sec:prelim} contains the proofs of Lemmas~\ref{lem:busy} and~\ref{lem:ravi}.

Let \(G=(V,E)\) be a digraph with \(n=|V|\) vertices and arc set \(E(G)=E\), and let \(s\in V\) be the source. For \DTM, \(K\subseteq V\setminus\{s\}\) is a set of \(k\) terminals, all reachable from \(s\); broadcast is the case \(K=V\setminus\{s\}\). For \(k\)-multicast, \(K\) instead contains \(t\ge k\) terminals, of which any \(k\) must be informed. An arc \((u,v)\) goes from its tail \(u\) to its head \(v\). We represent each undirected edge by two opposite arcs. The distance \(\operatorname{dist}(u,v)\) is the number of arcs in a shortest directed path from \(u\) to \(v\), or \(\infty\) if no such path exists.
All logarithms are natural unless a base is specified.

\paragraph{Schedules and arborescences.} 
From any feasible \DTM\ schedule, retain one call that first informs each informed vertex other than \(s\). These calls form an \emph{out-arborescence}: a directed tree rooted at \(s\), with all arcs pointing away from \(s\). Conversely, an out-arborescence and an ordering of each vertex’s children define a schedule: every vertex calls its children consecutively in that order, starting in the round after it becomes informed. %
If each vertex keeps its original child order but makes those calls consecutively, starting immediately after it becomes informed, every vertex receives the message no later than in the original schedule. We may therefore omit idle rounds and redundant calls.
We may also prune every branch containing no terminal. For a fixed remaining arborescence \(T\), an optimal child order can be computed in polynomial time by the algorithm of Slater, Cockayne and Hedetniemi~\cite{slater1981nptree}: process \(T\) bottom-up and order each vertex’s children by decreasing time needed to inform their subtrees. The difficulty of \DTM\ therefore lies in choosing the arborescence.

We use the following basic bounds. First, \(\mathrm{OPT}\ge\lceil\log_2(k+1)\rceil\), since the informed set can at most double per round, and \(\mathrm{OPT}\ge\max_{x\in K}\operatorname{dist}(s,x)\), since the message traverses at most one arc per round along any informing path. Second, the arborescence of an optimal schedule, pruned to the terminals, has height and maximum out-degree at most \(\mathrm{OPT}\): calls along each root-to-terminal path occur in strictly increasing rounds, and each vertex makes at most one call per round. %

\paragraph{Two schedules along a given arborescence.} 
We use two simple call orders with explicit bounds on the schedule length. The first gives a bound in terms of the tree’s height and number of leaves. We use it for the fallback schedule described below and for scheduling the gadgets in our hardness reductions.

\begin{restatable}[{after Elkin and Kortsarz~\cite[Lemma~1.1]{elkin2006directed}}]{lemma}{lembusy}\label{lem:busy}
Let $Q$ be an out-arborescence rooted at an informed vertex, with
height $h$ and leaf set $L$. In the \emph{busy} schedule, each informed
vertex calls one child it has not yet called in every round, until it
has called all its children. This schedule informs all vertices of $Q$
within $h+|L|-1$ rounds.
\end{restatable}

Fix a path from the root to a leaf. Until the leaf is informed, the deepest informed vertex on the path either calls the next vertex on the path or calls another child. The first type of call occurs at most \(h\) times. At each internal vertex \(v\), the second type occurs at most \(d(v)-1\) times, where \(d(v)\) is the number of children of \(v\). Since \(\sum_{v\text{ internal}}(d(v)-1)=|L|-1\), the leaf is informed within \(h+|L|-1\) rounds.

The second schedule gives a bound that depends only logarithmically on the number of leaves and is useful when the maximum out-degree is small. We use it to convert the trees found by \JumpGreedy\ into schedules (Section~\ref{sec:ovdtm}).

\begin{restatable}[{after Ravi~\cite[Claim~1]{ravi1994rapid}}]{lemma}{lemravi}\label{lem:ravi}
Let $T$ be an out-arborescence rooted at an informed vertex, with
height $h$, nonempty leaf set $L$, and maximum out-degree at most
$d\ge2$. If each vertex calls its children in non-increasing order
of the number of leaves in their subtrees, all vertices of $T$
are informed within $h+d\log|L|/\log d$ rounds. The same bound holds
with $|V(T)|$ in place of $|L|$ when children are ordered by
non-increasing subtree size.
\end{restatable}

Ravi proves the bound using vertex counts; the same induction works with leaf counts. The idea is to call children with more leaves below them first. If a vertex has \(m\) leaves below it, its \(i\)-th child has at most \(m/i\) leaves in its subtree. That child waits \(i-1\) rounds longer than the first child, but the smaller leaf count reduces its inductive completion-time bound by \(d\log i/\log d\). Since this reduction is at least \(i-1\) for \(1\le i\le d\), it offsets the delay. Thus, each child's later start is offset by its smaller subtree.

\paragraph{A fallback schedule for a large optimum.} 
When the optimum is at least \(k\), a simple schedule already gives a \(2\)-approximation: Construct a shortest-path out-arborescence rooted at \(s\) and prune all branches containing no terminal. Its height is at most \(\mathrm{OPT}\), and it has at most \(k\) leaves. By Lemma~\ref{lem:busy}, the busy schedule therefore informs all terminals within \(\mathrm{OPT}+k\) rounds (Lemma~\ref{lem:fallback}). For \(k\)-multicast, the same argument applies to the \(k\) terminals nearest to \(s\), whose distances are also at most \(\mathrm{OPT}\). This fallback gives a \(2\)-approximation whenever \(\mathrm{OPT}\ge k\), so it suffices to analyze our algorithms under the assumption \(\mathrm{OPT}<k\).

\section{Directed multicast on general digraphs}
\label{sec:ovdtm}

We present \JumpGreedy, the algorithm of Theorem~\ref{thm:dtmcomb}. For every integer parameter \(\ell\ge2\), it computes a schedule of length \(O(\ell^3\log k)(\mathrm{OPT}+k^{1/\ell})\) in time \(n^{O(\ell)}\). We first define jump arborescences and show that one with small depth and small vertex loads exists. We then describe how the algorithm searches for such a structure.
Throughout this section, \(\eta=2\ell-1\), and we assume \(\mathrm{OPT}<k\); otherwise, the fallback schedule gives a \(2\)-approximation (Lemma~\ref{lem:fallback}). Appendix~\ref{sec:dtm} contains the pseudocode (Algorithms~\ref{alg:greedy} and~\ref{alg:dtm}) and the full proofs.

\subsection{Jump arborescences and loads}
\label{sec:ovjump}

By Lemma~\ref{lem:ravi}, a tree reaching the terminals gives a short schedule if its height and maximum out-degree are small. Our algorithm searches for a shallow tree whose edges represent short walks in \(G\). For each vertex of \(G\), we add up the number of times these walks leave it, including departures from intermediate vertices. This total, called its load, is additive over the walks and bounds its out-degree in their union.

Fix an integer \(\theta\ge1\) bounding the length of each walk. The algorithm tries every \(\theta\in\{1,\ldots,k-1\}\) as a guess for \(\mathrm{OPT}\); the analysis considers the guess \(\theta=\mathrm{OPT}\).

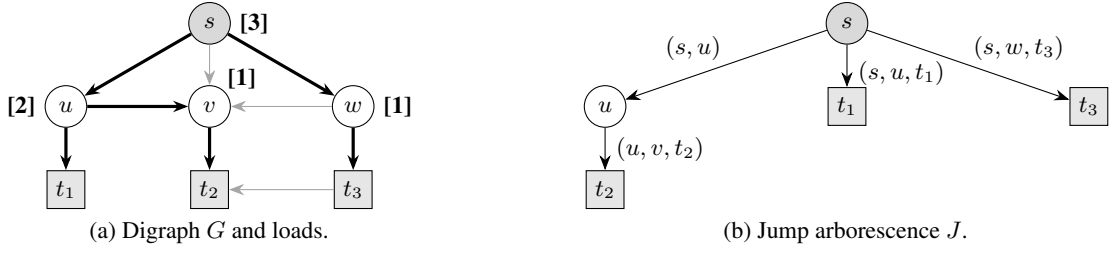
\begin{figure}[t]
\centering
\captionsetup[subfigure]{font=small,skip=4pt}
\tikzset{
  v/.style={circle,draw,inner sep=1pt,minimum size=5.5mm,font=\small},
  t/.style={rectangle,draw,fill=black!10,inner sep=2pt,minimum size=5mm,font=\small},
  ld/.style={font=\small\bfseries},
  jmp/.style={->,very thick},
  unused/.style={->,black!35},
  pathlabel/.style={font=\small,inner sep=1pt}}
\begin{subfigure}[t]{0.44\linewidth}
\centering
\begin{tikzpicture}[>={Stealth[length=2mm]}]
  \node[v,fill=black!15] (s) at (0,0) {$s$};
  \node[v] (u) at (-1.9,-1.1) {$u$};
  \node[v] (v) at (0,-1.1) {$v$};
  \node[v] (w) at (1.9,-1.1) {$w$};
  \node[t] (t1) at (-1.9,-2.2) {$t_1$};
  \node[t] (t2) at (0,-2.2) {$t_2$};
  \node[t] (t3) at (1.9,-2.2) {$t_3$};
  \draw[unused] (s) -- (v);
  \draw[unused] (w) -- (v);
  \draw[unused] (t3) -- (t2);
  \draw[jmp] (s) -- (u);
  \draw[jmp] (s) -- (w);
  \draw[jmp] (u) -- (t1);
  \draw[jmp] (u) -- (v);
  \draw[jmp] (v) -- (t2);
  \draw[jmp] (w) -- (t3);
  \node[ld,anchor=west] at (s.east) {[3]};
  \node[ld,anchor=east] at (u.west) {[2]};
  \node[ld,anchor=south west,inner sep=1pt] at (v.north east) {[1]};
  \node[ld,anchor=west] at (w.east) {[1]};
\end{tikzpicture}
\caption{Digraph $G$ and loads.}
\end{subfigure}\hfill
\begin{subfigure}[t]{0.54\linewidth}
\centering
\begin{tikzpicture}[>={Stealth[length=2mm]}]
  \node[v,fill=black!15] (S) at (0,0) {$s$};
  \node[v] (U) at (-3.2,-1.1) {$u$};
  \node[t] (T1) at (0,-1.1) {$t_1$};
  \node[t] (T3) at (3.2,-1.1) {$t_3$};
  \node[t] (T2) at (-3.2,-2.2) {$t_2$};
  \draw[->] (S) -- node[pathlabel,above left=1pt,pos=0.5] {$(s,u)$} (U);
  \draw[->] (S) -- node[pathlabel,right=3pt,pos=0.68] {$(s,u,t_1)$} (T1);
  \draw[->] (S) -- node[pathlabel,above right=1pt,pos=0.5] {$(s,w,t_3)$} (T3);
  \draw[->] (U) -- node[pathlabel,right=3pt] {$(u,v,t_2)$} (T2);
\end{tikzpicture}
\caption{Jump arborescence $J$.}
\end{subfigure}
\caption{A jump arborescence of depth $2$ with $\theta=2$.
Squares denote terminals. Edge labels in $J$ specify the jumps in $G$;
bold arcs are used by the jumps, and bracketed numbers are
vertex loads. The jumps $(s,u)$ and $(s,u,t_1)$ both traverse the arc
$(s,u)$. %
The maximum load is $\Lambda(J)=3$.}
\label{fig:jump}
\end{figure}

\begin{restatable}{definition}{defjump}\label{def:jump} A \emph{jump arborescence} is a rooted tree \(J\) of depth at most \(\eta\), with edges directed away from the root and nodes labelled by vertices of \(G\). The root is labelled \(s\), every terminal labels at least one node, and a vertex of \(G\) may label several nodes. Each edge whose parent and child have labels \(u\) and \(w\) carries a \emph{jump}: a directed walk from \(u\) to \(w\) with at most \(\theta\) arcs. When \(u=w\), this is the empty walk.
The \emph{load} \(L_J(v)\) of a vertex \(v\in V\) is the total number of departures from \(v\) along all jumps, counting repeated traversals both within a jump and across different jumps. We write $\Lambda(J)=\max_{v\in V}L_J(v)$ 
for the maximum load.
\end{restatable}
Note that loads belong to vertices of \(G\), rather than individual nodes of \(J\). Different jumps may share vertices and arcs, so their union need not be a tree. In Figure~\ref{fig:jump}, the maximum load is \(3\), attained at \(s\). The intermediate vertices \(v\) and \(w\) also have positive load, even though neither labels a node of \(J\).

The procedure \Schedule\ converts \(J\) into a schedule. It takes the union of the arcs traversed by the jumps, extracts a breadth-first out-arborescence rooted at \(s\), and prunes branches containing no terminal. The resulting tree has height at most \(\eta\theta\): each terminal is reached by at most \(\eta\) jumps, each of length at most \(\theta\). Its out-degree at each vertex \(v\) is at most \(L_J(v)\), and it has at most \(k\) leaves (Lemma~\ref{lem:schedtree}).

Apply Lemma~\ref{lem:ravi} with the degree bound
$
d=\max\{\Lambda(J),k^{1/\ell},2\}.
$
Since \(\log d\ge(\log k)/\ell\), we have \(d\log k/\log d\le\ell d\). Thus, the schedule informs all terminals within
$
\eta\theta+\ell\max\{\Lambda(J),k^{1/\ell},2\}
$
rounds (Lemma~\ref{lem:jumpsched}). For \(k\)-multicast, the same conversion applies after retaining any \(k\) covered terminals and pruning to them.

For the correct guess \(\theta=\mathrm{OPT}\), the height contribution is \(\eta\theta<2\ell\,\mathrm{OPT}\). To obtain Theorem~\ref{thm:dtmcomb}, it therefore suffices to find \(J\) with
$
\Lambda(J)=O(\ell^2\log k)(\mathrm{OPT}+k^{1/\ell}).
$
The next subsection shows that a jump arborescence with maximum load \(O(\ell)(\mathrm{OPT}+k^{1/\ell})\) exists. The recursive greedy algorithm finds one whose maximum load is within an \(O(\ell\log k)\) factor of this bound.

\subsection{Height reduction with bounded vertex loads}
\label{sec:ovhr}

An optimal schedule gives an arborescence whose height and maximum out-degree are at most \(\mathrm{OPT}\). We could regard each arc of this tree as a jump, but the resulting jump arborescence could have depth \(\mathrm{OPT}\). %
To obtain the desired running time, we need a jump arborescence with depth bounded by \(2\ell-1\), independently of \(\mathrm{OPT}\).
We obtain this jump arborescence by allowing a single jump to follow several arcs of the original tree. Fewer jumps are then needed along each source-to-terminal path. However, different jumps may reuse the same vertices, so reducing depth can increase their loads. The following height reduction lemma shows that we can make the tree shallow while keeping every vertex’s load small.
The construction starts from an optimal schedule, which the algorithm does not know. Its purpose is to establish that a good jump arborescence exists. Later, we compare the greedy algorithm’s choices with this arborescence to bound the loads of the solution it finds.

\begin{restatable}{lemma}{lemhr}\label{lem:hr}
Let \(T\) be an out-arborescence rooted at \(s\) that contains \(K\), with \(k=|K|\ge2\), height at most \(\theta\), and maximum out-degree at most \(d\). For every integer \(\ell\ge1\), let \(\rho=\lceil k^{1/\ell}\rceil\). There is a jump arborescence \(J\) of depth at most \(2\ell-1\), whose jumps are (possibly empty) directed paths in \(T\), such that
$
\Lambda(J)\le(\ell-1)\max\{d,\rho-1\}+\rho.
$ \end{restatable}

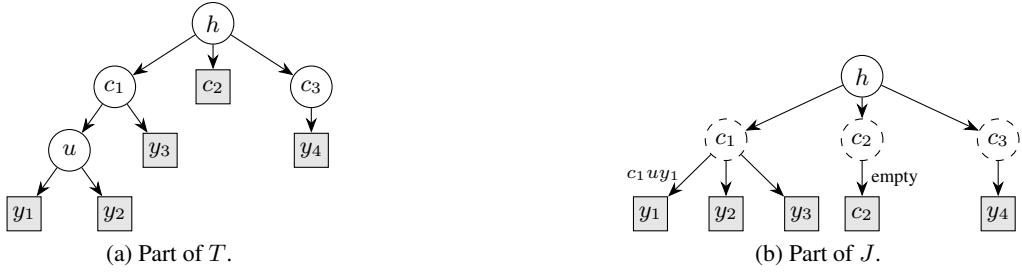
\begin{figure}[t]
\centering
\captionsetup[subfigure]{font=small,skip=4pt}
\begin{subfigure}[b]{0.4\linewidth}
\centering
\begin{tikzpicture}[yscale=0.7,>={Stealth[length=2mm]},
  v/.style={circle,draw,inner sep=1pt,minimum size=5.5mm,font=\small},
  it/.style={rectangle,draw,fill=black!10,inner sep=1.5pt,minimum size=4.5mm,font=\small}]
  \node[v] (h) at (-1.3,0) {$h$};
  \node[v] (c1) at (-2.6,-1.2) {$c_1$};
  \node[it] (c2) at (-1.3,-1.2) {$c_2$};
  \node[v] (c3) at (0,-1.2) {$c_3$};
  \node[v] (u) at (-3.2,-2.4) {$u$};
  \node[it] (y3) at (-2.0,-2.4) {$y_3$};
  \node[it] (y1) at (-3.8,-3.6) {$y_1$};
  \node[it] (y2) at (-2.6,-3.6) {$y_2$};
  \node[it] (y4) at (0,-2.4) {$y_4$};
  \draw[->] (h) -- (c1); \draw[->] (h) -- (c2); \draw[->] (h) -- (c3);
  \draw[->] (c1) -- (u); \draw[->] (c1) -- (y3);
  \draw[->] (u) -- (y1); \draw[->] (u) -- (y2);
  \draw[->] (c3) -- (y4);
\end{tikzpicture}
\caption{Part of $T$.}
\end{subfigure}\hfill
\begin{subfigure}[b]{0.56\linewidth}
\centering
\begin{tikzpicture}[yscale=0.7,>={Stealth[length=2mm]},
  v/.style={circle,draw,inner sep=1pt,minimum size=5.5mm,font=\small},
  it/.style={rectangle,draw,fill=black!10,inner sep=1.5pt,minimum size=4.5mm,font=\small}]
  \node[v] (H) at (5.4,0) {$h$};
  \node[v,dashed] (M1) at (3.6,-1.2) {$c_1$};
  \node[v,dashed] (M2) at (5.4,-1.2) {$c_2$};
  \node[v,dashed] (M3) at (7.2,-1.2) {$c_3$};
  \node[it] (Y1) at (2.6,-2.6) {$y_1$};
  \node[it] (Y2) at (3.6,-2.6) {$y_2$};
  \node[it] (Y3) at (4.6,-2.6) {$y_3$};
  \node[it] (C2) at (5.4,-2.6) {$c_2$};
  \node[it] (Y4) at (7.2,-2.6) {$y_4$};
  \draw[->] (H) -- (M1); \draw[->] (H) -- (M2); \draw[->] (H) -- (M3);
  \draw[->] (M1) -- node[left,font=\scriptsize]{$c_1uy_1$} (Y1);
  \draw[->] (M1) -- (Y2); \draw[->] (M1) -- (Y3);
  \draw[->] (M2) -- node[right,font=\scriptsize]{empty} (C2);
  \draw[->] (M3) -- (Y4);
\end{tikzpicture}
\caption{Part of $J$.}
\end{subfigure}
\caption{One stage of the height reduction with $\rho=5$.
Squares denote items from the previous stage, and dashed circles denote
intermediate nodes. The hub $h$ connects to its five assigned items
through its children $c_1,c_2,c_3$. %
Since $c_2$ is itself an item, its two occurrences in $J$ are connected
by an empty jump. The jumps $(c_1,u,y_1)$ and $(c_1,u,y_2)$ both leave
$u$, giving it load $2$; the three item paths from $c_1$ give it load $3$.
Both loads are below $\rho$.}
\label{fig:hr}
\end{figure}

The construction works in $\ell-1$ stages and keeps a set of
\emph{items}, initially the terminals. At each stage, we process \(T\) bottom-up. A vertex collects the items forwarded by its children, together with itself if it is an item. If it has collected at least \(\rho\) items, it becomes a \emph{hub}, retains those items, and forwards nothing. Otherwise, it forwards its collected items to its parent. The source is always a hub, regardless of how many items it receives. The hubs become the items for the next stage.
We represent each hub and the items it retains in \(J\) using at most two layers of jumps. Suppose a hub \(h\) receives items through a child \(c\) in \(T\). We add an intermediate node labelled \(c\), reached from the node labelled \(h\) by the one-arc jump \((h,c)\). From that intermediate node, we add jumps following the paths in \(T\) to the items forwarded by \(c\) (see Figure~\ref{fig:hr}). If the hub itself is an item, its item occurrence is attached by an empty jump. All jumps follow \(T\) away from the source and have length at most \(\theta\).
Every hub other than \(s\) receives at least \(\rho\) items, so a stage starting with \(x\) items produces at most \(x/\rho+1\) hubs. Since \(k\le\rho^\ell\), at most \(\rho+1\) hubs remain after \(\ell-1\) stages. We connect a root labelled \(s\) directly to them by jumps along \(T\). Each stage contributes at most two edges to a path in \(J\), and the final source connection contributes one. Hence the depth is at most \(2(\ell-1)+1=2\ell-1\).

The intermediate nodes keep the hubs' loads small: without them, a hub would start a separate jump to each item it retains, up to \(d(\rho-1)\) of them. At a given stage, a hub initiates at most \(d\) one-arc jumps to its children. The longer paths from those children to their items never leave another hub, because that hub would have retained the items. At a non-hub vertex, these paths cause at most \(\rho-1\) departures: each departure corresponds to an item forwarded through that vertex. Thus, a stage adds at most \(\max\{d,\rho-1\}\) to any vertex’s load. Among the final hubs is \(s\) itself, so the source connection adds at most \(\rho\) nonempty jumps. Summing gives the stated load bound; Appendix~\ref{sec:hr} contains the full proof.

The height-reduction results of~\cite{zelikovsky1997series,helvig2001improved} control the total cost of the reduced tree. In contrast, Lemma~\ref{lem:hr} bounds how often the jumps leave each vertex of the original graph, including vertices inside the jumps. This simultaneous per-vertex guarantee prevents the load from becoming concentrated at a single caller.
 
Consider the tree \(T^*\) formed by an optimal schedule, after removing branches containing no terminal. Its height and maximum out-degree are both at most \(\mathrm{OPT}\). Applying the height-reduction lemma with \(\theta=\mathrm{OPT}\) gives a jump arborescence \(J^*\) in which every vertex has load at most \(\ell(\mathrm{OPT}+\rho+1)\), where \(\rho=\lceil k^{1/\ell}\rceil\) as in Lemma~\ref{lem:hr}. We denote this upper bound by \(\Lambda\).
The jumps in $J^*$ use only vertices of \(T^*\). Since \(T^*\) is the union of \(k\) source-to-terminal paths, each of length at most \(\mathrm{OPT}<k\), it has at most
$
1+k\,\mathrm{OPT}\le k^2
$ vertices. Consequently, \(J^*\) has positive load on at most \(k^2\) vertices. This support bound will allow the potential analysis to depend on \(k\), rather than \(n\).

\subsection{The algorithm}
\label{sec:ovgreedy}
\label{sec:ovunfold}

In this subsection we describe the algorithm: the tree on which
it searches, the potential that prices its choices, the recursive greedy
procedure \Greedy{} (Algorithm~\ref{alg:greedy}) and the main algorithm
\JumpGreedy\ (Algorithm~\ref{alg:dtm}).
\paragraph{The unfolding.}
To search for a jump arborescence, we organize the possible sequences of jump endpoints into a rooted tree, called the \emph{unfolding} \(\mathcal U=\mathcal U_{\eta,\theta}\). Its nodes are sequences \((s,w_1,\ldots,w_j)\), where \(0\le j\le\eta\) and each vertex can reach the next within \(\theta\) arcs. The parent of a sequence is obtained by deleting its last vertex, and that last vertex is the \emph{label} of the sequence. Thus, a path in \(\mathcal U\) records a possible sequence of jump endpoints, while the actual walks connecting them remain to be chosen. The unfolding has at most \(2n^\eta=n^{O(\ell)}\) nodes.

We may assume that siblings in a jump arborescence have distinct labels. Otherwise, merge siblings with the same label, retaining one incoming jump and attaching their children to the merged node. This increases neither depth nor loads. After these merges, the jump arborescence is a rooted subtree of \(\mathcal U\), with a jump assigned to each edge. Finding a jump
arborescence is therefore a \emph{group Steiner tree} problem on the tree
$\mathcal U$: all occurrences of a terminal form a group, and we must select at least one occurrence from each group. Our objective, however, is to minimize the maximum vertex load. %
Appending chains of nodes with empty jumps, we may
also assume that the terminals label leaves of $\mathcal U$.

\paragraph{Structures and the potential.} 
The algorithm builds partial solutions called structures. A structure \(S\) rooted at \(\sigma\) follows parent–child edges of \(\mathcal U\), with a jump assigned to each edge. It may contain multiple copies of an unfolding node. Its load vector \(L_S\) records the departures made by all its jumps. It covers a terminal \(t\) if it contains a leaf of \(\mathcal U\) labelled \(t\); multiple occurrences of the same terminal count only once.

Let \(L\) denote the current loads of the vertices of \(G\), which we call the \emph{state}. We measure the cost of adding \(S\) by the increase
\[
\Delta_L(S)=\Phi(L+L_S)-\Phi(L),
\qquad
\Phi(L)=\sum_{v\in V}\bigl(e^{L(v)/\Gamma}-1\bigr),
\]
where the main algorithm fixes the scale \(\Gamma>0\). As in online load balancing~\cite{aspnes1997online}, a departure from a heavily loaded vertex carries a larger penalty, encouraging subsequent choices to use other routes. The potential controls every individual load: since each vertex’s contribution is at most the total potential,
$e^{L(v)/\Gamma}-1\le\Phi(L),$ so, $L(v)\le\Gamma\ln(1+\Phi(L)).$
Thus, to keep every load at \(O(\Gamma\log k)\), it suffices to show that the potential remains polynomial in \(k\).
For fixed \(L\), we can compute a jump between two endpoints that minimizes the potential increase. Removing cycles cannot increase its loads, so a \emph{cheapest jump} can be taken to be a path. Along a path, each departure from \(v\) contributes \(e^{L(v)/\Gamma}(e^{1/\Gamma}-1)\). These vertex weights are positive, so a minimum-weight walk with at most \(\theta\) arcs, which dynamic programming over the number of arcs finds, is a path, and it is a cheapest jump.

\paragraph{The recursive greedy procedure.} 

A call \Greedy{}\((\sigma,z,L,R)\) receives an unfolding node \(\sigma\), a demand \(z\), a state \(L\), and a set \(R\) of terminals still available for coverage. Its goal is to cover \(z\) terminals from \(R\). For \DTM, the demand is \(z=|R|\); for \(k\)-multicast, the demand may be smaller. The call returns its chosen structures as an ordered list, allowing its parent to use only an initial portion of that list.

Each iteration generates candidates by considering every child \(\tau\) of \(\sigma\). First, it computes a cheapest jump \(P\) between their labels. If only one level remains, \(\tau\) is a leaf of \(\mathcal U\), and this jump alone forms a candidate. Otherwise, it makes one recursive call at \(\tau\), with the current residual set and demand, and with state \(L+L_P\). Each nonempty prefix of the returned list, together with the incoming jump \(P\), forms a candidate at \(\sigma\).

A candidate \(C\) receives credit for the distinct residual terminals it covers, capped at the number still needed:
$\mathrm{cnt}(C)=\min\{|\mathrm{cov}(C)\cap R|,z\},
$
where \(\mathrm{cov}(C)\) is the set of terminals that \(C\) covers, and \(R\) and \(z\) now denote the current residual set and demand. Among candidates with positive count, the procedure chooses one minimizing
$\frac{\Delta_L(C)}{\mathrm{cnt}(C)}.$
It appends this candidate to its list, adds its loads to the state, removes its covered terminals from \(R\), and reduces \(z\) by its count. It repeats until the demand is met or no candidate covers a residual terminal.
Considering every prefix lets the parent choose how much of a recursive solution to take. Unlike the recursive greedy algorithms of~\cite{charikar1999dst} and of Chekuri, Even and Kortsarz~\cite{chekuri2006greedy}, our procedure does not guess separate coverage demands for its recursive calls.

\paragraph{
The main algorithm.} \JumpGreedy\ first computes the fallback schedule. It then tries each guess \(\theta\in\{1,\ldots,k-1\}\). For each guess, it sets
$\Lambda=\ell(\theta+\rho+1)$ with $\rho=\lceil k^{1/\ell}\rceil$, $\mu=20\eta\Lambda$ and $\Gamma=1/\ln(1+1/\mu)$. Then $e^{L(v)/\Gamma}=(1+1/\mu)^{L(v)}$, so the increases and the ratios that the algorithm compares are rational numbers, computed exactly, and $\mu\le\Gamma\le\mu+1$.
When \(\theta=\mathrm{OPT}\), the height-reduction lemma guarantees a comparison solution with maximum load at most \(\Lambda\). The algorithm calls \Greedy\ at the root of \(\mathcal U_{\eta,\theta}\), with demand \(k\), zero initial loads, and terminal set \(K\).
The returned structures are joined at their common root. If together they cover at least \(k\) terminals, \JumpGreedy\ converts them into a schedule using \Schedule. For \DTM, this means covering all terminals; for \(k\)-multicast, \Schedule\ retains any \(k\) covered terminals. The algorithm returns the shortest schedule obtained, including the fallback.
Each call makes at most \(k\) iterations, since every iteration reduces its remaining demand by at least one. In each iteration, it examines at most \(n\) children and makes one shortest-path computation and at most one recursive call per child. The recursion has depth \(\eta\), giving a bound of \((kn)^\eta\operatorname{poly}(n)=n^{O(\ell)}\) on the arithmetic operations. They act on rational numbers with \(n^{O(\ell)}\) bits (Appendix~\ref{sec:comb}), so the running time is \(n^{O(\ell)}\).

The analysis of \JumpGreedy\ is in Appendix~\ref{sec:comb}.

\section{Planar digraphs and root-path separable classes}
\label{sec:ovplanar}

In this section we present our algorithm for planar digraphs, \LPSep, in its general
form, for every class of graphs with balanced separators made of a few
paths from the root of any spanning tree. Appendix~\ref{sec:planar} gives the pseudocode, as
Algorithms~\ref{alg:pllevelsep} and~\ref{alg:planar}, and proves
Theorems~\ref{cor:plplanar} and~\ref{thm:plclass},
Corollary~\ref{cor:pltw} and the lemmas.

\subsection{Root-path separable classes}
\label{sec:ovplclass}

In this subsection we define the classes of graphs on which the algorithm
works and state its guarantee. The \emph{underlying graph} of a digraph is
the undirected graph on its vertices in which two distinct vertices are
adjacent if an arc joins them. A class of graphs is \emph{minor-closed} if
it contains every graph isomorphic to a minor of one of its graphs, that
is, to a graph obtained from one of them by deleting vertices and edges
and contracting edges. For a spanning tree $A$ of a graph, rooted at a
vertex $r$, the \emph{root path} $A[r,v]$ is the vertex set of the path in
$A$ from $r$ to $v$, and $A[r,Z]$ is the union of the root paths to the
vertices of a set $Z$; \emph{vertex weights} $\mu$ are nonnegative reals on
the vertices, and $\mu(V')$ is their sum over a set $V'$. Our algorithm for
planar digraphs uses planarity only through balanced separators made of
three root paths of any spanning tree, in the minors of the underlying
graph that its recursion creates. It therefore works for every class with
the following property:

\begin{restatable}{definition}{defplclass}\label{def:plclass}
Let $p\ge1$ be an integer. A connected graph $H$ is \emph{$p$-root-path
separable} if for every vertex $r$ of $H$, every spanning tree $A$ of $H$ rooted at $r$ and all
vertex weights $\mu$ on $H$, there is a nonempty set $Z$ of at most $p$
vertices of $H$ such that every component of $H-A[r,Z]$ has weight at most
$\mu(V(H))/2$. A class $\mathcal G$ of graphs is \emph{$p$-root-path
separable} if it is minor-closed, every connected graph in $\mathcal G$ is
$p$-root-path separable,
and there is a deterministic polynomial-time algorithm that computes such a
set $Z$ for every connected graph $H\in\mathcal G$, vertex $r$ of $H$,
spanning tree $A$ of $H$ rooted at $r$ and rational vertex weights $\mu$.
\end{restatable}

For fixed $p$, trying all sets $Z$ of at most $p$ vertices gives such an
algorithm, in time $n^{p+O(1)}$ on graphs with $n$ vertices. Our algorithm
achieves the following ratio on every such class:

\begin{restatable}{theorem}{thmplclass}\label{thm:plclass}
Let $p\ge1$ be an integer and $\mathcal G$ a $p$-root-path separable class.
When run with the separator algorithm of Definition~\ref{def:plclass} for $\mathcal G$,
\LPSep\ is a deterministic polynomial-time algorithm that, given a digraph
whose underlying graph lies in $\mathcal G$, a source $s$ and a set $K$ of
$k\ge2$ terminals reachable from $s$, computes a \DTM\ schedule of length
at most $14\min\{p,k\}(\bar d+1)^2\,\OPT$, where $\bar d=\lfloor\log_2k\rfloor$.
\end{restatable}

Since $(\bar d+1)^2\le4\log_2^2k$ for $k\ge2$, the ratio is $O(p\log^2k)$, and,
as in Theorem~\ref{thm:dtmcomb}, it depends only on $k$ and on the class.
The planar graphs form a $3$-root-path separable class, by a lemma of
Thorup~\cite[Lemma~2.3]{thorup2004compact} (Lemma~\ref{lem:plplanarclass}),
so Theorem~\ref{thm:plclass} gives Theorem~\ref{cor:plplanar}. The graphs of
treewidth at most $w$ (the least width of a tree decomposition, defined in
Appendix~\ref{sec:plplanartw}) form a $(w+1)$-root-path separable class
(Lemma~\ref{lem:pltwsep}), so Theorem~\ref{thm:plclass} also gives:

\begin{restatable}{corollary}{corpltw}\label{cor:pltw}
For every fixed integer $w\ge1$, a deterministic polynomial-time algorithm
computes, for every \DTM\ instance with $k\ge2$ terminals whose underlying
graph has treewidth at most $w$, a schedule of length at most
$14\min\{w+1,k\}(\lfloor\log_2k\rfloor+1)^2\,\OPT$.
\end{restatable}

\subsection{The algorithm}
\label{sec:ovplrec}
\label{sec:ovpllp}

In this subsection we describe \LPSep\ completely. It follows the recursion
of Friggstad and Mousavi~\cite{friggstad2023planar} for directed Steiner
tree on planar digraphs, which rests on a lemma of
Thorup~\cite[Lemma~2.3]{thorup2004compact}: for every spanning tree of a
connected planar graph, rooted at any vertex, some three paths of the tree
from its root form a balanced separator. The recursion adds these paths to the solution,
contracts them into the root, and recurses on the components, the
\emph{instances} of the next level; with weight $1$ on the terminals, every
level halves the number of terminals of an instance, so there are at most
$\bar d+1$ levels. For multicast, what matters is how often a vertex enters
the separators of one level, and a linear program guides these choices. The
fallback schedule covers the case $\OPT\ge k$ (Lemma~\ref{lem:fallback}),
so the algorithm only tries the guesses $\theta\in\{1,\dots,k-1\}$ for $\OPT$.
Separators consist of at most $\sigma=\min\{p,k\}$ root paths, and the
algorithm uses the parameters $\epsilon=1/(2(\bar d+1))$ and
$\beta=7/(4\epsilon)$.

\paragraph{The linear program.} For a guess $\theta$, the linear program
$\LP(\theta)$ (Definition~\ref{def:pllp}) routes a unit flow from $s$ to
every terminal along paths with at most $\theta$ arcs, under arc capacities
$x_e$ whose sum out of every vertex is at most $\theta$. It is feasible for
$\theta\ge\OPT$, as the arcs of an optimal schedule show. It has a variable for
every such path, but a linear program of polynomial size, in which the flow
of every terminal is a flow in a copy of $G$ with $\theta$ layers, has the same
feasible capacities, so $\LP(\theta)$ can be solved in polynomial time
(Lemma~\ref{lem:pllp}). The algorithm takes the least $\theta$ for which
$\LP(\theta)$ is feasible, so $\theta\le\OPT$, and a solution $x$.

\paragraph{Instances.} An instance consists of a \emph{root set} $R$ of
vertices, which the schedule informs in earlier phases, and a
\emph{region} still to be reached, with the terminals in it
(Definition~\ref{def:plinst}). The arcs from $R$ into the region are its
\emph{entry arcs}, and contracting $R$ into a root $r$ gives the
\emph{instance digraph}, whose underlying graph is a minor of that of $G$
(Lemma~\ref{lem:plstruct}\ref{lem:plstruct:minor}); this is why
Definition~\ref{def:plclass} asks for minor-closed classes. The first
instance has the root set $\{s\}$ and all terminals. An instance at depth
$d$ of the recursion gets the capacities $x$ scaled by $(1-\epsilon)^{-d}$; as
no level is deeper than $\bar d$, the scaling costs a factor at most
$e^{1/2}$. The \emph{entry mass} $m_I(u)$ of a vertex $u$ of $G$ in an
instance $I$ is the scaled capacity of the entry arcs from $u$ into $I$.

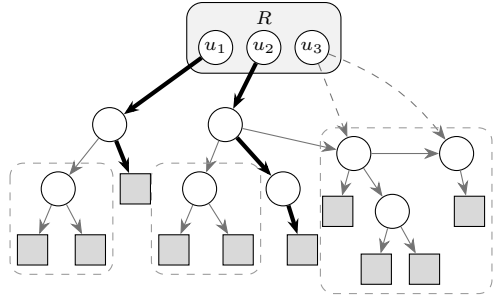
\begin{figure}[t]
\centering
\begin{tikzpicture}[scale=0.85,>={Stealth[length=2mm]},
  v/.style={circle,draw,inner sep=0pt,minimum size=4.5mm,font=\scriptsize},
  t/.style={rectangle,draw,fill=black!15,inner sep=0pt,minimum size=4mm,font=\scriptsize},
  piece/.style={->,line width=1.6pt},
  tree/.style={->,black!55},
  exp/.style={->,dashed,black!55},
  lab/.style={font=\scriptsize}]
  \draw[rounded corners=6pt,fill=black!5] (-1.2,-0.55) rectangle (1.2,0.55);
  \node[lab] at (0,0.32) {$R$};
  \node[v,fill=white] (u1) at (-0.75,-0.15) {$u_1$};
  \node[v,fill=white] (u2) at (0,-0.15) {$u_2$};
  \node[v,fill=white] (u3) at (0.75,-0.15) {$u_3$};
  \node[v] (a) at (-2.4,-1.35) {};
  \node[v] (b) at (-0.6,-1.35) {};
  \node[v] (c) at (1.4,-1.8) {};
  \node[v] (dd) at (3.0,-1.8) {};
  \node[v] (a1) at (-3.2,-2.35) {};
  \node[t] (a2) at (-2.0,-2.35) {};
  \node[v] (b1) at (-1.0,-2.35) {};
  \node[v] (b2) at (0.3,-2.35) {};
  \node[t] (c1) at (1.15,-2.7) {};
  \node[v] (c2) at (2.0,-2.7) {};
  \node[t] (d1) at (3.2,-2.7) {};
  \node[t] (a3) at (-3.6,-3.3) {};
  \node[t] (a4) at (-2.7,-3.3) {};
  \node[t] (b3) at (-1.4,-3.3) {};
  \node[t] (b4) at (-0.4,-3.3) {};
  \node[t] (b5) at (0.6,-3.3) {};
  \node[t] (c3) at (1.75,-3.6) {};
  \node[t] (c4) at (2.5,-3.6) {};
  \draw[piece] (u1) -- (a);
  \draw[piece] (u2) -- (b);
  \draw[exp] (u3) -- (c);
  \draw[exp] (u3) to[bend left=15] (dd);
  \draw[tree] (b) -- (c);
  \draw[tree] (c) -- (dd);
  \draw[tree] (a) -- (a1);
  \draw[piece] (a) -- (a2);
  \draw[tree] (b) -- (b1);
  \draw[piece] (b) -- (b2);
  \draw[tree] (c) -- (c1);
  \draw[tree] (c) -- (c2);
  \draw[tree] (a1) -- (a3);
  \draw[tree] (a1) -- (a4);
  \draw[tree] (b1) -- (b3);
  \draw[tree] (b1) -- (b4);
  \draw[piece] (b2) -- (b5);
  \draw[tree] (c2) -- (c3);
  \draw[tree] (c2) -- (c4);
  \draw[tree] (dd) -- (d1);
  \draw[black!40,rounded corners=5pt,dashed] (-3.95,-1.95) rectangle (-2.35,-3.68);
  \draw[black!40,rounded corners=5pt,dashed] (-1.75,-1.95) rectangle (-0.05,-3.68);
  \draw[black!40,rounded corners=5pt,dashed] (0.88,-1.4) rectangle (3.55,-3.98);
\end{tikzpicture}
\caption{One instance, drawn with its root set $R$ (the shaded box) in
place of the root $r$; the arcs from $R$ are the entry arcs. Bold arcs form
the piece, the union of two root paths of $A_j$; grey arcs are the other
arcs of $A_j$; dashed arcs are entry arcs that are not cheap; squares are
terminals; and the dashed boxes are the children. The vertex $u_3$ has a large
weight, so its entry arcs are not cheap; their heads still belong to
$H'_j$, as $A_j$ reaches them via the entry arc from $u_2$. The separator
$S_j$ consists of two root paths of $A_j$. Removing it leaves three
components with terminals, each with at most half of them: the children,
whose root sets contain $R$ and the vertices of the piece.}
\label{fig:plstep}
\end{figure}

\paragraph{Processing a level.} The algorithm processes the instances
level by level, and the instances of a level one after the other. During a
level, every vertex $u$ of $G$ has a \emph{load} $\ell(u)$, the number of
arcs from $u$ into the pieces chosen so far in the level, and a \emph{future
mass} $\bar m(u)$, its entry mass in the instances of the level that are
not processed yet; its \emph{weight} is
$\omega(u)=\exp(\ell(u)/\sigma+\beta\bar m(u))$. To process the $j$-th instance
$I_j$ of the level, the algorithm removes the masses of $I_j$ from the
future masses and computes the \emph{weighted entry mass}
$\Omega_j=\sum_u\omega(u)m_{I_j}(u)$; an entry arc of $I_j$ is \emph{cheap}
if its tail has weight at most $\Omega_j/\epsilon$, so that the instance may be
entered only through vertices that are not much busier than its other
possible entries. Then it takes three steps (Figure~\ref{fig:plstep}).
First, it keeps the part $H'_j$ of the instance digraph that $r$ reaches
within $\theta$ arcs through a cheap entry arc, and a breadth-first
out-arborescence $A_j$ of $H'_j$ from $r$ that enters only through cheap
entry arcs; its root paths have at most $\theta$ arcs. Second, it chooses a
\emph{separator} $S_j$ made of at most $\sigma$ root paths of $A_j$: if $I_j$
has a single terminal, the root path to it; otherwise, if $p\le k$, the root
paths to the vertices that the algorithm of Definition~\ref{def:plclass}
returns for the underlying graphs of $H'_j$ and $A_j$ with weight $1$ on the
terminals, and if $p>k$, the root paths to all terminals of $I_j$. Third, the arcs of
$A_j$ into $S_j$ form the \emph{piece} of the instance, a tree with at most
$\sigma$ leaves (Lemma~\ref{lem:plstruct}\ref{lem:plstruct:pieces}), whose arcs
from $r$ are added to the loads of their tails in $G$; and the components of
$H'_j-S_j$ that contain terminals become the children of the instance, whose
root sets also contain the vertices of the separator. The algorithm
computes the weights as rational numbers that are larger by a factor at most
$e^{1/100}$, which is why $\beta$ is $7/(4\epsilon)$ rather than $(e-1)/\epsilon$
(Appendix~\ref{sec:plalg}).

\paragraph{The schedule.} The pieces are informed level by level, in the
\emph{phase schedule} (Definition~\ref{def:plphase}).
The \emph{per-level load} $\ell_d(u)$ of a vertex $u$ of $G$ is the number
of arcs from $u$ into the pieces of the instances of level $d$, the value
of $\ell(u)$ at the end of level $d$. In phase $d$, every vertex first
calls, one per round, the heads of its arcs into the pieces of depth $d$;
below these heads, every piece is informed by the busy schedule of
Lemma~\ref{lem:busy}, as it is a tree of height less than $\theta$ with at most
$\sigma$ leaves. The vertices that make the first calls of phase $d$ lie in
root sets, so earlier phases have informed them, and the other callers lie
in one piece each. So the phase schedule is a valid schedule that informs
all terminals within
\[
  \sum_d\Bigl(\max_{u\in V(G)}\ell_d(u)+\theta+\sigma-2\Bigr)
\]
rounds (Lemma~\ref{lem:plphase}), the sum over the levels that have an
instance. \LPSep\ returns the shorter of the phase schedule and the
fallback schedule.

The analysis of \LPSep\ is in Appendix~\ref{sec:planalysis}.

\section{Open problems}
\label{sec:conclusion}

We close with the main questions that remain open.
The main open question is whether \DTM\ admits a polynomial-time \(\operatorname{polylog}(k)\)-approximation. \JumpGreedy\ achieves an \(O(\log^4 k)\) approximation in quasi-polynomial time. In polynomial time, it gives an \(O(\log k)\) approximation whenever \(\mathrm{OPT}\ge k^\epsilon\) for a fixed \(\epsilon\in(0,1)\). The remaining goal is therefore to obtain a \(\operatorname{polylog}(k)\) guarantee in polynomial time without this lower bound on \(\mathrm{OPT}\).

For planar digraphs, can the approximation ratio be improved from \(O(\log^2 k)\) to \(O(\log k)\), or even to a constant? Other directions are to obtain the \(O(\log^2 k)\) guarantee with a combinatorial algorithm and to extend the separator approach to \(k\)-multicast.
For directed broadcast, the gap between the \(O(\log n)\) approximation and the conditional \(\Omega(\sqrt{\log n})\) hardness bound of Elkin and Kortsarz~\cite{elkin2005combinatorial} remains open. Improving either bound would clarify how the difficulty of broadcast compares with that of multicast.

\section*{Acknowledgments}

We used Claude (Anthropic) to assist with developing and refining the general-digraph and planar algorithms, their potential-function analyses, and their proofs. A detailed account of this assistance appears in the AI methodology statement at the end of the paper.

\clearpage
\appendix
\section{Proofs for the preliminaries}
\label{sec:prelim}

This appendix proves Lemmas~\ref{lem:busy} and~\ref{lem:ravi}, the two
schedules along a given out-arborescence of Section~\ref{sec:ovsched}, and
states the form of the hardness of Set Cover from which our reductions
start. We use the notation of Section~\ref{sec:ovsched}.

The busy schedule is good when the arborescence has few leaves; we use it
for the fallback schedule of Appendix~\ref{sec:dtm}
(Lemma~\ref{lem:fallback}) and inside the gadgets of
Appendix~\ref{sec:hard}. We restate Lemma~\ref{lem:busy}:

\lembusy*

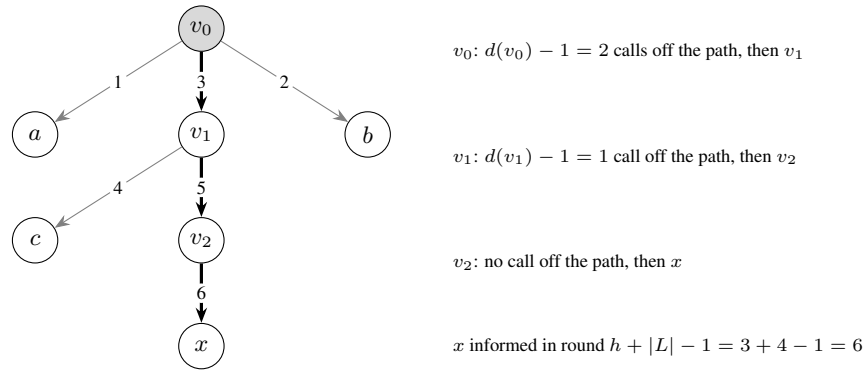
\begin{figure}[htb]
\centering
\begin{tikzpicture}[>={Stealth[length=2mm]},
  v/.style={circle,draw,inner sep=1pt,minimum size=6mm,font=\small},
  r/.style={circle,fill=white,inner sep=0.5pt,font=\scriptsize}]
  \node[v,fill=black!15] (v0) at (0,0) {$v_0$};
  \node[v] (v1) at (0,-1.4) {$v_1$};
  \node[v] (v2) at (0,-2.8) {$v_2$};
  \node[v] (x) at (0,-4.2) {$x$};
  \node[v] (a) at (-2.2,-1.4) {$a$};
  \node[v] (b) at (2.2,-1.4) {$b$};
  \node[v] (c) at (-2.2,-2.8) {$c$};
  \draw[->,very thick] (v0) -- node[r]{3} (v1);
  \draw[->,very thick] (v1) -- node[r]{5} (v2);
  \draw[->,very thick] (v2) -- node[r]{6} (x);
  \draw[->,black!50] (v0) -- node[r,text=black]{1} (a);
  \draw[->,black!50] (v0) -- node[r,text=black]{2} (b);
  \draw[->,black!50] (v1) -- node[r,text=black]{4} (c);
  \node[font=\scriptsize,anchor=west,align=left] at (3.2,-0.3)
    {$v_0$: $d(v_0)-1=2$ calls off the path, then $v_1$};
  \node[font=\scriptsize,anchor=west,align=left] at (3.2,-1.7)
    {$v_1$: $d(v_1)-1=1$ call off the path, then $v_2$};
  \node[font=\scriptsize,anchor=west,align=left] at (3.2,-3.1)
    {$v_2$: no call off the path, then $x$};
  \node[font=\scriptsize,anchor=west,align=left] at (3.2,-4.2)
    {$x$ informed in round $h+|L|-1=3+4-1=6$};
\end{tikzpicture}
\caption{The worst case in the proof of Lemma~\ref{lem:busy} for the leaf
$x$; numbers are the rounds of the calls. The arborescence has height
$h=3$ and the leaves $a$, $b$, $c$ and $x$. In each round the deepest
informed vertex $v$ on the path from the root to $x$ (bold) either calls
its child on the path, which happens at most $h$ times, or another child
(grey), which happens at most $d(v)-1$ times at each $v$ on the path.}
\label{fig:busy}
\end{figure}

\begin{proof}
Elkin and Kortsarz state the bound $h+|L|$; their argument gives one round
less. Fix a leaf $x$ and consider, after each round, the deepest informed vertex
$v$ on the path from the root to $x$ (Figure~\ref{fig:busy}). In each round, $v$ either calls its
child towards $x$, which happens at most $h$ times, or calls another
child. The second case happens at most $d(v)-1$ times for each $v$ on the
path. Since $\sum_v(d(v)-1)$ over the internal vertices of $Q$ equals
$|L|-1$, the leaf $x$ is informed within $h+|L|-1$ rounds.
\end{proof}

The second schedule is good when the arborescence has small out-degree,
however many leaves it has. It turns the jump arborescences of
Appendix~\ref{sec:dtm} into schedules (Lemma~\ref{lem:jumpsched}). We
restate Lemma~\ref{lem:ravi}:

\lemravi*

Ravi proves the bound with vertex counts; the version with leaves follows
by the same induction.

\begin{figure}[htb]
\centering
\begin{tikzpicture}[>={Stealth[length=2mm]},
  v/.style={circle,draw,inner sep=1pt,minimum size=6mm,font=\small},
  r/.style={circle,fill=white,inner sep=0.5pt,font=\scriptsize}]
  \node[v,fill=black!15] (root) at (0,0) {};
  \foreach \i/\x/\w in {1/-3.0/0.95,2/-1.0/0.75,3/0.8/0.55,4/2.3/0.4} {
    \node[v] (c\i) at (\x,-1.4) {$c_\i$};
    \draw[->] (root) -- node[r]{\i} (c\i);
    \draw (\x,-1.72) -- (\x-\w,{-1.72-2*\w}) -- (\x+\w,{-1.72-2*\w}) -- cycle;
    \node[font=\scriptsize] at (\x,{-1.95-2*\w}) {$w_\i\le w(r)/\i$};
  }
  \node[font=\scriptsize,align=center] at (-0.3,-4.4)
    {child $c_i$ is informed in round $i$, and its subtree\\
     needs at most $d\log(w(r)/i)/\log d+h-1$ further rounds};
  \begin{scope}[shift={(5.0,-3.6)}]
    \draw[->] (0,0) -- (3.4,0) node[right,font=\scriptsize] {$i$};
    \draw[->] (0,0) -- (0,3.1);
    \foreach \i in {1,2,3,4}
      \draw ({\i-1},0.06) -- ({\i-1},-0.06) node[below,font=\scriptsize] {\i};
    \draw[thick] (0,0) -- (3,{0.7*3}) node[right,font=\scriptsize] {$i-1$};
    \draw[thick,domain=1:4,samples=40] plot ({\x-1},{0.7*2*ln(\x)/ln(2)});
    \node[font=\scriptsize,anchor=west] at (3,{0.7*4+0.05}) {$d\log i/\log d$};
    \node[font=\scriptsize] at (1.5,3.2) {$d=4$};
  \end{scope}
\end{tikzpicture}
\caption{The inductive step in the proof of Lemma~\ref{lem:ravi}. Left: the
root calls its children in non-increasing order of their numbers of
leaves, so the subtree of the $i$-th child has $w_i\le w(r)/i$ leaves. Right, for $d=4$: the
delay $i-1$ of the $i$-th child (line) is at most the saving $d\log i/\log
d$ from its smaller subtree (curve), since the curve is concave, the two
agree at $i=1$, and at $i=d$ the curve is at $d\ge d-1$.}
\label{fig:ravi}
\end{figure}
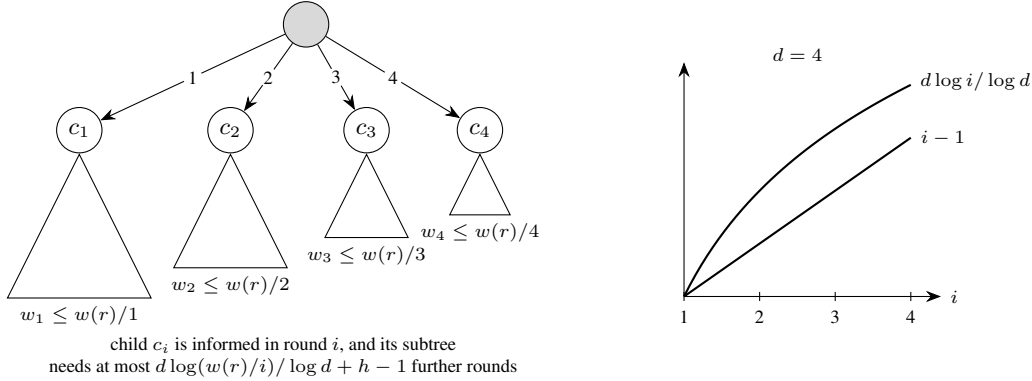

\begin{proof}
For a vertex $v$ of $T$ let $w(v)$ be the number of leaves of the subtree
of $v$ for the first statement of Lemma~\ref{lem:ravi}, and the number of
its vertices for the second.
In both cases $w(v)\ge1$, and $w(v)\ge w(c_1)+\dots+w(c_m)$ for the
children $c_1,\dots,c_m$ of $v$ (for vertex counts, $w(v)=1+w(c_1)+\dots+
w(c_m)$). Let every vertex call its children in non-increasing order of
$w$, and let the root $r$ be informed in round $0$. We show by induction on
$h$ that every vertex of $T$ is informed within $d\log w(r)/\log d+h$
rounds; this is Lemma~\ref{lem:ravi}, as $w(r)$ is $|L|$ or $|V(T)|$.

\emph{Base case, $h=0$.} Then $T$ is the single vertex $r$, informed in
round $0$, and the bound is at least $0$, as $w(r)\ge1$.

\emph{Inductive step, $h\ge1$} (Figure~\ref{fig:ravi}). Let the root have
children $c_1,\dots,c_m$ in the calling order, where $1\le m\le d$, and let
$w_i=w(c_i)$. As $w_1\ge\dots\ge w_m$ and $w_1+\dots+w_m\le w(r)$, we have
$iw_i\le w_1+\dots+w_i\le w(r)$, so $w_i\le w(r)/i$. The vertex $c_i$ is
informed in round $i$, and its subtree has out-degree at most $d$ and
height at most $h-1$. By the induction hypothesis, every vertex of this
subtree is informed within $d\log w_i/\log d+h-1$ further rounds, hence by
round
\[
  i+\frac{d\log(w(r)/i)}{\log d}+h-1\;=\;\frac{d\log w(r)}{\log d}+h+
  \Bigl(i-1-\frac{d\log i}{\log d}\Bigr)\;\le\;\frac{d\log w(r)}{\log d}+h,
\]
because $i-1\le d\log i/\log d$ for $1\le i\le d$. Indeed, the right side
is concave in $i$, the two sides are equal at $i=1$, and at $i=d$ the right
side is $d\ge d-1$.
\end{proof}

\paragraph{Set Cover.} An instance consists of a universe of $m$ elements
and sets $S_1,\dots,S_\nu$; a \emph{cover} is a family of sets whose union
is the universe. Our reductions start from the form of the hardness of
Set Cover in Theorem~\ref{thm:setcover}.
Feige~\cite[Lemma~4.1 and Theorem~4.4]{feige1998threshold} proved it under
the assumption $\mathrm{NP}\not\subseteq\mathrm{DTIME}(n^{O(\log\log n)})$, and the
reduction of Moshkovitz~\cite[Theorem~1.2]{moshkovitz2015projection} with
the parallel repetition theorem of Dinur and
Steurer~\cite{dinur2014analytical} makes it NP-hard. This form is not
stated in~\cite{moshkovitz2015projection,dinur2014analytical}, but it
follows from the construction. The reduction goes through Label Cover on a
bipartite graph with sides $X$ and $Y$, for a parameter $\alpha>0$. In the
completeness case the cover has one set for every vertex of $X$, and in the
soundness case every cover has more than $(1-2\alpha)|X|\ln|P|$
sets~\cite[Claim~4.10]{moshkovitz2015projection}, where the universe is
$Y\times P$ for a set $P$ with $|P|=\Theta(|Y|^{1/\alpha})$ (proof
of~\cite[Theorem~1.2]{moshkovitz2015projection}). As the Label Cover instances of~\cite[Corollary~4.8]{moshkovitz2015projection}
(Theorem~2.3 there, followed by Lemma~4.4) have bi-regular graphs with
polylogarithmic degrees and alphabets of polynomial size, the number of
sets, the number of elements and $|X|$ are polynomially related.

\begin{theorem}[{\cite[Theorem~1.2 and Claim~4.10]{moshkovitz2015projection}; \cite{dinur2014analytical}}]\label{thm:setcover}
For every $\epsilon\in(0,1)$ there is a constant $C$ such that the following is
NP-hard. Given a Set Cover instance with $m$ elements and $\nu\le m^C$ sets,
and an integer $c\ge m^{1/C}$, distinguish the case that a cover of size
$c$ exists from the case that every cover has more than $(1-\epsilon)c\ln
m$ sets.
\end{theorem}

\section{Directed multicast on general digraphs}
\label{sec:dtm}

This appendix presents our algorithm for directed multicast and proves
Theorem~\ref{thm:dtmcomb}, which we restate:

\thmdtmcomb*

The choice of $\ell$ trades the running time against the ratio, as
Corollary~\ref{cor:dtm}, restated, shows:

\cordtm*

The ratios depend on $k$ only; $n$ enters only the running time. The
proof uses height reduction as in Zelikovsky~\cite{zelikovsky1997series} and~\cite{charikar1999dst}, an unfolding into a tree, and
a variant of the recursive greedy algorithm of~\cite{charikar1999dst} and of Chekuri, Even and Kortsarz~\cite{chekuri2006greedy}, combined with the
exponential potential of online load balancing~\cite{aspnes1997online}.
Three points are new. We bound \emph{loads}, the number of times jumps leave a vertex of $G$,
instead of out-degrees; loads are additive, so greedy covering applies.
Our height reduction bounds the load of \emph{every} vertex of $G$, not only the
total cost. And our analysis of the greedy algorithm loses only a factor
$O(\ell\log k)$ in the loads; Appendix~\ref{sec:comb} explains the two
observations that make this possible.

Appendix~\ref{sec:dtmoverview} recalls the roles of the parameters of the
algorithm. Appendices~\ref{sec:jump}--\ref{sec:alg} define it precisely,
in pseudocode as Algorithms~\ref{alg:greedy} and~\ref{alg:dtm},
Appendices~\ref{sec:hr} and~\ref{sec:comb} analyse it, and
Appendix~\ref{sec:timeexp} proves Theorem~\ref{thm:te} and
Corollary~\ref{cor:te}, which give a second algorithm, through
degree-bounded Steiner trees.

\subsection{Overview of the algorithm}
\label{sec:dtmoverview}

Section~\ref{sec:ovdtm} explains the ideas behind the algorithm. Here we
recall the roles of its parameters, which
Appendices~\ref{sec:jump}--\ref{sec:comb} define and analyse. The
algorithm tries every guess $\theta\in\{1,\dots,k-1\}$ for $\OPT$ and searches
for a \emph{jump arborescence}, a tree of depth $\eta=2\ell-1$ whose edges are walks
in $G$ of at most $\theta$ arcs; the \emph{load} of a vertex of $G$ counts the
walks that leave it. For $\theta=\OPT$, the height reduction
(Lemma~\ref{lem:hr}) gives a jump arborescence with maximum load at most
$\Lambda=\ell(\theta+\rho+1)$, where $\rho=\lceil k^{1/\ell}\rceil$. The search runs on the unfolding of $G$, a
tree with $n^{O(\ell)}$ nodes, which gives the running time. The recursive
greedy algorithm charges every choice the increase of the potential
$\Phi(L)=\sum_{v\in V}(e^{L(v)/\Gamma}-1)$ of the loads $L$. The scale
$\Gamma=\Theta(\eta\Lambda)$ is large compared with $\Lambda$, so that adding
a partial jump arborescence of load at most $\Lambda$ changes the potential by at most a
constant factor, and a factor $\eta$ larger still, so that the error caused by
the changing potential is at most a factor $e^{1/\eta}$ per level of the
recursion. The final bound $O(\ell^3\log k)(\OPT+k^{1/\ell})$ collects
four factors: $\ell$ in the loads of the height-reduced optimal jump arborescence,
$\eta$ in the scale $\Gamma$, $\log k$ from greedy covering, and $\ell$ when
the jump arborescence is turned into a schedule.

\subsection{Jump arborescences}
\label{sec:jump}

In this subsection we recall jump arborescences, the structures that the algorithm
searches for instead of schedules, and define the procedure that turns a
jump arborescence into a schedule. Fix an integer $\theta\ge1$, which bounds the number
of arcs of every jump. Jump arborescences are defined in
Section~\ref{sec:ovjump}; we restate Definition~\ref{def:jump}:

\defjump*

If two children of a node have the same label, we may keep one of them,
with its jump, and move the subtree of the other below it. This does not
increase loads or depth. So we may assume that siblings have distinct
labels. Figure~\ref{fig:jump} shows an example.

\paragraph{From a jump arborescence to a schedule.} For a jump arborescence $J$, the
procedure \Schedule{}$(J)$ lets $F$ be the set of arcs used by the
jumps of $J$, takes a breadth-first out-arborescence of $(V,F)$ from $s$,
deletes its branches that contain no terminal, and lets every vertex call
its children in non-increasing order of the number of leaves in their
subtrees, starting in the round after it is informed. For $k$-multicast it
keeps $k$ of the terminals that $J$ covers and deletes the branches that
contain none of them. It runs in time polynomial in $|J|$ and $n$.

\subsection{The unfolding}
\label{sec:unfold}

To search for a jump arborescence of depth $\eta$, we unfold $G$ into a tree whose
nodes are the possible sequences of labels on a path from the root.

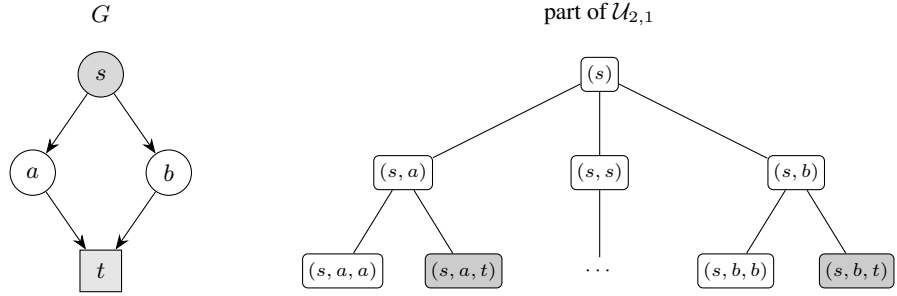
\begin{figure}[t]
\centering
\begin{tikzpicture}[>={Stealth[length=2mm]},
  v/.style={circle,draw,inner sep=1pt,minimum size=6mm,font=\small},
  t/.style={rectangle,draw,fill=black!10,inner sep=2pt,minimum size=5.5mm,font=\small},
  q/.style={rectangle,rounded corners=2pt,draw,inner sep=2pt,font=\scriptsize,minimum height=4.5mm},
  g/.style={q,fill=black!20}]
  \node[font=\small] at (0,0.8) {$G$};
  \node[v,fill=black!15] (s) at (0,0) {$s$};
  \node[v] (a) at (-0.9,-1.3) {$a$};
  \node[v] (b) at (0.9,-1.3) {$b$};
  \node[t] (t) at (0,-2.6) {$t$};
  \draw[->] (s) -- (a); \draw[->] (s) -- (b);
  \draw[->] (a) -- (t); \draw[->] (b) -- (t);
  \begin{scope}[shift={(6.6,0)}]
    \node[font=\small] at (0,0.8) {part of $\mathcal U_{2,1}$};
    \node[q] (r) at (0,0) {$(s)$};
    \node[q] (sa) at (-2.6,-1.3) {$(s,a)$};
    \node[q] (ss) at (0,-1.3) {$(s,s)$};
    \node[q] (sb) at (2.6,-1.3) {$(s,b)$};
    \node[q] (saa) at (-3.4,-2.6) {$(s,a,a)$};
    \node[g] (sat) at (-1.8,-2.6) {$(s,a,t)$};
    \node[font=\scriptsize] (dots) at (0,-2.6) {$\cdots$};
    \node[q] (sbb) at (1.8,-2.6) {$(s,b,b)$};
    \node[g] (sbt) at (3.4,-2.6) {$(s,b,t)$};
    \draw (r) -- (sa); \draw (r) -- (ss); \draw (r) -- (sb);
    \draw (sa) -- (saa); \draw (sa) -- (sat); \draw (ss) -- (dots);
    \draw (sb) -- (sbb); \draw (sb) -- (sbt);
  \end{scope}
\end{tikzpicture}
\caption{The unfolding $\mathcal U_{2,1}$ (depth $2$, $\theta=1$) of the digraph
on the left, partly drawn. A node is a sequence starting at $s$ in which
consecutive vertices are equal or within distance $\theta$; its label is its
last vertex. The group of the terminal $t$ is shaded. A jump arborescence of depth
2 whose siblings have distinct labels is a subtree of $\mathcal U$ that
contains the root and meets every group.}
\label{fig:unfold}
\end{figure}

Let $\eta=2\ell-1$. The \emph{unfolding}
$\mathcal U=\mathcal U_{\eta,\theta}$ is the rooted tree whose nodes are the
sequences $\sigma=(w_0,\dots,w_j)$ with $w_0=s$, $0\le j\le \eta$ and
$\dist(w_{i-1},w_i)\le \theta$ for all $i$ (we allow $w_i=w_{i-1}$). The parent
of $\sigma$ is obtained by dropping $w_j$, and the label of $\sigma$ is
$\lambda(\sigma)=w_j$. It has at most $2n^\eta$ nodes. A jump arborescence of depth
$\eta$ whose siblings have distinct labels is, as a labelled tree, a subtree
of $\mathcal U$ containing the root, together with a jump for every
non-root node. Finding a jump arborescence is therefore a \emph{group Steiner tree}
problem on $\mathcal U$: choose a subtree containing the root and, for
every terminal $t$, a node of the group $\{\sigma:\lambda(\sigma)=t\}$.
Instead of a cost, the objective is the maximum load
(Figure~\ref{fig:unfold}). Appending to a node labelled $t$ a chain of
nodes labelled $t$, with empty jumps, down to depth $\eta$ changes no
load and keeps the depth at most $\eta$. So we may assume that every terminal labels a node at depth exactly
$\eta$, that is, a leaf of $\mathcal U$, and we let only leaves cover
terminals.

\subsection{The algorithm}
\label{sec:alg}

We now present the algorithm. A recursive greedy procedure builds a jump
arborescence on the unfolding (Algorithm~\ref{alg:greedy}), and an outer loop tries
the parameters and turns the jump arborescences into schedules
(Algorithm~\ref{alg:dtm}). We first define the partial jump arborescences that the
procedure builds and the potential that guides it.

\paragraph{Structures.} A \emph{structure} rooted at a node $\sigma$ of
$\mathcal U$ is a finite rooted tree with root $\sigma$ whose nodes are
nodes of $\mathcal U$ (a node of $\mathcal U$ may occur several times);
each non-root node is a child in $\mathcal U$ of its parent and carries a
jump from the label of its parent to its own label: a path with at most
$\theta$ arcs, or the empty path if the labels agree. For a path $P$ in $G$ and
a vertex $v\in V$, let $L_P(v)$ be the number of arcs of $P$ that leave
$v$. The loads of a structure $S$ are defined as for jump arborescences: $L_S(v)$
is the sum of $L_P(v)$ over the jumps $P$ of $S$. Its set
$\mathrm{cov}(S)$ of \emph{covered} terminals consists of the terminals
among the labels of its nodes that are leaves of $\mathcal U$. For a node $\tau$ of $S$, the subtree of $S$ rooted
at $\tau$, with the same jumps, is again a structure, the
\emph{substructure} of $S$ rooted at $\tau$. A structure rooted at the root
of $\mathcal U$ that covers all terminals is a jump arborescence of depth $\eta$.

\paragraph{Potential.} Fix a scale $\Gamma>0$. For a load vector
$L=(L(v))_{v\in V}$, which gives every vertex of $G$ a load, the
\emph{state}, let $\Phi(L)=\sum_{v\in V}\bigl(e^{L(v)/\Gamma}-1\bigr)$, so that
$\Phi(0)=0$, and for a structure $S$ let $\Delta_L(S)=\Phi(L+L_S)-\Phi(L)$
be its \emph{increase}. For a path $P$, the increase $\Delta_L(P)$ is the total weight of
the vertices of $G$ that $P$ leaves, where a vertex $v\in V$ has the weight
$e^{L(v)/\Gamma}(e^{1/\Gamma}-1)$. So a path with at most $\theta$ arcs between
two given vertices of $G$ that minimizes $\Delta_L(P)$ is a shortest path for these weights
among paths with at most $\theta$ arcs. Dynamic programming over the number
of arcs finds a minimum-weight walk with at most $\theta$ arcs; as the weights
are positive, removing a cycle would lower its weight, so this walk is such
a path. As $e^{mx}-1\ge m(e^x-1)$ for $x>0$ and $m\ge1$, the increase of a
walk is at least its weight, so this path is also a cheapest jump among all
walks.

\begin{figure}[t]
\centering
\begin{tikzpicture}[>={Stealth[length=2mm]},
  q/.style={rectangle,rounded corners=2pt,draw,inner sep=2pt,font=\small,minimum height=5.5mm},
  pc/.style={draw,fill=black!8},
  lab/.style={font=\scriptsize}]
  \node[q] (sig) at (0,0) {$\sigma$};
  \node[lab,anchor=west] at (0.45,0) {height $h\ge2$, state $L_{\mathrm{cur}}$,
    residual set $R_{\mathrm{cur}}$, residual demand $z_{\mathrm{cur}}$};
  \node[q] (t1) at (-4.2,-1.6) {$\tau_1$};
  \node[q] (t2) at (0,-1.6) {$\tau_2$};
  \node[q] (t3) at (4.2,-1.6) {$\tau_3$};
  \draw[->] (sig) -- node[lab,above left]{$P_1$} (t1);
  \draw[->] (sig) -- node[lab,right]{$P_2$} (t2);
  \draw[->] (sig) -- node[lab,above right]{$P_3$} (t3);
  \node[lab,align=center] at (-4.2,-2.6) {list of\\ the call at $\tau_1$};
  \node[lab,align=center] at (4.2,-2.6) {list of\\ the call at $\tau_3$};
  \foreach \i/\x in {1/-1.5,2/0,3/1.5} {
    \draw[pc] (t2.south) -- (\x-0.5,-3.3) -- (\x+0.5,-3.3) -- cycle;
    \node[lab] at ($(t2.south)!0.82!(\x,-3.3)$) {$C_\i$};
  }
  \node[lab] at (0,-5.45) {$C_1,C_2,C_3$: the structures returned by
    $\Greedy{}(\tau_2,z_{\mathrm{cur}},L_{\mathrm{cur}}+L_{P_2},
    R_{\mathrm{cur}})$, in the order chosen};
  \draw[decorate,decoration={brace,mirror,amplitude=3pt}] (-2.0,-3.45) -- (-1.0,-3.45)
    node[midway,below=3pt,lab] {$i=1$};
  \draw[decorate,decoration={brace,mirror,amplitude=3pt}] (-2.0,-4.0) -- (0.5,-4.0)
    node[midway,below=3pt,lab] {$i=2$};
  \draw[decorate,decoration={brace,mirror,amplitude=3pt}] (-2.0,-4.55) -- (2.0,-4.55)
    node[midway,below=3pt,lab] {$i=3$};
\end{tikzpicture}
\caption{The candidates in one iteration of the while loop in
line~\ref{line:while} of Algorithm~\ref{alg:greedy} at a node $\sigma$ of height $h\ge2$. For every child $\tau$ the procedure
takes the cheapest jump $P$ and calls itself once at $\tau$; the call
returns the structures it chose, in the order in which it chose them (for $\tau_2$:
$C_1,C_2,C_3$). Each prefix of such a list, below $\tau$ and $P$, is a
candidate (braces). The count of a candidate is the number of residual
terminals that it covers, capped at $z_{\mathrm{cur}}$, and the candidate
with the least increase per unit of count is chosen.}
\label{fig:cands}
\end{figure}

\paragraph{The recursive greedy algorithm.} The function \Greedy{}
(Algorithm~\ref{alg:greedy}) takes a node $\sigma$ of $\mathcal U$ of
height $h\ge1$, a demand $z\ge1$, a state $L$ and a set $R$ of terminals
still to be covered. It returns the list of the structures rooted at
$\sigma$ that it chooses, in the order in which it chooses them. It keeps a
current state $L_{\mathrm{cur}}$, a residual set $R_{\mathrm{cur}}$ and a
residual demand $z_{\mathrm{cur}}$, initially $L$, $R$ and $z$, and
repeats the following iteration while $z_{\mathrm{cur}}>0$
(line~\ref{line:while}). First it collects the \emph{candidates}. For
every child $\tau$ of $\sigma$ it computes the cheapest jump $P$ from the
label of $\sigma$ to the label of $\tau$, that is, a path with at most $\theta$
arcs that minimizes $\Delta_{L_{\mathrm{cur}}}(P)$
(line~\ref{line:path}). If $h=1$, the node $\tau$ with the jump $P$ is a
candidate. If $h\ge2$, the function calls itself once at $\tau$, with the
demand $z_{\mathrm{cur}}$, the state $L_{\mathrm{cur}}+L_P$ and the set
$R_{\mathrm{cur}}$ (line~\ref{line:recurse}), and for every $i\ge1$ the
node $\tau$ with the jump $P$ and the first $i$ returned structures below it is
a candidate (Figure~\ref{fig:cands}). The \emph{count}
$\mathrm{cnt}(C)$ of a candidate $C$ is the number of terminals of
$R_{\mathrm{cur}}$ that it covers, capped at $z_{\mathrm{cur}}$. If no
candidate has positive count, the function returns its list. Otherwise
it chooses a candidate $C$ with the least increase per unit of count,
$\Delta_{L_{\mathrm{cur}}}(C)/\mathrm{cnt}(C)$ (line~\ref{line:choose}),
appends it to its list, adds its loads to $L_{\mathrm{cur}}$, removes
the terminals it covers from $R_{\mathrm{cur}}$ and subtracts its count
from $z_{\mathrm{cur}}$. The increase of a candidate is computed by
building it, which gives its loads, integers; as $e^{y/\Gamma}=q^y$ for the
rational number $q=1+1/\mu$ that Algorithm~\ref{alg:dtm} fixes, the increase
is a rational number, and the iteration compares the ratios exactly. When $z_{\mathrm{cur}}$ reaches $0$, the
function returns its list. Unlike the recursive greedy algorithms of~\cite{charikar1999dst}, which try every demand for the
recursive calls, and of Chekuri, Even and Kortsarz~\cite{chekuri2006greedy},
which try a geometric sequence of demands and keep either the whole list
of a call or its first prefix that covers a $1/h$ fraction of the demand,
it guesses no demands: every prefix of the list that a call returns is a
candidate.

For \DTM\ the demand always equals the number of residual terminals: the
top call, made in line~\ref{line:topcall} of Algorithm~\ref{alg:dtm}, has
$z=k=|R|$, a recursive call receives
$z_{\mathrm{cur}}$ and $R_{\mathrm{cur}}$, and if $z_{\mathrm{cur}}=
|R_{\mathrm{cur}}|$, the count of a candidate is the number of residual
terminals it covers, so the update keeps $z_{\mathrm{cur}}=
|R_{\mathrm{cur}}|$; the cap never applies. The demand is needed for
$k$-multicast, where the top call receives all given terminals but only the
demand $k$: a call stops once it has covered its demand, and a candidate
gets credit only for the terminals still needed.

\paragraph{The main algorithm.} The function \JumpGreedy{}
(Algorithm~\ref{alg:dtm}) solves both problems. It receives the
instance with its set $K$ of terminals, the number $k$ of terminals to
inform, and the integer $\ell\ge2$; for \DTM\ $k=|K|$, and for
$k$-multicast $K$ is the set of all $t\ge k$ given terminals. It sets
$\eta=2\ell-1$. It first computes the busy schedule along a shortest-path
out-arborescence from $s$ to the $k$ terminals of $K$ nearest to $s$ (all
of $K$ for \DTM, line~\ref{line:fallback}) and keeps it as the best
schedule found so far. Then it tries every value $\theta\in\{1,\dots,k-1\}$ for
the optimum (line~\ref{line:loopT}). For each $\theta$ it takes jumps of at
most $\theta$ arcs, the load bound $\Lambda=\ell\,(\theta+\rho+1)$, where
$\rho=\lceil k^{1/\ell}\rceil$, and the scale $\Gamma=1/\ln q$, where $q=1+1/\mu$
and $\mu=20\eta\Lambda$. Then $e^{y/\Gamma}=q^y$ for every $y$, so the potential,
the weights and all increases are rational numbers. As $x/(1+x)\le\ln(1+x)\le
x$ for $x=1/\mu$, we have $\mu\le\Gamma\le\mu+1$; in particular
$\Gamma\ge20\eta\Lambda\ge4e(e-1)\eta\Lambda$, as $4e(e-1)<19$. It calls \Greedy{} at the root of
$\mathcal U_{\eta,\theta}$ with demand $k$, the zero state and the set $K$ (the
\emph{top call}, line~\ref{line:topcall}). The returned structures share the root, and
together they form a structure $J$. If $J$ covers $k$ terminals (all
terminals, for \DTM), the function turns it into the schedule
\Schedule{}$(J)$, which for $k$-multicast keeps $k$ of the covered
terminals, and keeps the shorter of this schedule and the best one
(line~\ref{line:best}). Finally it returns the best schedule. Larger values
of $\theta$ are not needed: when $\OPT\ge k$, the first schedule is already a
2-approximation (Lemma~\ref{lem:fallback}).

\begin{algorithm}[htb]
\caption{The recursive greedy algorithm on $\mathcal U=\mathcal U_{\eta,\theta}$,
with scale $\Gamma$}
\label{alg:greedy}
\begin{algorithmic}[1]
\Require
  \IOitem{$\sigma$: a node of $\mathcal U$ of height $h\ge1$}
  \IOitem{$z\ge1$: the demand, the number of terminals of $R$ still to
    cover}
  \IOitem{$L$: the state, the current load of every vertex of $G$}
  \IOitem{$R$: the set of terminals not covered yet}
  \IOitem{fixed throughout: the unfolding $\mathcal U=\mathcal U_{\eta,\theta}$ and the
    scale $\Gamma$}
\Ensure the list of the chosen structures, rooted at $\sigma$, in the order
  in which they are chosen
\Function{\Greedy}{$\sigma,z,L,R$}
  \State $S\gets$ the empty list, \ $L_{\mathrm{cur}}\gets L$, \
    $R_{\mathrm{cur}}\gets R$, \ $z_{\mathrm{cur}}\gets z$
  \While{$z_{\mathrm{cur}}>0$}\label{line:while}
    \State $\mathcal C\gets\emptyset$\Comment{the candidates}
    \ForAll{children $\tau$ of $\sigma$ in $\mathcal U$}
      \State $P\gets$ a path from $\lambda(\sigma)$ to $\lambda(\tau)$ with
        at most $\theta$ arcs that minimizes $\Delta_{L_{\mathrm{cur}}}(P)$
        \label{line:path}
      \If{$h=1$}
        \State add to $\mathcal C$ the node $\tau$ with the jump $P$
      \Else
        \State $(C_1,\dots,C_q)\gets{}$\Call{\Greedy}{$\tau,z_{\mathrm{cur}},
          L_{\mathrm{cur}}+L_P,R_{\mathrm{cur}}$}\label{line:recurse}
        \State for $i=1,\dots,q$, add to $\mathcal C$ the node $\tau$ with
          the jump $P$ and the structures $C_1,\dots,C_i$ below it
      \EndIf
    \EndFor
    \State $\mathrm{cnt}(C)\gets\min\{|\mathrm{cov}(C)\cap
      R_{\mathrm{cur}}|,z_{\mathrm{cur}}\}$ for $C\in\mathcal C$; remove
      the candidates with $\mathrm{cnt}(C)=0$
    \If{$\mathcal C=\emptyset$}
      \State \Return $S$
    \EndIf
    \State $C\gets$ a candidate in $\mathcal C$ that minimizes
      $\Delta_{L_{\mathrm{cur}}}(C)/\mathrm{cnt}(C)$\label{line:choose}
    \State append $C$ to $S$; \ $L_{\mathrm{cur}}\gets L_{\mathrm{cur}}+
      L_C$; \ $R_{\mathrm{cur}}\gets R_{\mathrm{cur}}\setminus\mathrm{cov}(C)$;
      \ $z_{\mathrm{cur}}\gets z_{\mathrm{cur}}-\mathrm{cnt}(C)$
  \EndWhile
  \State \Return $S$
\EndFunction
\end{algorithmic}
\end{algorithm}

\begin{algorithm}[htb]
\caption{\JumpGreedy: directed multicast and $k$-multicast, for an integer $\ell\ge2$}
\label{alg:dtm}
\begin{algorithmic}[1]
\Require
  \IOitem{$G$: a digraph with $n$ vertices; $s$: the source}
  \IOitem{$K$: the set of terminals}
  \IOitem{$k$: the number of terminals to inform; $k=|K|$ for \DTM, and
    $k\le|K|$ for $k$-multicast}
  \IOitem{$\ell\ge2$: the depth parameter; jump arborescences have depth $2\ell-1$}
\Ensure a schedule that informs $k$ terminals of $K$
\Function{\JumpGreedy}{$G,s,K,k,\ell$}
  \State $\mathit{best}\gets$ the busy schedule along a shortest-path
    out-arborescence from $s$ to the $k$ terminals of $K$ nearest to
    $s$\label{line:fallback}
  \State $\eta\gets2\ell-1$
  \For{$\theta=1,\dots,k-1$}\label{line:loopT}
    \State $\Lambda\gets\ell\,(\theta+\lceil k^{1/\ell}\rceil+1)$, \ $\mu\gets20\eta\Lambda$, \ $\Gamma\gets1/\ln(1+1/\mu)$
    \State $J\gets$ the union of the structures returned by
      \Call{\Greedy}{$\text{root of }\mathcal U_{\eta,\theta},\,k,\,0,\,K$}
      \Comment{Algorithm~\ref{alg:greedy}}\label{line:topcall}
    \If{$J$ covers $k$ terminals}
      \State $\mathit{best}\gets$ the shorter of $\mathit{best}$ and
        \Call{\Schedule}{$J$}\label{line:best}
    \EndIf
  \EndFor
  \State \Return $\mathit{best}$
\EndFunction
\end{algorithmic}
\end{algorithm}

\subsection{Height reduction}
\label{sec:hr}

In this subsection we show that every out-arborescence of small height and
out-degree, in particular that of an optimal schedule, can be cut into a
jump arborescence of depth $2\ell-1$ with small loads (Lemma~\ref{lem:hr}). The
analysis compares the algorithm with the jump arborescence obtained from an
optimal schedule; the algorithm does not compute it.
Section~\ref{sec:ovhr} explains the construction, and we restate
Lemma~\ref{lem:hr}:

\lemhr*

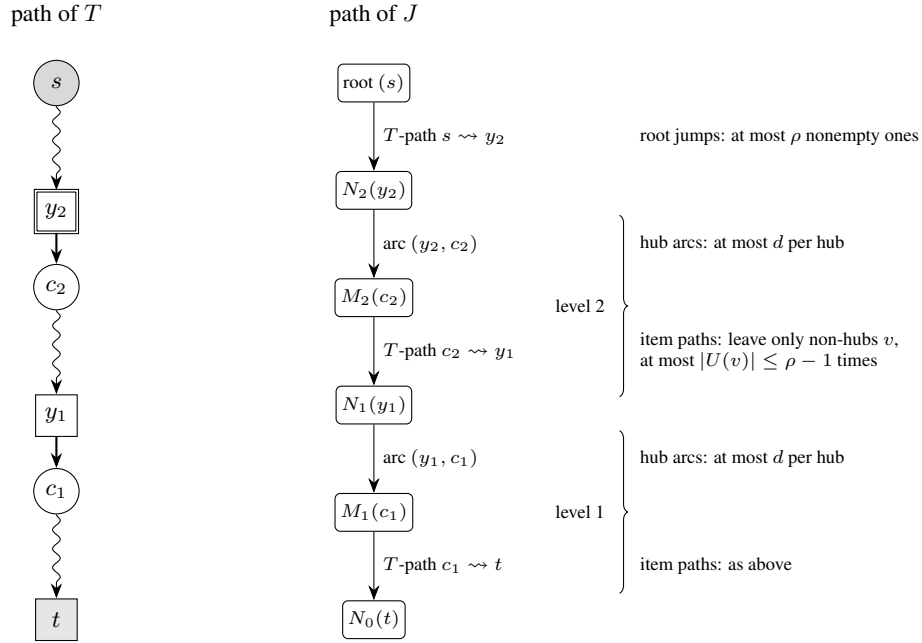
\begin{figure}[t]
\centering
\begin{tikzpicture}[>={Stealth[length=2mm]},
  v/.style={circle,draw,inner sep=1pt,minimum size=6mm,font=\small},
  hub/.style={rectangle,draw,inner sep=1.5pt,minimum size=5.5mm,font=\small},
  hubb/.style={rectangle,draw,double,inner sep=1.5pt,minimum size=5.5mm,font=\small},
  t/.style={rectangle,draw,fill=black!10,inner sep=2pt,minimum size=5.5mm,font=\small},
  jn/.style={rectangle,rounded corners=2pt,draw,inner sep=2pt,font=\scriptsize,minimum height=5mm},
  tp/.style={->,decorate,decoration={snake,amplitude=0.5mm,segment length=2.5mm,post length=1.5mm}},
  lab/.style={font=\scriptsize,anchor=west}]
  \node[font=\small] at (0,0.9) {path of $T$};
  \node[v,fill=black!15] (s) at (0,0) {$s$};
  \node[hubb] (y2) at (0,-1.7) {$y_2$};
  \node[v] (c2) at (0,-2.7) {$c_2$};
  \node[hub] (y1) at (0,-4.4) {$y_1$};
  \node[v] (c1) at (0,-5.4) {$c_1$};
  \node[t] (t) at (0,-7.1) {$t$};
  \draw[tp] (s) -- (y2);
  \draw[->,thick] (y2) -- (c2);
  \draw[tp] (c2) -- (y1);
  \draw[->,thick] (y1) -- (c1);
  \draw[tp] (c1) -- (t);
  \node[font=\small] at (4.2,0.9) {path of $J$};
  \node[jn] (R) at (4.2,0) {root $(s)$};
  \node[jn] (N2) at (4.2,-1.42) {$N_2(y_2)$};
  \node[jn] (M2) at (4.2,-2.84) {$M_2(c_2)$};
  \node[jn] (N1) at (4.2,-4.26) {$N_1(y_1)$};
  \node[jn] (M1) at (4.2,-5.68) {$M_1(c_1)$};
  \node[jn] (N0) at (4.2,-7.1) {$N_0(t)$};
  \draw[->] (R) -- node[lab,right]{$T$-path $s\leadsto y_2$} (N2);
  \draw[->] (N2) -- node[lab,right]{arc $(y_2,c_2)$} (M2);
  \draw[->] (M2) -- node[lab,right]{$T$-path $c_2\leadsto y_1$} (N1);
  \draw[->] (N1) -- node[lab,right]{arc $(y_1,c_1)$} (M1);
  \draw[->] (M1) -- node[lab,right]{$T$-path $c_1\leadsto t$} (N0);
  \node[lab,align=left] at (7.6,-0.71)
    {root jumps: at most $\rho$ nonempty ones};
  \node[lab,align=left] at (7.6,-2.13)
    {hub arcs: at most $d$ per hub};
  \node[lab,align=left] at (7.6,-3.55)
    {item paths: leave only non-hubs $v$,\\ at most $|U(v)|\le\rho-1$ times};
  \node[lab,align=left] at (7.6,-4.97)
    {hub arcs: at most $d$ per hub};
  \node[lab,align=left] at (7.6,-6.39)
    {item paths: as above};
  \draw[decorate,decoration={brace,amplitude=3pt}] (7.45,-1.75) -- (7.45,-4.15)
    node[midway,left=2pt,font=\scriptsize] {level 2};
  \draw[decorate,decoration={brace,amplitude=3pt}] (7.45,-4.6) -- (7.45,-6.75)
    node[midway,left=2pt,font=\scriptsize] {level 1};
\end{tikzpicture}
\caption{A root-to-leaf path of the jump arborescence $J$ of Lemma~\ref{lem:hr}
for $\ell=3$ (middle) and the path of $T$ from $s$ to the terminal $t$
that its jumps follow (left). The vertex $y_2$ is a level-2 hub (double
square), $c_2$ is the child of $y_2$ that passed the item $y_1$ up, $y_1$ is
a level-1 hub, and $c_1$ is its child that passed $t$ up. Nodes $N_i$
alternate with intermediate nodes $M_i$, so the depth is $2\ell-1=5$. The
three kinds of jumps give the three terms of the bound $\Lambda(J)\le
(\ell-1)\max\{d,\rho-1\}+\rho$ (right).}
\label{fig:hrpath}
\end{figure}

\begin{proof}
\emph{Items.} Let $I_0=K$. For $i=1,\dots,\ell-1$ we obtain $I_i$ from
$I_{i-1}$ by one pass over $T$ from the leaves up. Every vertex $v$ of $T$ forms
the set $U(v)$, which consists of $v$ itself if $v\in I_{i-1}$ and of the
sets passed up by its children. If $|U(v)|\ge\rho$ or $v=s$, then $v$
becomes a \emph{level-$i$ hub}: the items of $U(v)$ are \emph{assigned} to
$v$, and $v$ passes nothing up. Otherwise $v$ passes $U(v)$ to its parent.
Let $I_i$ be the set of level-$i$ hubs. Every item of $I_{i-1}$ is assigned
to exactly one hub, and every hub other than $s$ receives at least $\rho$
items, so $|I_i|\le|I_{i-1}|/\rho+1$. Since $|I_0|=k\le\rho^\ell$ and
$\rho\ge2$, this gives $|I_i|\le\rho^{\ell-i}+\sum_{j<i}\rho^{-j}<
\rho^{\ell-i}+2$, and in particular $|I_{\ell-1}|\le\rho+1$.

\emph{The jump arborescence} (Figure~\ref{fig:hr}). For $0\le i\le\ell-1$ and
$x\in I_i$, $J$ has a node $N_i(x)$ labelled $x$, with the following
children.
\begin{itemize}
  \item $N_0(t)$, for $t\in K$, is a leaf.
  \item For $i\ge1$ and a level-$i$ hub $h$: the child $N_{i-1}(h)$ with the
    empty jump if $h\in I_{i-1}$; and, for every child $c$ of $h$ in $T$
    that passed up a nonempty set $U(c)$, an \emph{intermediate} child
    $M_i(c)$ labelled $c$, with the one-arc jump $(h,c)$. The children of
    $M_i(c)$ are the nodes $N_{i-1}(y)$ for $y\in U(c)$, each with the jump
    along the $T$-path from $c$ to $y$ (empty if $y=c$).
  \item The root has the children $N_{\ell-1}(y)$, $y\in I_{\ell-1}$, each
    with the jump along the $T$-path from $s$ to $y$.
\end{itemize}
The items assigned to a hub $h$ are $U(h)=(\{h\}\cap I_{i-1})\cup\bigcup_c
U(c)$, so every item of $I_{i-1}$ gets a parent node at level $i$. So every
node $N_i(x)$ belongs to $J$, in particular every terminal labels a node,
and the depth is at most $2(\ell-1)+1$. The jumps are $T$-paths and hence
have at most $\theta$ arcs.

\emph{Loads} (Figure~\ref{fig:hrpath}). Fix a level $i\in\{1,\dots,\ell-1\}$
and a vertex $v$ of $G$. The
level-$i$ jumps are the arcs $(h,c)$ from a hub to its children, at most
$d$ per hub, and the $T$-paths from an intermediate node $c$ to the items
$y\in U(c)$. Such a path leaves $v$ only if $v$ lies on it and $v\ne y$;
then $y$ was passed up through $v$, so $y\in U(v)$ and $v$ is not a hub.
Hence these paths leave no hub, and they leave any other vertex at most
$|U(v)|\le\rho-1$ times. So the level-$i$ jumps leave $v$ at most
$\max\{d,\rho-1\}$ times. At most $\rho$ jumps from the root are nonempty:
$|I_{\ell-1}|-1\le\rho$ of them if $\ell\ge2$, since the jump to $s\in
I_{\ell-1}$ is empty, and $k\le\rho$ if $\ell=1$. Summing over the levels
gives the bound.
\end{proof}

Apply Lemma~\ref{lem:hr} to the arborescence $T^*$ of an optimal schedule,
pruned to the paths from $s$ to the terminals. Its out-degree is
$d^*\le\OPT$ and its height is $h^*\le\OPT$. Lemma~\ref{lem:hr} gives a jump arborescence,
for $\theta=\OPT$, as $h^*\le\OPT$, with
\begin{equation}\label{eq:lambdastar}
  \Lambda^*:=(\ell-1)\max\{d^*,\rho-1\}+\rho\;\le\;\ell\,\bigl(\OPT+\rho+1\bigr).
\end{equation}
The loads of this jump arborescence are positive only on vertices of $T^*$, of
which there are at most $1+kh^*$.

\subsection{Analysis of the algorithm}
\label{sec:comb}

In this subsection we prove Theorem~\ref{thm:dtmcomb} and
Corollary~\ref{cor:dtm}. We bound the length
of the two kinds of schedules that Algorithm~\ref{alg:dtm} compares, show
that one iteration of the recursive greedy algorithm chooses a cheap
structure (Lemma~\ref{lem:greedy}) and that the top call finds a jump arborescence
with small loads (Lemma~\ref{lem:topcall}), and combine these bounds.

As for the recursive greedy algorithm of~\cite{charikar1999dst}, which Chekuri,
Even and Kortsarz~\cite{chekuri2006greedy} analyse for group Steiner tree
on trees, a candidate of a node of height $h$ has density within a factor
$O(h)$ of the optimum (Lemma~\ref{lem:greedy}). The objective, the
maximum load, is a maximum over the vertices of $G$ rather than a sum,
which is why a candidate costs the increase of
an exponential potential, as in online load balancing~\cite{aspnes1997online}.
The loads end up at most $O(\Gamma\log k)$, and two observations allow the
scale $\Gamma=\Theta(\eta\Lambda)$. In the prefix of a recursive call that
the analysis uses, the structures before the last one cover only a $1/h$
fraction of what the optimum covers there, and spend only $O(1)$ times its
cost; only they raise the cost of a later structure. At the top, comparing
every iteration with the current potential instead of the final one turns
the harmonic sum of greedy covering into a factor $k^{O(1)}$ inside the
potential, which costs only $O(\log k)$ in the loads.

\paragraph{The two kinds of schedules.} Algorithm~\ref{alg:dtm} compares
the schedules \Schedule{}$(J)$ with the schedule of
line~\ref{line:fallback}. We bound both. First, the arborescence that
\Schedule{} builds inherits small out-degree and height from $J$. Let
$Y$ be the set of terminals that \Schedule{}$(J)$ keeps: all
terminals for \DTM, and $k$ of the terminals that $J$ covers for
$k$-multicast; so $|Y|=k$.

\begin{lemma}\label{lem:schedtree}
Let $J$ be a structure rooted at the root of $\mathcal U_{\eta,\theta}$ that covers
every terminal of $Y$. The out-arborescence that
\Schedule{}$(J)$ builds contains $Y$, has height at most
$\eta\theta$ and out-degree at most $\Lambda(J)$, and has at most $|Y|$ leaves.
\end{lemma}
\begin{proof}
Let $F$ be the set of arcs used by the jumps of $J$. Every arc of $F$ that
leaves a vertex $v$ of $G$ is an arc of a jump of $J$ that leaves $v$, so
$v$ has out-degree at most $L_J(v)\le\Lambda(J)$ in the digraph $(V,F)$,
and hence in every subgraph of it. For $t\in Y$, some node of $J$ is
labelled $t$, and the jumps on the path from the root of $J$ to this node
form a walk from $s$ to $t$ in $(V,F)$. The path has at most $\eta$ edges and
every jump at most $\theta$ arcs, so the walk has at most $\eta\theta$ arcs, and $t$
lies at depth at most $\eta\theta$ in the breadth-first out-arborescence of
$(V,F)$ from $s$. Deleting the branches that contain no terminal of $Y$
keeps the paths from $s$ to $Y$, so every remaining vertex has depth at most
$\eta\theta$, and every leaf of the result is in $Y$.
\end{proof}

The schedule along this arborescence is short:

\begin{lemma}\label{lem:jumpsched}
Let $J$ be as in Lemma~\ref{lem:schedtree}, and let $d\ge2$ with
$\Lambda(J)\le d$. Then \Schedule{}$(J)$ informs $Y$
within $d\log_2k/\log_2d+\eta\theta$ rounds.
\end{lemma}
\begin{proof}
By Lemma~\ref{lem:schedtree}, the arborescence contains $Y$, has out-degree
at most $d$, height at most $\eta\theta$, and at most $|Y|=k$ leaves. Every
vertex calls its children in non-increasing order of their numbers of
leaves, so Lemma~\ref{lem:ravi}, with leaf counts, gives
the bound.
\end{proof}

For $d=\max\{\Lambda(J),k^{1/\ell},2\}$ we have $\log_2d\ge
\log_2k/\ell$, so the length is at most
\begin{equation}\label{eq:ell}
  \ell\cdot\max\{\Lambda(J),k^{1/\ell},2\}+\eta\theta .
\end{equation}

When $\OPT$ is large compared with $k$, jump arborescences are not needed: the
fallback schedule of line~\ref{line:fallback} of Algorithm~\ref{alg:dtm} is already a
2-approximation.

\begin{lemma}\label{lem:fallback}
Let $Y$ be the set of terminals (for $k$-multicast, the $k$ terminals
nearest to $s$). The busy schedule along a shortest-path out-arborescence
from $s$ to $Y$, computed in line~\ref{line:fallback} of
Algorithm~\ref{alg:dtm}, informs $Y$ within $\OPT+k$ rounds. In particular it is a
2-approximation when $\OPT\ge k$.
\end{lemma}
\begin{proof}
Every terminal of $Y$ is within distance $\OPT$ of $s$. For $k$-multicast,
an optimal schedule informs $k$ terminals within distance $\OPT$, so the
$k$ nearest are also within $\OPT$. The pruned arborescence therefore has
height at most $\OPT$ and at most $k$ leaves, and Lemma~\ref{lem:busy}
applies.
\end{proof}

\paragraph{Linear cost.} For a structure $S$ and a state $L$ let
\[
  c_L(S)=\frac1\Gamma\sum_{v\in V}e^{L(v)/\Gamma}L_S(v)
\]
be its \emph{linear cost}; for a path $P$, the linear cost $c_L(P)$ is
defined in the same way, with $L_P$ in place of $L_S$. Here $L$ is the state, the loads that the
structures chosen so far put on the vertices of $G$, and $L_S$ are the loads
of $S$ alone. Every unit of load that $S$ adds at a vertex $v\in V$ is
priced at $e^{L(v)/\Gamma}/\Gamma$, the derivative of $\Phi$ at $L$ in the
coordinate $v$; so $c_L(S)$ is the linear approximation of the increase
$\Delta_L(S)=\Phi(L+L_S)-\Phi(L)$. Since the loads of a structure are the
sums of the loads of its jumps, its linear cost is the sum of the linear
costs of its jumps, and it is nondecreasing in $L$. Lemma~\ref{lem:linear} bounds the increase of
a structure with small loads by a constant times its linear cost, and
shows that when the state grows, the linear cost of such a structure grows
at most $\Lambda/\Gamma$ times as much as the potential.

\begin{lemma}\label{lem:linear}
Let $Q$ be a structure with maximum load at most $\Lambda\le\Gamma$, and
let $L\le L'$ be states (coordinatewise). Then
\begin{enumerate}[label=(\alph*),ref=(\alph*)]
  \item\label{lem:linear:increase} $\Delta_L(Q)\le(e-1)\,c_L(Q)$, and
  \item\label{lem:linear:growth} $c_{L'}(Q)-c_L(Q)\le(\Lambda/\Gamma)\bigl(\Phi(L')-\Phi(L)\bigr)$.
\end{enumerate}
\end{lemma}
\begin{proof}
Both parts are checked vertex by vertex. \ref{lem:linear:increase} The increase is $\Delta_L(Q)=
\sum_{v\in V}e^{L(v)/\Gamma}(e^{L_Q(v)/\Gamma}-1)$. For every $v\in V$ the
exponent $y=L_Q(v)/\Gamma$ lies in $[0,1]$, since $L_Q(v)\le\Lambda\le
\Gamma$, and on $[0,1]$ the convex function $e^y$ lies below its chord from
$(0,1)$ to $(1,e)$, so $e^y-1\le(e-1)y$. Hence the term of $v$ is at most
$(e-1)e^{L(v)/\Gamma}L_Q(v)/\Gamma$, and summing over $v$ gives~\ref{lem:linear:increase}.
\ref{lem:linear:growth} Every difference $e^{L'(v)/\Gamma}-e^{L(v)/\Gamma}$ is nonnegative, as
$L\le L'$, and $L_Q(v)\le\Lambda$, so
\[
  c_{L'}(Q)-c_L(Q)=\frac1\Gamma\sum_{v\in V}\bigl(e^{L'(v)/\Gamma}-
  e^{L(v)/\Gamma}\bigr)L_Q(v)\le\frac\Lambda\Gamma\sum_{v\in V}
  \bigl(e^{L'(v)/\Gamma}-e^{L(v)/\Gamma}\bigr)=\frac\Lambda\Gamma
  \bigl(\Phi(L')-\Phi(L)\bigr).\qedhere
\]
\end{proof}

The core of the analysis bounds the candidate that one iteration of the
while loop in line~\ref{line:while} of Algorithm~\ref{alg:greedy} chooses: its increase per counted terminal is at
most $O(h)$ times the linear cost per terminal of any structure with small
loads. This is the counterpart, for loads, of the density bound of
Chekuri, Even and Kortsarz~\cite[Lemma~3.4]{chekuri2006greedy}.

\begin{lemma}\label{lem:greedy}
Let $b_h=(e-1)e^{(h-1)/\eta}$ and $\beta_h=h\,b_h$ for $1\le h\le\eta$, so that $\beta_h<
e(e-1)h$, and suppose $\Gamma\ge4e(e-1)\eta\Lambda$. Consider an iteration of
the while loop in line~\ref{line:while} of Algorithm~\ref{alg:greedy} at a
node $\sigma$ of height
$h\ge1$, with state $L_{\mathrm{cur}}$, residual set $R_{\mathrm{cur}}$ and
residual demand $z_{\mathrm{cur}}\ge1$. If some structure $Q$ rooted at
$\sigma$ with maximum load at most $\Lambda$ covers $x\ge1$ terminals of
$R_{\mathrm{cur}}$, then some candidate has positive count, and the chosen
candidate $C$ satisfies
\[
  \Delta_{L_{\mathrm{cur}}}(C)\le\beta_h\,\mathrm{cnt}(C)\,
  c_{L_{\mathrm{cur}}}(Q)/\bar x,\qquad\bar x=\min\{x,z_{\mathrm{cur}}\}.
\]
\end{lemma}

\begin{figure}[t]
\centering
\begin{tikzpicture}[>={Stealth[length=2mm]},
  q/.style={rectangle,rounded corners=2pt,draw,inner sep=2pt,font=\small,minimum height=5.5mm},
  lab/.style={font=\scriptsize}]
  \node[q] (sig) at (0,0) {$\sigma$};
  \node[q,very thick] (t1) at (-1.9,-1.4) {$\tau_1$};
  \node[q] (t2) at (0,-1.4) {$\tau$};
  \node[q] (t3) at (1.9,-1.4) {$\tau'$};
  \draw[->,very thick] (sig) -- node[lab,above left]{$P^*$} (t1);
  \draw[->] (sig) -- (t2);
  \draw[->] (sig) -- (t3);
  \draw[very thick,fill=black!8] (t1.south) -- (-2.75,-3.1) -- (-1.05,-3.1) -- cycle;
  \node[lab] at (-1.9,-2.7) {$Q'$};
  \draw[dotted,thick,rounded corners=6pt] (-0.18,-0.42) -- (-1.25,0.12) --
    (-2.65,-0.95) -- (-3.1,-3.32) -- (-0.72,-3.32) -- (-1.33,-1.5) -- cycle;
  \node[lab] at (-3.05,-0.55) {$Q_{\tau_1}$};
  \draw[fill=black!8] (t2.south) -- (-0.5,-2.6) -- (0.5,-2.6) -- cycle;
  \draw[fill=black!8] (t3.south) -- (1.3,-2.8) -- (2.5,-2.8) -- cycle;
  \node[lab,align=center] at (-1.9,-3.6) {covers $y$ of\\ the residual terminals};
  \node[lab] at (0,-4.5) {$c(Q_{\tau_1})/y\le c(Q)/\bar x$,\quad
    branch $Q_{\tau_1}$ = jump $P^*$ and substructure $Q'$};
\end{tikzpicture}
\caption{The proof of Lemma~\ref{lem:greedy}. The structure $Q$ splits
into branches, one for each child of its root, and the branch $Q_{\tau_1}$
of some child $\tau_1$ (dotted), the jump $P^*$ together with the
substructure $Q'$, is at least as cheap per covered terminal as $Q$.}
\label{fig:greedyproof}
\end{figure}

\begin{proof}
We prove Lemma~\ref{lem:greedy} by induction on the height $h$ of $\sigma$. Throughout
we write $c=c_{L_{\mathrm{cur}}}$, $\Delta=\Delta_{L_{\mathrm{cur}}}$ and
$z=z_{\mathrm{cur}}$.

\emph{A branch of $Q$ that is as cheap as $Q$.} Split $Q$ into
\emph{branches} $Q_\tau$, one for each child $\tau$ of the root of $Q$ (two
children may be copies of the same node of $\mathcal U$; $\tau$ also denotes
that node): the
branch $Q_\tau$ consists of the jump $P^*_\tau$ that $Q$ uses from $\sigma$
to $\tau$, followed by the substructure $Q'_\tau$ of $Q$ rooted at $\tau$.
By additivity of the linear cost, $c(Q_\tau)=c(P^*_\tau)+c(Q'_\tau)$ and
$c(Q)=\sum_\tau c(Q_\tau)$. Let $y_\tau=\min\{|\mathrm{cov}(Q_\tau)\cap
R_{\mathrm{cur}}|,z\}$. Then $\sum_\tau y_\tau\ge\bar x$: if some branch has
$y_\tau=z$, this term alone is at least $\bar x$, and otherwise $\sum_\tau
y_\tau=\sum_\tau|\mathrm{cov}(Q_\tau)\cap R_{\mathrm{cur}}|\ge x$, since
every terminal that $Q$ covers is covered by one of its branches. Hence
some child $\tau_1$ with $y:=y_{\tau_1}\ge1$ satisfies $c(Q_{\tau_1})/y\le
c(Q)/\bar x$ (Figure~\ref{fig:greedyproof}): otherwise $c(Q_\tau)>y_\tau
c(Q)/\bar x$ for every child $\tau$ with $y_\tau\ge1$, and summing over
these children gives $c(Q)>c(Q)$, as costs are nonnegative. We write
$P^*=P^*_{\tau_1}$ and $Q'=Q'_{\tau_1}$, so that $Q_{\tau_1}$ is the jump
$P^*$ followed by $Q'$, and $c(Q_{\tau_1})=c(P^*)+c(Q')$. Note that $Q'$
covers at least $y$ terminals of $R_{\mathrm{cur}}$, and that $y\le z$.

\emph{The jump to $\tau_1$.} The candidates at $\tau_1$ do not use $P^*$,
but the jump $P$ that line~\ref{line:path} of Algorithm~\ref{alg:greedy}
chooses for $\tau_1$, the one with the least increase. So
$\Delta(P)\le\Delta(P^*)$, and $\Delta(P^*)\le(e-1)c(P^*)$ by
Lemma~\ref{lem:linear}\ref{lem:linear:increase}, applied to the jump $P^*$, whose loads are at
most those of $Q$.

\emph{It suffices to find one good candidate.} We show that some candidate
$\hat C$ has $\mathrm{cnt}(\hat C)>0$ and
\begin{equation}\label{eq:goodcand}
  \frac{\Delta(\hat C)}{\mathrm{cnt}(\hat C)}\le\beta_h\,
  \frac{c(Q_{\tau_1})}y .
\end{equation}
This proves Lemma~\ref{lem:greedy}. Indeed, $\hat C$ shows that some candidate has
positive count, so line~\ref{line:choose} of Algorithm~\ref{alg:greedy}
chooses a candidate $C$, and as $C$ minimizes the ratio among the
candidates with positive count,
\[
  \frac{\Delta(C)}{\mathrm{cnt}(C)}\le\frac{\Delta(\hat C)}{\mathrm{cnt}(\hat
  C)}\le\beta_h\frac{c(Q_{\tau_1})}y\le\beta_h\frac{c(Q)}{\bar x};
\]
multiplying by $\mathrm{cnt}(C)$ gives the inequality of
Lemma~\ref{lem:greedy}.

\emph{Height $1$.} If $h=1$, then $\tau_1$ is a leaf of $\mathcal U$
labelled by a terminal of $R_{\mathrm{cur}}$, so $y=1$, and $Q_{\tau_1}$ is
the jump $P^*$. The node $\tau_1$ with the jump $P$ is a candidate that
covers this terminal; its count is $1$, as $z\ge1$, and its ratio is
$\Delta(P)\le(e-1)c(P^*)=\beta_1c(Q_{\tau_1})/y$, since $\beta_1=b_1=e-1$.

\emph{Height $h\ge2$: the recursive call at $\tau_1$.} Let $h\ge2$, assume
Lemma~\ref{lem:greedy} for height $h-1$, and write $b=b_{h-1}$; note that $e-1\le
b<e(e-1)$. The candidates at $\tau_1$ come from one recursive call, made in
line~\ref{line:recurse} of Algorithm~\ref{alg:greedy} at $\tau_1$ with the
demand $z$, the state $L_{\mathrm{cur}}+L_P$ and the set $R_{\mathrm{cur}}$.
It returns structures $C_1,\dots,C_q$, in the order in which it chooses
them, and for every $i\le q$ the node $\tau_1$ with the jump $P$ and the
structures $C_1,\dots,C_i$ is a candidate at $\sigma$. The structure $C_j$
is the one that the recursive call chooses in the $j$-th iteration of its
own while loop, in line~\ref{line:while} of Algorithm~\ref{alg:greedy} run
at $\tau_1$. For this iteration let $L^{(j)}$ be the state of the recursive
call at its start, $c_j$ the count of $C_j$ and
$\Delta_j=\Delta_{L^{(j)}}(C_j)$ its increase, and let
\[
  S_j=\Phi(L^{(j)})-\Phi(L_{\mathrm{cur}}),\qquad m_j=c_1+\dots+c_j ,
\]
the increase of the potential so far, measured from the state at $\sigma$,
and the count so far, with $m_0=0$. The recursive call starts in the state
$L^{(1)}=L_{\mathrm{cur}}+L_P$, so $S_1=\Delta(P)$, and every iteration adds
the loads of the chosen structure, so $S_{j+1}=S_j+\Delta_j$. In particular
$S_1\le S_2\le\cdots$, as increases are nonnegative.

\emph{While fewer than $y/h$ terminals are counted, the induction applies.}
Let $j\ge1$ with $m_{j-1}<y/h$. In the recursive call, a count is capped
only when the chosen structure covers at least as many residual terminals
as the residual demand; then the count equals the residual demand, the
residual demand drops to $0$, and the call ends. The first $j-1$ counts add
up to $m_{j-1}<y/h\le y\le z$, so the residual demand is still
$z-m_{j-1}>0$, and none of these counts was capped. Hence each of them is
the number of terminals that its structure removed from the residual set
of the call, and the call has removed exactly $m_{j-1}$ terminals so far.
As $Q'$ covers at least $y$ terminals of $R_{\mathrm{cur}}$, it still covers
$x'>y-y/h=y(h-1)/h$ terminals of the residual set of the call, and the
residual demand $z-m_{j-1}$ is also more than $y(h-1)/h$, as $y\le z$. So
Lemma~\ref{lem:greedy} for height $h-1$ applies to iteration $j$ of the recursive call,
with the state $L^{(j)}$ and the structure $Q'$, which is rooted at
$\tau_1$ and has maximum load at most $\Lambda$; its $\bar x$ is $\bar
x'=\min\{x',z-m_{j-1}\}>y(h-1)/h$. Lemma~\ref{lem:greedy} for height $h-1$
shows that some candidate of
this iteration has positive count, so the iteration takes place and
chooses $C_j$ with $c_j\ge1$, and that
\[
  \Delta_j\le\beta_{h-1}\,\frac{c_j}{\bar x'}\,c_{L^{(j)}}(Q')
  \le\beta_{h-1}\,\frac{c_j}{y(h-1)/h}\,c_{L^{(j)}}(Q')
  =h\,b\,\frac{c_j}y\,c_{L^{(j)}}(Q'),
\]
since $\beta_{h-1}=(h-1)b$. Loads only grow during the call, so
$L_{\mathrm{cur}}\le L^{(j)}$, and $\Lambda\le\Gamma$; so
Lemma~\ref{lem:linear}\ref{lem:linear:growth} gives $c_{L^{(j)}}(Q')\le c(Q')+(\Lambda/\Gamma)\bigl(\Phi(L^{(j)})-\Phi(L_{\mathrm{cur}})\bigr)
=c(Q')+(\Lambda/\Gamma)S_j$. Together,
\begin{equation}\label{eq:step}
  \Delta_j\le h\,b\,\frac{c_j}y\Bigl(c(Q')+\frac\Lambda\Gamma S_j\Bigr)
  \qquad\text{whenever }m_{j-1}<y/h.
\end{equation}

\emph{The candidate $\hat C$.} As long as $m_{j-1}<y/h$, iteration $j$ takes
place and adds $c_j\ge1$ to the count, so the counts reach $y/h$ after at
most $\lceil y/h\rceil$ iterations: there is a first index $p$ with $m_p\ge
y/h$. Let $\hat C$ be the candidate
consisting of $\tau_1$, the jump $P$ and the structures $C_1,\dots,C_p$, and
let $m=m_p$ and $m^\circ=m_{p-1}$; so $m^\circ<y/h\le m$ and $m=m^\circ+c_p$
(with $m^\circ=0$ if $p=1$). The counted terminals are distinct terminals
of $R_{\mathrm{cur}}$ that $\hat C$ covers, and $m\le z$, since the counts
of the call add up to at most its demand $z$. So $\mathrm{cnt}(\hat C)$, the
number of terminals of $R_{\mathrm{cur}}$ that $\hat C$ covers, capped at
$z$, is at least $m\ge y/h>0$. The increase of $\hat C$ is the increase of
the potential from $L_{\mathrm{cur}}$ to the state after the $p$-th
iteration of the call, that is, $\Delta(\hat C)=S_p+\Delta_p$.

\emph{The structures before the last one.} Let $S=S_p$ and
$\kappa=b\Lambda/\Gamma$. As $b<e(e-1)$ and $\Gamma\ge4e(e-1)\eta\Lambda$, we
have $\kappa<1/(4\eta)$. The increase $S=\Delta(P)+\sum_{j<p}\Delta_j$ is due
to the jump $P$ and to the structures before $C_p$. We bound every
$\Delta_j$ with $j<p$ by~\eqref{eq:step}, which applies as $m_{j-1}\le
m^\circ<y/h$, and use $S_j\le S$, $\sum_{j<p}c_j=m^\circ$ and $hm^\circ/y<1$:
\[
  S\le\Delta(P)+hb\,\frac{m^\circ}y\,c(Q')+hb\,\frac{m^\circ}y\cdot
  \frac\Lambda\Gamma\,S\le\Delta(P)+hb\,\frac{m^\circ}y\,c(Q')+\kappa S .
\]
We solve for $S$, and then use $\Delta(P)\le(e-1)c(P^*)\le b\,c(P^*)$ and
$hm^\circ/y<1$ once more:
\begin{equation}\label{eq:Sbound}
  S\le\frac1{1-\kappa}\Bigl(\Delta(P)+hb\,\frac{m^\circ}y\,c(Q')\Bigr)\le
  \frac b{1-\kappa}\bigl(c(P^*)+c(Q')\bigr).
\end{equation}
So the jump $P$ and the structures before the last one raise the potential
by at most a constant times the cost of the branch $Q_{\tau_1}$.

\emph{The last structure.} By~\eqref{eq:step} for $j=p$, which applies as
$m_{p-1}=m^\circ<y/h$, and by the second bound in~\eqref{eq:Sbound}, with
$(\Lambda/\Gamma)\cdot b/(1-\kappa)=\kappa/(1-\kappa)$,
\[
  \Delta_p\le hb\,\frac{c_p}y\Bigl(c(Q')+\frac\Lambda\Gamma\,S\Bigr)\le
  hb\,\frac{c_p}y\Bigl(c(Q')+\frac\kappa{1-\kappa}\bigl(c(P^*)+c(Q')\bigr)
  \Bigr)=\frac{hb}y\cdot c_p\cdot\frac{\kappa\,c(P^*)+c(Q')}{1-\kappa}.
\]

\emph{The ratio of $\hat C$.} We divide $\Delta(\hat C)=S+\Delta_p$ by $m$,
using the first bound in~\eqref{eq:Sbound} for $S$. As $m\ge y/h$, we have
$\Delta(P)/m\le h\Delta(P)/y\le hb\,c(P^*)/y$, so
\[
  \frac Sm\le\frac{hb}y\cdot\frac{c(P^*)+c(Q')\,m^\circ/m}{1-\kappa},
  \qquad
  \frac{\Delta_p}m\le\frac{hb}y\cdot\frac{c_p}m\cdot
  \frac{\kappa\,c(P^*)+c(Q')}{1-\kappa}.
\]
Adding the two bounds, and using $m^\circ+c_p=m$ and $c_p\le m$, gives
\[
  \frac{\Delta(\hat C)}m\le\frac{hb}y\cdot\frac{(1+\kappa c_p/m)\,c(P^*)+
  c(Q')}{1-\kappa}\le\frac{hb}y\cdot\frac{1+\kappa}{1-\kappa}\,
  \bigl(c(P^*)+c(Q')\bigr)=\frac{hb}y\cdot\frac{1+\kappa}{1-\kappa}\,
  c(Q_{\tau_1}).
\]
Finally $(1+\kappa)/(1-\kappa)=1+2\kappa/(1-\kappa)\le1+(24/11)\kappa
<1+1/\eta\le e^{1/\eta}$, as $\kappa<1/(4\eta)\le1/12$ (recall
$\eta=2\ell-1\ge3$). Since $\mathrm{cnt}(\hat C)\ge m$, the ratio of $\hat C$ is
at most $\Delta(\hat C)/m\le hb_{h-1}e^{1/\eta}c(Q_{\tau_1})/y=hb_hc(Q_{\tau_1})
/y=\beta_hc(Q_{\tau_1})/y$, which is~\eqref{eq:goodcand}.
\end{proof}

Lemma~\ref{lem:topcall} applies Lemma~\ref{lem:greedy} to all iterations of the top
call. It compares the greedy covering with any structure $J'$ of small
loads, and bounds the loads of the result in terms of the number of
vertices of $G$ on which $J'$ has positive loads. Theorem~\ref{thm:dtmcomb}
uses it.

\begin{lemma}[top call]\label{lem:topcall}
Let $\theta\ge1$ and $\Lambda>0$, let $\Gamma\ge4e(e-1)\eta\Lambda$, and let the
demand $k$ be at most the number of terminals in $K$. Suppose that a
structure $J'$ rooted at the root of $\mathcal U_{\eta,\theta}$ covers at least $k$
terminals, has maximum load at most $\Lambda$, and has positive loads on a
set $V'$ of at most $\nu$ vertices of $G$. Then the structures returned by
\Greedy{}$(\text{root of }\mathcal U_{\eta,\theta},k,0,K)$ together
form a structure $J$ that covers at least $k$ terminals, and every load of
$J$ is at most $\Gamma\bigl(\ln \nu+(1+\ln k)/4\bigr)$.
\end{lemma}

\begin{proof}
Let $L_j$, $z_j$ and $C_j$ be the state, the residual demand and the chosen
candidate in the $j$-th iteration of the while loop
(line~\ref{line:while} of Algorithm~\ref{alg:greedy}) of the call, and
$\Phi_j=\Phi(L_j)$, so that $\Phi_1=\Phi(0)=0$.

\emph{Every iteration finds a candidate.} A capped count makes the residual
demand $0$ and ends the call. So as long as the call runs, no count has
been capped, the call has removed exactly $k-z_j$ terminals from its
residual set before iteration $j$, and at least $z_j$ of the terminals that
$J'$ covers are still residual. Hence in every iteration
Lemma~\ref{lem:greedy} applies with $Q=J'$ and $\bar x=z_j$: the call is at
the root of $\mathcal U$, of height $\eta$, and $\Gamma\ge4e(e-1)\eta\Lambda$. In
particular every iteration has a candidate with positive count, so the call
ends only when $z_j$ reaches $0$, after counting $k$ terminals. The
returned structures share the root, and together they form a structure $J$
that covers these $k$ terminals.

\emph{The linear cost of $J'$ at the current state.} By definition,
$c_{L_j}(J')=(1/\Gamma)\sum_{v\in V}e^{L_j(v)/\Gamma}L_{J'}(v)$. Only
the vertices of $V'$ contribute, and $L_{J'}(v)\le\Lambda$. Write
$e^{L_j(v)/\Gamma}=(e^{L_j(v)/\Gamma}-1)+1$; the potential $\Phi_j$ is the
sum of the nonnegative terms $e^{L_j(v)/\Gamma}-1$ over all $v\in V$. So
\[
  c_{L_j}(J')\le\frac\Lambda\Gamma\sum_{v\in V'}e^{L_j(v)/\Gamma}\le
  \frac\Lambda\Gamma\bigl(\Phi_j+|V'|\bigr).
\]

\emph{The potential stays small.} We apply Lemma~\ref{lem:greedy} with the
current state $L_j$, not with the final one; this is what allows us to
bound $c_{L_j}(J')$ in terms of the current potential $\Phi_j$. Since
$\beta_\eta=\eta(e-1)e^{(\eta-1)/\eta}<e(e-1)\eta$ and $\Gamma\ge4e(e-1)\eta\Lambda$, we have
$\beta_\eta\Lambda/\Gamma<1/4$, so
\[
  \Phi_{j+1}-\Phi_j=\Delta_{L_j}(C_j)\le\beta_\eta\frac{\mathrm{cnt}(C_j)}
  {z_j}\,c_{L_j}(J')\le\frac{\mathrm{cnt}(C_j)}{4z_j}\bigl(\Phi_j+|V'|
  \bigr).
\]
Adding $\Phi_j+|V'|$ to both sides and using $1+y\le e^y$ gives
$\Phi_{j+1}+|V'|\le(\Phi_j+|V'|)\exp(\mathrm{cnt}(C_j)/(4z_j))$.
Multiplying over all iterations, starting from $\Phi_1+|V'|=|V'|$, the
final loads $L_f$, which are the loads of $J$, satisfy
$\Phi(L_f)+|V'|\le|V'|\exp\bigl((1/4)\sum_j\mathrm{cnt}(C_j)/z_j
\bigr)$. The residual demand starts at $z_1=k$, drops by
$\mathrm{cnt}(C_j)$ in iteration $j$, and ends at $0$. Each term
$\mathrm{cnt}(C_j)/z_j$ is at most the sum of $1/i$ over the
$\mathrm{cnt}(C_j)$ integers $i$ with $z_{j+1}<i\le z_j$, since each of them
is at most $z_j$. Summing over $j$ gives $\sum_j\mathrm{cnt}(C_j)/z_j\le
\sum_{i=1}^k1/i\le1+\ln k$. The source $s$ is in
$V'$, since a jump path from $s$ to a covered terminal leaves $s$; so
$1\le|V'|\le \nu$ and
\[
  1+\Phi(L_f)\le\Phi(L_f)+|V'|\le|V'|\,e^{(1+\ln k)/4}\le \nu\,e^{(1+\ln k)/4}.
\]

\emph{The loads of $J$.} For every $v\in V$ the term $e^{L_f(v)/\Gamma}-1$
is at most $\Phi(L_f)$, so $L_f(v)\le\Gamma\ln(1+\Phi(L_f))\le\Gamma\bigl(
\ln \nu+(1+\ln k)/4\bigr)$.
\end{proof}

\begin{proof}[Proof of Theorem~\ref{thm:dtmcomb}]
We show that \JumpGreedy\ (Algorithm~\ref{alg:dtm}) has the stated guarantee.

\emph{The case $\OPT\ge k$.} The schedule of
line~\ref{line:fallback} has length at most $2\OPT$ by
Lemma~\ref{lem:fallback}, and the returned schedule is not longer.

\emph{The comparison jump arborescence.} Otherwise $\OPT<k$, so $k\ge2$, and the
loop in line~\ref{line:loopT} has an iteration with $\theta=\OPT$; we consider
it. There $\theta=\OPT\ge h^*$ and, by~\eqref{eq:lambdastar},
$\Lambda=\ell(\OPT+\rho+1)\ge\Lambda^*$. For $k$-multicast, $T^*$ is
the arborescence of an optimal schedule pruned to the $k$ terminals that it
informs. Let $J^*$ be the jump arborescence that Lemma~\ref{lem:hr} gives for $T^*$,
with every node identified with the sequence of labels on its path from the
root, and with a chain of nodes with the same label and empty jumps
appended down to depth $\eta$ below every node labelled by a terminal, as
in Appendix~\ref{sec:unfold}. As its depth is at most $\eta$ and its jumps
are paths of $T^*$ (Lemma~\ref{lem:hr}), with at most $h^*\le\theta$ arcs, it is
a structure rooted at the root of $\mathcal U_{\eta,\theta}$. It covers the $k$ terminals of $T^*$, and its
maximum load is at most $\Lambda^*\le\Lambda$, as the chains change no
load. Let $V^*$ be the set of
vertices of $G$ with positive load in $J^*$. Every jump of $J^*$ is a path
of $T^*$, so it leaves only vertices of $T^*$. The arborescence $T^*$ is
the union of $k$ paths from $s$ with at most $h^*$ arcs each, so it has at
most $1+kh^*$ vertices, and $h^*\le\OPT\le k-1$. Hence $|V^*|\le1+kh^*\le
1+k(k-1)\le k^2$.

\emph{The loads of $J$ and the schedule.} By Lemma~\ref{lem:topcall} with
$J'=J^*$ and $\nu=k^2$, the top call returns a structure $J$ that covers $k$
terminals (all terminals for \DTM, so $J$ is then a jump arborescence of depth
$\eta$), and every load of $J$ is at most $\Gamma(2\ln k+(1+\ln k)/4)\le
3\Gamma\ln k$, as $k\ge2$. Since $\Gamma\le\mu+1=20(2\ell-1)\ell\,(\OPT+\rho+1)+1$, $\rho<k^{1/\ell}+1$
and $\OPT\ge1$, this gives $\Lambda(J)=O(\ell^2\log k)
(\OPT+k^{1/\ell})$. By Lemma~\ref{lem:jumpsched} and~\eqref{eq:ell}, with
$\eta\theta=(2\ell-1)\OPT$, \Schedule{}$(J)$ informs $k$ terminals within
$\ell\max\{\Lambda(J),k^{1/\ell},2\}+(2\ell-1)\OPT=O(\ell^3\log k)(\OPT+
k^{1/\ell})$ rounds. Line~\ref{line:best} keeps the shortest schedule
found, so the returned schedule is not longer.

\emph{Running time.} Every iteration of the while loop
(line~\ref{line:while} of Algorithm~\ref{alg:greedy}) counts at least one
terminal and the demand is at most $k$, so a call makes at most $k$
iterations. The node $\sigma$ has at most $n$ children in $\mathcal U$,
and in an iteration each of them needs one shortest-path computation and,
at height $h\ge2$,
one call at height $h-1$, whose list has at most $k$ structures, since
each counts at least one terminal; so the child gives at most $k$
candidates. The call at height $h-1$ returns its structures together with
their loads and the terminals that each of them removed from the residual
set, so the increases and counts of these candidates take time
$\mathrm{poly}(n)$. Hence a call at height $h$ takes time $t_h\le kn(t_{h-1}+
\mathrm{poly}(n))$, line~\ref{line:topcall} of Algorithm~\ref{alg:dtm}
makes $(kn)^\eta\mathrm{poly}(n)=n^{O(\ell)}$ arithmetic operations, and the loop in
line~\ref{line:loopT} of Algorithm~\ref{alg:dtm} makes $k-1$ iterations.
These operations act on rational numbers: the loads are integers, every
weight is $q^{L(v)}=(\mu+1)^{L(v)}/\mu^{L(v)}$, and every increase and every
ratio that an iteration compares is obtained from such numbers by
additions, subtractions, multiplications and divisions. Every jump has at
most $\theta<k$ arcs, and the run builds at most as many jumps as it makes
operations, so every load that arises is at most $n^{O(\ell)}$. As
$\mu\le20\eta\ell(2k+1)$, every number has a numerator and a denominator of
$n^{O(\ell)}$ bits, every operation takes time $n^{O(\ell)}$, and the total
running time is $n^{O(\ell)}$.
\end{proof}

\begin{proof}[Proof of Corollary~\ref{cor:dtm}]
Divide the bound by $\OPT\ge\log_2(k+1)$: the term $k^{1/\ell}/\OPT$ is at
most $k^{1/\ell}/\log_2k$. For~\ref{cor:dtm:poly} take $\ell=\max\{2,\lceil1/\epsilon
\rceil\}$: then $\ell$ is a constant, $k^{1/\ell}\le k^\epsilon$ and
$\log k=O(k^\epsilon)$. For~\ref{cor:dtm:quasi}, $k^{1/\ell}\le2$ when $\ell=\max\{2,\lceil\log_2k\rceil\}$.
For~\ref{cor:dtm:largeopt}, let $\ell=\max\{2,\lceil1/\epsilon\rceil\}$.
Then $\ell\ge1/\epsilon$, so $k^{1/\ell}\le k^\epsilon\le\OPT$, and
$\ell\le\max\{2,1/\epsilon+1\}\le2/\epsilon$, as $\epsilon\le1$. So the
schedule of Theorem~\ref{thm:dtmcomb} has length
$O(\ell^3\log k)\cdot(\OPT+k^{1/\ell})\le O(\ell^3\log k)\cdot2\OPT
=O(\epsilon^{-3}\log k)\cdot\OPT$, and it is computed in time
$n^{O(\ell)}=n^{O(1/\epsilon)}$. For a fixed $\epsilon<1$,
Proposition~\ref{prop:largeopt} shows that, unless $\mathrm{P}=\mathrm{NP}$, no
polynomial-time algorithm has ratio $(1-\delta)(1-\epsilon)\ln k$ on the
instances with $\OPT\ge k^\epsilon$, for any fixed $\delta\in(0,1)$. The
ratio $O(\epsilon^{-3}\log k)$ is therefore best possible up to a factor
that depends only on $\epsilon$.

For~\ref{cor:dtm:pw}, the \emph{pathwidth} of a graph is the least width of a tree
decomposition (Appendix~\ref{sec:plplanartw}) whose tree is a path. Let $X$
be the set of vertices that an optimal schedule informs within $\OPT$
rounds, and let $U$ be the underlying undirected graph of $G[X]$. The
vertex that informs a vertex of $X$ is informed earlier, so it is in $X$;
hence the calls of the schedule inform all of $X$ along edges of $U$
within $\OPT$ rounds, $U$ is connected, and broadcasting from $s$ in $U$
takes at most $\OPT$ rounds. The set $X$ contains $s$ and the $k$
terminals, so $|X|\ge k+1\ge2$. Restricting the bags
of a path decomposition of the underlying undirected graph of $G$ to $X$
shows that $U$ has pathwidth $q\le w$, and $q\ge1$, as $U$ is connected
and has at least two vertices. By
Corollary~21 of~\cite{egami2025broadcasting}, $\OPT^2\ge
(|X|/(q+1))^{1/(q+1)}/(2q)$, so
\[
  k\le|X|\le(q+1)\,(2q\OPT^2)^{q+1}\le(4q\OPT^2)^{q+1}\le
  (4w\OPT^2)^{w+1},
\]
as $q+1\le2^{q+1}$, $q\le w$ and $4q\OPT^2\ge1$. Hence $k^{1/(2w+2)}\le
(4w\OPT^2)^{1/2}=2\sqrt w\,\OPT$. With $\ell=2w+2$, the schedule of
Theorem~\ref{thm:dtmcomb} has length $O(\ell^3\log k)\cdot
(\OPT+k^{1/\ell})\le O(w^3\log k)\cdot(1+2\sqrt w)\,\OPT=O(w^{7/2}\log
k)\cdot\OPT$, and it is computed in time $n^{O(\ell)}=n^{O(w)}$.

Finally, Theorem~\ref{thm:dtmcomb} gives the same bound for directed
$k$-multicast, so the arguments for~\ref{cor:dtm:poly} and~\ref{cor:dtm:quasi} apply to it verbatim; an
undirected graph is a symmetric digraph.
\end{proof}

\subsection{The reduction to degree-bounded Steiner trees}
\label{sec:timeexp}

In this subsection we show that \DTM\ is a special case of a Steiner tree
problem with degree bounds (Theorem~\ref{thm:te}), and we derive what
algorithms for that problem give for \DTM\ (Corollary~\ref{cor:te}). The
ratio of Corollary~\ref{cor:te} depends on $n$ rather than on $k$, so it
complements Theorem~\ref{thm:dtmcomb}.

In the \emph{degree-bounded directed Steiner tree} problem (DB-DST) the
input is a digraph with arc costs $c\ge0$, a root, a set of terminals and a
bound $d_v\ge0$ for every vertex $v$. A solution is an out-arborescence
rooted at the root that contains all terminals and in which every vertex
$v$ has at most $d_v$ children, and its cost, the sum of the costs of its
arcs, is to be minimized. An $(\alpha,\beta)$-bicriteria algorithm returns,
whenever a solution exists, an out-arborescence rooted at the root that
contains all terminals, has cost (for a randomized algorithm, expected
cost) at most $\alpha$ times the optimum and gives every vertex $v$ at most
$\beta d_v$ children. Guo, Kortsarz, Laekhanukit, Li, Vaz and
Xian~\cite{guo2022degree} give a randomized
$(O(\log\nu\log k),O(\log^2\nu))$-bicriteria algorithm with running time
$\nu^{O(\log\nu)}$ on instances with $\nu$ vertices. We show that \DTM\ is
the special case of DB-DST with zero costs on a time-expanded digraph, and
that a violation of the degree bounds by a factor $\beta$ costs at most a
factor $\beta$ in the number of rounds.

\paragraph{The idea.} For a guess $\theta$ of the optimum, the time-expanded
digraph has a copy of every vertex for every round, and a node for every
call that a vertex can make in a round. A schedule then becomes an
out-arborescence in which, in every round, every informed vertex passes
the message to its copy in the next round and may make one call. This gives the degree bounds $2$ and $1$, and the
costs play no role, so an algorithm for DB-DST can search for the
schedule. Such an algorithm may exceed the degree bounds by a factor
$\beta$, that is, ask a vertex for $\beta$ calls in one round. The
procedure \Convert{} spreads these calls over consecutive rounds,
which stretches every informing time by the factor $\beta$ and no more.

The time-expanded digraph is defined as follows:

\begin{definition}\label{def:te}
For an integer $\theta\ge1$ the \emph{time-expanded digraph} $G_\theta$ has a
\emph{state node} $(v,t)$ for every $v\in V$ and $0\le t\le \theta$ (vertex $v$
at the end of round $t$), a \emph{call node} $c(v,t)$ for every $v\in V$
and $0\le t<\theta$ (the call of $v$ in round $t+1$), and a \emph{sink}
$\hat x$ for every terminal $x\in K$. Its arcs are $(v,t)\to(v,t+1)$ and
$(v,t)\to c(v,t)$ for every $v\in V$ and $t<\theta$, $c(v,t)\to(w,t+1)$ for
every arc $(v,w)\in E(G)$ and $t<\theta$, and $(x,\theta)\to\hat x$ for every $x\in K$.
The root is $(s,0)$, the terminals are the sinks, all costs are $0$, state
nodes have degree bound $2$ and call nodes have degree bound $1$.
\end{definition}

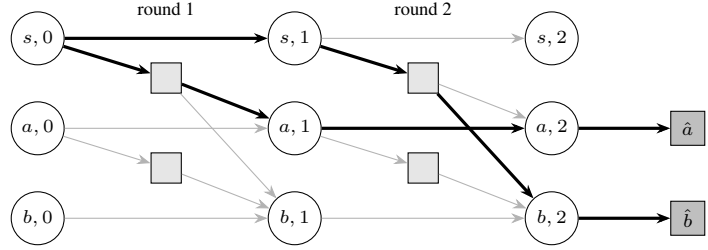
\begin{figure}[htb]
\centering
\begin{tikzpicture}[yscale=0.7,>={Stealth[length=1.8mm]},
  st/.style={circle,draw,inner sep=0.5pt,minimum size=7mm,font=\scriptsize},
  cl/.style={rectangle,draw,fill=black!10,inner sep=1pt,minimum size=4mm},
  sk/.style={rectangle,draw,fill=black!25,inner sep=1.5pt,minimum size=4.5mm,font=\scriptsize},
  faint/.style={->,black!30},
  used/.style={->,very thick}]
  \foreach \v/\y in {s/0,a/-1.7,b/-3.4} {
    \foreach \t/\x in {0/0,1/3.4,2/6.8} {
      \node[st] (\v\t) at (\x,\y) {$\v,\t$};
    }
  }
  \foreach \v/\y in {s/0,a/-1.7} {
    \foreach \t/\x in {0/1.7,1/5.1} {
      \node[cl] (c\v\t) at (\x,\y-0.75) {};
    }
  }
  \node[sk] (ha) at (8.6,-1.7) {$\hat a$};
  \node[sk] (hb) at (8.6,-3.4) {$\hat b$};
  \foreach \v in {s,a,b} { \draw[faint] (\v0) -- (\v1); \draw[faint] (\v1) -- (\v2); }
  \foreach \v in {s,a} { \draw[faint] (\v0) -- (c\v0); \draw[faint] (\v1) -- (c\v1); }
  \draw[faint] (cs0) -- (a1); \draw[faint] (cs0) -- (b1);
  \draw[faint] (cs1) -- (a2); \draw[faint] (cs1) -- (b2);
  \draw[faint] (ca0) -- (b1); \draw[faint] (ca1) -- (b2);
  \draw[faint] (a2) -- (ha); \draw[faint] (b2) -- (hb);
  \draw[used] (s0) -- (cs0); \draw[used] (cs0) -- (a1);
  \draw[used] (s0) -- (s1); \draw[used] (s1) -- (cs1); \draw[used] (cs1) -- (b2);
  \draw[used] (a1) -- (a2); \draw[used] (a2) -- (ha); \draw[used] (b2) -- (hb);
  \node[font=\scriptsize] at (1.7,0.55) {round 1};
  \node[font=\scriptsize] at (5.1,0.55) {round 2};
\end{tikzpicture}
\caption{The time-expanded digraph $G_2$ for the digraph with arcs $sa$,
$sb$ and $ab$ and terminals $a$ and $b$ (the call nodes of $b$, which has no
out-arcs, are omitted). Grey squares are call nodes, with degree bound $1$.
Bold: the arborescence of the schedule in which $s$ calls $a$ in round~1
and $b$ in round~2. The state node $(s,0)$ uses its degree bound $2$: it
waits and calls.}
\label{fig:te}
\end{figure}

If layer $t$ consists of the state nodes $(v,t)$ and the call nodes
$c(v,t)$, and the sinks form layer $\theta+1$, then every arc goes from a layer
$t$ to the layer $t$ or $t+1$. The digraph $G_\theta$ has $O(n\theta)$ vertices and
$O((n+|E(G)|)\theta)$ arcs. Figure~\ref{fig:te} shows an example.

\paragraph{Reading a schedule off an arborescence.} The procedure
\Convert{} takes an out-arborescence $A'$ of $G_\theta$, rooted at
$(s,0)$, that contains all sinks, and returns a schedule. For every vertex
$v$ of $G$ that has a state node in $A'$, let $\tau(v)$ be the least $t$
with $(v,t)\in A'$; this is the round by which $A'$ intends $v$ to be
informed. For $v\ne s$, the parent of $(v,\tau(v))$ in $A'$ is a call node
$c(u,\tau(v)-1)$, and the vertex $u$ becomes the \emph{parent} $p(v)$ of
$v$. In the schedule, every vertex $v$ of $G$ calls the vertices $w$ with
$p(w)=v$ in non-decreasing order of $\tau(w)$, one per round, starting in
the round after it is informed. Algorithm~\ref{alg:te} states \Convert{} in
pseudocode.

\paragraph{The algorithm.} The function \MulticastDBDST{}
(Algorithm~\ref{alg:te}) takes as a parameter any algorithm
\DBDST{} for DB-DST. It first computes the busy schedule along a
shortest-path out-arborescence from $s$ to $K$, as in
Algorithm~\ref{alg:dtm}, and keeps it as the best schedule found so far
(line~\ref{line:tefallback}). Then, for every $\theta=1,\dots,k$, it builds
the time-expanded digraph $G_\theta$ of Definition~\ref{def:te}
(line~\ref{line:buildgt}) and calls \DBDST{} on it
(line~\ref{line:dbdst}). If the
result $A'$ is an out-arborescence rooted at $(s,0)$ that contains all
sinks, the function turns it into a schedule with \Convert{} and
keeps the shorter of this schedule and the best one
(line~\ref{line:convert}). Finally it returns the best schedule.

\begin{algorithm}[H]
\caption{Directed multicast through degree-bounded Steiner trees, for a
given DB-DST algorithm \DBDST{}}
\label{alg:te}
\begin{algorithmic}[1]
\Require
  \IOitem{$A'$: an out-arborescence of the time-expanded digraph $G_\theta$,
    rooted at $(s,0)$, that contains all sinks}
\Ensure a schedule that informs all terminals
\Function{\Convert}{$A'$}
  \ForAll{vertices $v$ of $G$ with a state node in $A'$}
    \State $\tau(v)\gets\min\{t:(v,t)\in A'\}$\label{line:tau}
  \EndFor
  \ForAll{such vertices $v\ne s$}
    \State $p(v)\gets$ the vertex $u$ such that $c(u,\tau(v)-1)$ is the
      parent of $(v,\tau(v))$ in $A'$\label{line:parent}
  \EndFor
  \State \Return the schedule in which every vertex $v$ of $G$ calls the
    vertices $w$ with $p(w)=v$ in non-decreasing order of $\tau(w)$, one per
    round, starting in the round after it is informed\label{line:order}
\EndFunction
\Statex
\Require
  \IOitem{$G$: a digraph with $n$ vertices; $s$: the source; $K$: the set of
    $k$ terminals}
  \IOitem{fixed throughout: an algorithm \DBDST{} for DB-DST, as in
    Corollary~\ref{cor:te}}
\Ensure a schedule that informs all terminals
\Function{\MulticastDBDST}{$G,s,K$}
  \State $\mathit{best}\gets$ the busy schedule along a shortest-path
    out-arborescence from $s$ to $K$\label{line:tefallback}
  \For{$\theta=1,\dots,k$}\label{line:teloop}
    \State build the time-expanded digraph $G_\theta$ of
      Definition~\ref{def:te}\label{line:buildgt}
    \State $A'\gets{}$\Call{\DBDST}{$G_\theta$}\label{line:dbdst}
    \If{$A'$ is an out-arborescence rooted at $(s,0)$ that contains all
      sinks}
      \State $\mathit{best}\gets$ the shorter of $\mathit{best}$ and
        \Call{\Convert}{$A'$}\label{line:convert}
    \EndIf
  \EndFor
  \State \Return $\mathit{best}$
\EndFunction
\end{algorithmic}
\end{algorithm}

The time-expanded digraph captures \DTM\ exactly when the degree bounds
hold, and an arborescence that violates them by a factor $\beta$ still
gives a schedule that is only a factor $\beta$ longer:

\begin{theorem}\label{thm:te}
Let $\theta\ge1$.
\begin{enumerate}[label=(\roman*),ref=(\roman*)]
  \item\label{thm:te:solution} If $\OPT\le \theta$, then $G_\theta$ has a solution of DB-DST, that is, an
    out-arborescence rooted at $(s,0)$ that contains all sinks and
    respects all degree bounds.
  \item\label{thm:te:convert} If $A'$ is an out-arborescence of $G_\theta$ rooted at $(s,0)$ that
    contains all sinks and in which every call node has at most
    $\beta\ge1$ children, then \Convert{}$(A')$ is a schedule of
    length at most $\beta \theta$, computed in polynomial time.
\end{enumerate}
In particular, $\OPT$ is the least $\theta$ for which $G_\theta$ has a solution.
\end{theorem}

\begin{proof}
\ref{thm:te:solution} Take a schedule of length at most $\theta$ and let $\tau_v$ be the round in
which $v$ is informed, with $\tau_s=0$. For every informed vertex $v$ of $G$ take
the arcs $(v,t)\to(v,t+1)$ for $\tau_v\le t<\theta$. For every call by which
$v$ informs $w$ in round $r$ (calls to informed vertices are ignored) take
$(v,r-1)\to c(v,r-1)\to(w,r)$; here $\tau_v\le r-1$ and $r=\tau_w\le \theta$.
Finally take $(x,\theta)\to\hat x$ for every terminal $x$. Each chosen node
other than $(s,0)$ has exactly one entering arc: $(w,\tau_w)$ from the call
node of its caller, $(v,t)$ with $t>\tau_v$ from $(v,t-1)$, a call node
from its state node, and a sink from $(x,\theta)$. Every arc goes from layer $t$
to layer $t+1$, or within a layer from a state node to a call node, so the
chosen arcs form an out-arborescence rooted at $(s,0)$, and it contains all
sinks. A state node has at most two children, the next state node and one
call node, because $v$ calls at most once per round, and every call node
has one child.

\ref{thm:te:convert} As $(s,0)$ is the root of $A'$, we have $\tau(s)=0$. For $v\ne s$, the arc of $A'$
that enters $(v,\tau(v))$ does not come from $(v,\tau(v)-1)$, by the choice
of $\tau(v)$, so it comes from a call node $c(u,\tau(v)-1)\in A'$ with
$(u,v)\in E(G)$; hence the parent $p(v)=u$ is well defined. The
only arc into this call node comes from $(u,\tau(v)-1)$, which is therefore
in $A'$, so $\tau(p(v))\le\tau(v)-1$. Since $\tau$ increases from parent to
child, the arcs $(p(v),v)$ form an out-arborescence $T'$ of $G$ rooted at
$s$, and the calls of \Convert{} form a schedule along $T'$.
The arborescence $T'$ contains every terminal $x$: the sink $\hat x$ is in $A'$, hence so is
$(x,\theta)$.

Fix a vertex $v$ of $T'$ and an integer $j\ge1$. A child $w$ of $v$ in $T'$
with $\tau(w)\le\tau(v)+j$ is a child in $A'$ of the call node
$c(v,\tau(w)-1)$, and this call node is in $A'$ only if $(v,\tau(w)-1)$ is,
so $\tau(v)\le\tau(w)-1\le\tau(v)+j-1$. These are $j$ call nodes with at
most $\beta$ children each, so
\begin{equation}\label{eq:prefix}
  \bigl|\{w\text{ child of }v\text{ in }T':\ \tau(w)\le\tau(v)+j\}\bigr|
  \;\le\;\beta j .
\end{equation}
We show by induction on $\tau$ that every vertex $w$ of $T'$ is informed by
round $\beta\tau(w)$. For $s$ this holds. Let $w$ be the $r$-th child that
its parent $v$ calls in the schedule of \Convert{}, and
$j=\tau(w)-\tau(v)\ge1$.
The first $r$ children called have $\tau$ at most $\tau(w)$, so $r\le\beta
j$ by~\eqref{eq:prefix}. As $v$ is informed by round $\beta\tau(v)$, the
child $w$ is informed by round $\beta\tau(v)+r\le\beta(\tau(v)+j)=
\beta\tau(w)\le\beta \theta$. \Convert{} clearly runs in time polynomial in
the size of $G_\theta$. The last claim of Theorem~\ref{thm:te} is~\ref{thm:te:solution} together with~\ref{thm:te:convert}
for $\beta=1$.
\end{proof}

By Theorem~\ref{thm:te}, every algorithm for DB-DST that violates the
degree bounds by a factor $\beta$, at any cost, yields a
$\beta$-approximation for \DTM\ when it is run on $G_\theta$ for
$\theta=1,\dots,n-1$, as $\OPT\le n-1$. Algorithm~\ref{alg:te} needs only
$\theta\le k$, because its fallback schedule handles the case $\OPT\ge k$
(Lemma~\ref{lem:fallback}). With a bicriteria algorithm for DB-DST, this
gives an approximation algorithm for \DTM:

\begin{corollary}\label{cor:te}
Suppose \DBDST{} runs in time $R(\nu)$ on every DB-DST instance with $\nu$
vertices and, if the instance has a solution, returns an out-arborescence
rooted at its root that contains all terminals and gives every vertex $v$ at
most $\beta(\nu)d_v$ children, of any cost, where $R$ and $\beta\ge1$ are
nondecreasing functions. Then
Algorithm~\ref{alg:te} is a $\max\{2,\beta(Cnk)\}$-approximation for \DTM\
with running time $O(k\cdot R(Cnk))+n^{O(1)}$, for some constant $C$ such
that every $G_\theta$ with $\theta\le k$ has at most $Cnk$ vertices. With the algorithm
of~\cite{guo2022degree}, \DTM\ has a randomized
$O(\log^2n)$-approximation with running time $n^{O(\log n)}$.
\end{corollary}

\begin{proof}
If $\OPT\ge k$, the schedule of line~\ref{line:tefallback} of
Algorithm~\ref{alg:te} has length at most $2\OPT$ by
Lemma~\ref{lem:fallback}, and the returned schedule is not longer. If
$\OPT<k$, consider the iteration $\theta=\OPT$ of the loop in
line~\ref{line:teloop}, and let $\nu_\theta$ be the number of vertices of $G_\theta$. By
Theorem~\ref{thm:te}\ref{thm:te:solution} the instance $G_\theta$ has a solution, so
line~\ref{line:dbdst} returns an out-arborescence rooted at $(s,0)$ that
contains all sinks and in which every call node has at most
$\beta(\nu_\theta)$ children, and by
Theorem~\ref{thm:te}\ref{thm:te:convert} line~\ref{line:convert} yields a schedule of length
at most $\beta(\nu_\theta)\,\OPT$.

By Definition~\ref{def:te}, $G_\theta$ has $n(\theta+1)$ state nodes, $n\theta$ call nodes
and $k$ sinks, so $\nu_\theta=n(2\theta+1)+k$. For $\theta\le k$ this is at most $4nk$, as
$n+k\le2nk$ for $n,k\ge1$; so $C=4$ suffices. As $\beta$ is nondecreasing,
the schedule above has length at most $\beta(Cnk)\,\OPT$, and so does the
returned schedule. In both cases, the returned schedule has length at most
$\max\{2,\beta(Cnk)\}\,\OPT$. The loop calls \DBDST{} $k$ times, each time in
time $R(\nu_\theta)\le R(Cnk)$, as $R$ is nondecreasing, and building $G_\theta$,
checking $A'$ and \Convert{} take time polynomial in $n$, as $k\le n$.
This gives the running time $O(k\cdot R(Cnk))+n^{O(1)}$.

For the algorithm of~\cite{guo2022degree}, $\beta(\nu)=O(\log^2\nu)$ and
$R(\nu)=\nu^{O(\log \nu)}$, and $\log(Cnk)=O(\log n)$, since $k\le n$. Their
algorithm meets the terminal and degree requirements with constant
probability; as both can be checked, $O(\log n)$ repetitions succeed with
high probability.
\end{proof}

The ratio of Corollary~\ref{cor:te} depends on $n$, and the algorithm is
randomized and solves linear programs. It is better than
Corollary~\ref{cor:dtm}\ref{cor:dtm:quasi} when $\log n$ is small compared with $\log^2k$.
The reduction uses only zero costs, degree bounds $1$ and $2$ and layered
digraphs, so any algorithm for DB-DST with a smaller degree violation, even
one that works only on such instances, gives \DTM\ the same ratio.

Corollary~\ref{cor:te} also turns polynomial-time algorithms for DB-DST
into algorithms for \DTM. Bamas, Li and
Rohwedder~\cite{bamas2026randomized} solve problems whose solutions follow
the recurrence of a dynamic program under additional packing constraints,
which they satisfy approximately, by randomized rounding of a linear
program. They remark, without details, that their framework handles the
degree bounds of DB-DST and gives a polynomial-time
$O(\nu^\epsilon)$-approximation for DB-DST on $\nu$ vertices, for every fixed
$\epsilon>0$. This follows from their algorithm for robust directed Steiner
tree cover~\cite[Theorem~1.5]{bamas2026randomized}, with one packing
constraint for the out-degree of every vertex, repeated until all
terminals are covered: the degree bounds are violated by a factor
$\nu^{O(\epsilon)}$, in time $\nu^{O(1/\epsilon)}$. In
Corollary~\ref{cor:te}, after rescaling $\epsilon$, this means
$\beta(\nu)=O(\nu^\epsilon)$ and $R(\nu)=\nu^{O(1/\epsilon)}$. As $Cnk\le Cn^2$ for
the constant $C$ of Corollary~\ref{cor:te}, Corollary~\ref{cor:te} then
gives \DTM\ a randomized ratio $\max\{2,\beta(Cnk)\}=O(n^{2\epsilon})$ in
time $O(k\cdot R(Cnk))+n^{O(1)}=n^{O(1/\epsilon)}$, that is, ratio
$O(n^\epsilon)$ in time $n^{O(1/\epsilon)}$ for every fixed $\epsilon>0$.
Corollary~\ref{cor:dtm}\ref{cor:dtm:poly} has ratio $O(k^\epsilon)$ in the same time, with
a deterministic combinatorial algorithm, and by Theorem~\ref{thm:dtmcomb}
the ratio is even $O(\ell^3\log k)$ when $\OPT\ge k^{1/\ell}$, because the
term $k^{1/\ell}$ there is additive; this gives
Corollary~\ref{cor:dtm}\ref{cor:dtm:largeopt}. The introduction compares the bounds in
more detail, in particular when $k$ is small compared with $n$.

\section{Planar digraphs and other graph classes}
\label{sec:planar}

This appendix presents a deterministic polynomial-time algorithm with ratio
$O(\log^2k)$ for directed multicast on planar digraphs, and more generally
with ratio $O(p\log^2k)$ on digraphs whose underlying graphs lie in a
minor-closed class with balanced separators made of $p$ root paths of any
spanning tree, and proves Theorems~\ref{thm:plclass} and~\ref{cor:plplanar}
and Corollary~\ref{cor:pltw}. Planar digraphs are the natural first class
beyond trees for problems of this kind: for directed Steiner tree, which
has no polylogarithmic ratio in polynomial time known in general, Friggstad
and Mousavi obtained ratio $O(\log k)$ on planar
digraphs~\cite[Theorem~1]{friggstad2023planar}. On general digraphs,
the ratio of Corollary~\ref{cor:dtm}\ref{cor:dtm:poly} is polynomial in $k$.

\paragraph{Graph classes.} Graphs in this appendix are undirected and have
no loops or parallel edges. The \emph{underlying graph} of a digraph has
the same vertices, and two distinct vertices are adjacent in it if an arc
joins them, in either direction; so the weakly connected components of a
digraph are the components of its underlying graph. A digraph is
\emph{planar} if its underlying graph is planar. For a set $V'$ of vertices
of a graph $H$ that induces a connected subgraph, the graph $H/V'$ arises
from $H$ by \emph{contracting} $V'$: we delete $V'$ and add a new vertex
adjacent to every vertex outside $V'$ that has a neighbour in $V'$. A
\emph{minor} of $H$ is a graph obtained from a subgraph of $H$ by
contracting the two ends of edges, one edge at a time; so $H/V'$ is a minor
of $H$ (contract the edges of a spanning tree of $H[V']$ one at a time),
and so is every subgraph of a minor. A class of graphs
is \emph{minor-closed} if it contains every graph isomorphic to a minor of
one of its graphs. For a tree or an out-arborescence $A$ rooted at a vertex
$r$ and a vertex $v$ of $A$, the \emph{root path} $A[r,v]$ is the vertex set
of the path in $A$ from $r$ to $v$, and for a set $Z$ of vertices of $A$,
$A[r,Z]:=\bigcup_{v\in Z}A[r,v]$. \emph{Vertex weights} on $H$ are a function
$\mu\colon V(H)\to\mathbb R_{\ge0}$, with $\mu(V'):=\sum_{v\in V'}\mu(v)$.

The algorithm needs balanced separators made of few root paths, for every
root and every spanning tree, as in Definition~\ref{def:plclass}, which we
restate:

\defplclass*

For fixed $p$ the algorithm is automatic: trying
all sets $Z$ of at most $p$ vertices takes time $n^{p+O(1)}$ on graphs with
$n$ vertices. Abraham and Gavoille~\cite[Definition~1]{abraham2006object}
call a weighted graph with $n$ vertices $q$-path separable (they write $k$)
if removing at most $q$ paths, taken in groups, each a shortest path in the
graph left after removing the earlier groups, leaves components with at
most $n/2$ vertices that are again $q$-path separable; Diot and
Gavoille~\cite{diot2010path} extend this to vertex weights and require such
a separator, leaving components of at most half the weight, in every
induced subgraph. Definition~\ref{def:plclass} asks instead, for every root
and every spanning tree, for one balanced separator made of root paths of
that tree. The main result of this appendix is an algorithm for
every such class, Theorem~\ref{thm:plclass}, restated:

\thmplclass*

Since $(\bar d+1)^2\le(1+\log_2k)^2\le4\log_2^2k$ for $k\ge2$, the ratio is
$O(p\log^2k)$; for $k=1$ a shortest path from $s$ to the terminal is an
optimal schedule. By Lemma~\ref{lem:plplanarclass}, the planar graphs form
a $3$-root-path separable class, so Theorem~\ref{thm:plclass} gives
Theorem~\ref{cor:plplanar}, restated:

\corplplanar*

The \emph{treewidth} of a graph, defined in Appendix~\ref{sec:plplanartw},
measures how close the graph is to a tree. By Lemma~\ref{lem:pltwsep}, for
every fixed $w$ the graphs of treewidth at most $w$ form a
$(w+1)$-root-path separable class, so Theorem~\ref{thm:plclass} also
gives Corollary~\ref{cor:pltw}, restated:

\corpltw*

For bounded treewidth, Hathcock, Kortsarz and Ravi give a randomized
polynomial-time $O(\log^2n)$-approximation for degree-bounded directed
Steiner tree with degree violation
$O(\log^2n)$~\cite[Theorem~1.4]{hathcock2026telephone}. It does not bound the height of
the tree, so it gives no schedule; nor does it give one through
Theorem~\ref{thm:te}, as the time-expanded digraph of Definition~\ref{def:te}
need not have bounded treewidth.

As for Theorem~\ref{thm:dtmcomb}, the ratios depend only on $k$ and on the
class, and $n$ enters only the running time. Unlike the algorithm
of Theorem~\ref{thm:dtmcomb}, it solves a linear program; like that
algorithm, it is deterministic. Appendices~\ref{sec:ploverview}--\ref{sec:planalysis}
present and analyse the algorithm, and Appendix~\ref{sec:plplanartw} proves
Theorem~\ref{cor:plplanar} and Corollary~\ref{cor:pltw}.

\subsection{Overview of the algorithm}
\label{sec:ploverview}

Section~\ref{sec:ovplanar} explains the ideas behind the algorithm; here
we only recall its plan, which Appendices~\ref{sec:pldefs}
and~\ref{sec:plalg} define formally. Like the algorithm of Friggstad and
Mousavi for directed Steiner tree on planar digraphs, it buys a balanced
separator made of root paths, contracts it into the root, and recurses on
the components, the \emph{instances} of the next level; there are at most
$\lfloor\log_2k\rfloor+1$ levels. For multicast what matters is the
\emph{per-level load} of a vertex of $G$, the number of arcs from it into
the separators of one level, as a schedule that informs the levels one
after the other takes the sum over the levels of the largest per-level
load, plus $O(\theta+\sigma)$ rounds per level for a guess $\theta$ of $\OPT$ and
separators made of at most $\sigma$ root paths (Lemma~\ref{lem:plphase}).
A linear program guides the choices: an instance may be entered only
through \emph{cheap} arcs, whose tails have small weight compared with the
instance's weighted entry capacity, which by Markov's
inequality~\cite[Theorem~3.2]{motwani1995randomized} keeps most
of every terminal's flow (Lemma~\ref{lem:plsurvive}); and the weights form a
pessimistic estimator that bounds every per-level load by
$O(\sigma \theta\log k)$ (Lemma~\ref{lem:plloads}).

\subsection{The linear program, instances and separators}
\label{sec:pldefs}

In this subsection we define the linear program that guides the algorithm,
the instances of its recursion, and the separators that it uses.
Throughout this appendix, $\bar d:=\lfloor\log_2k\rfloor$. For an integer
$\theta\ge1$, let $V_{\le \theta}$ be the set of vertices $v$ with $\dist(s,v)\le \theta$,
and let $G_{\le \theta}$ be the subdigraph of $G$ induced by $V_{\le \theta}$, with
the arcs into $s$ deleted. For a terminal $t\in K\cap V_{\le \theta}$, let
$\mathcal P_t$ be the set of directed paths from $s$ to $t$ in $G_{\le \theta}$
with at most $\theta$ arcs.

\begin{definition}\label{def:pllp}
For an integer $\theta\ge1$ with $K\subseteq V_{\le \theta}$, the linear program
$\LP(\theta)$ has a variable $x_e\ge0$ for every arc $e$ of $G_{\le \theta}$ and a
variable $f^t_P\ge0$ for every terminal $t$ and every path
$P\in\mathcal P_t$, and the constraints
\begin{gather*}
  \sum_{P\in\mathcal P_t}f^t_P=1\quad(t\in K),\qquad
  \sum_{P\in\mathcal P_t:\,e\in P}f^t_P\le x_e\quad(t\in K,\ e\in
  E(G_{\le \theta})),\\
  \sum_{e\in E(G_{\le \theta})\text{ leaving }v}x_e\le \theta\quad(v\in V_{\le \theta}).
\end{gather*}
It has no objective. If $K\not\subseteq V_{\le \theta}$, we call $\LP(\theta)$
infeasible.
\end{definition}

Lemma~\ref{lem:pllp} shows that $\LP(\theta)$ is a relaxation of the problem and
can be solved in polynomial time.

\begin{lemma}\label{lem:pllp}
Let $\theta\ge1$ be an integer.
\begin{enumerate}[label=(\alph*),ref=(\alph*)]
  \item\label{lem:pllp:feasible} If $\theta\ge\OPT$, then $\LP(\theta)$ is feasible.
  \item\label{lem:pllp:solve} In time polynomial in the size of $G$ and in $\theta$ one can decide
    whether $\LP(\theta)$ is feasible and, if it is, compute a vector $x$ of
    rationals of polynomial bit size for which some $f$ makes $(x,f)$
    feasible.
\end{enumerate}
\end{lemma}

\begin{proof}
\ref{lem:pllp:feasible} Let $T^*$ be the out-arborescence of an optimal schedule, pruned to the
paths from $s$ to the terminals (Section~\ref{sec:ovsched}). Its height and
out-degree are at most $\OPT\le \theta$, so all its vertices lie in $V_{\le \theta}$,
and it has no arc into its root $s$. Hence $x_e:=1$ for the arcs $e$ of
$T^*$, $x_e:=0$ for the other arcs, and $f^t_P:=1$ for the path $P$ of $T^*$
from $s$ to $t$ (and $0$ otherwise) satisfy the constraints of
Definition~\ref{def:pllp}: every $f^t$ uses only arcs of $T^*$, and the
capacity out of a vertex is its out-degree in $T^*$.

\ref{lem:pllp:solve} If $K\not\subseteq V_{\le \theta}$ the answer is no. Otherwise consider the
layered digraph with a node $(v,h)$ for every $v\in V_{\le \theta}$ and $0\le
h\le \theta$, and an arc from $(u,h)$ to $(v,h+1)$ for every arc $(u,v)$ of
$G_{\le \theta}$ and $0\le h<\theta$. The \emph{layered program}, a linear program
of polynomial size, has the variables $x_e$,
and for every terminal $t$ a flow $g^t\ge0$ in the layered digraph of value
$1$ from $(s,0)$ to the nodes $(t,h)$, $1\le h\le \theta$, such that the sum of
$g^t$ over the $\theta$ copies of an arc $e$ is at most $x_e$; the out-degree
constraints are those of Definition~\ref{def:pllp}. It has $O(k\theta|E(G)|)$
variables and constraints, so linear programming decides it and finds a
basic solution, whose entries are rationals of polynomial bit size. The
layered program and $\LP(\theta)$ have the same feasible vectors $x$. Given
$(x,f)$, layering every path of $\mathcal P_t$ (its $i$-th arc goes from
layer $i-1$ to layer $i$) gives $g^t$. Conversely, a path decomposition of
$g^t$ projects to walks from $s$ to $t$ in $G_{\le \theta}$ with at most $\theta$
arcs; cutting each walk at its first visit of $t$ and removing its cycles
gives a path of $\mathcal P_t$ whose arcs are a sub-multiset of the walk's
arcs, so the flow on no arc increases, and the result is an $f^t$ with
$\sum_{P\ni e}f^t_P\le x_e$.
\end{proof}

Let $\TLP$ denote the least integer $\theta\ge1$ for which $\LP(\theta)$ is feasible. As
the terminals are reachable from $s$, $\OPT$ is finite, so $\TLP$ exists and
$\TLP\le\OPT$ by Lemma~\ref{lem:pllp}\ref{lem:pllp:feasible}.

The algorithm works on instances: a region of $G_{\le \theta}$ still to be
reached and the vertices from which it may be entered.

\begin{definition}\label{def:plinst}
Fix $\theta$. An \emph{instance} is a triple $I=(R,W,K_I)$ of disjoint subsets
$R,W$ of $V_{\le \theta}$ with $s\in R$ and $G_{\le \theta}[R]$ weakly connected, and
with $K_I=K\cap W$. We call $R$ the \emph{root set}, $W$ the \emph{region}
and $K_I$ the \emph{terminals} of $I$, and also write $R_I:=R$ and
$W_I:=W$. The \emph{instance digraph} $H(I)$ has as vertices the vertices
of $W$ and a new vertex $r$, its \emph{root}. For every arc $(u,v)$ of
$G_{\le \theta}$ with $u,v\in W$ it has an \emph{internal arc} from $u$ to $v$,
and for every arc $(u,v)$ of $G_{\le \theta}$ with $u\in R$ and $v\in W$ an
\emph{entry arc} from $r$ to $v$; parallel arcs are kept. Every arc $e$ of
$H(I)$ thus comes from an arc of $G_{\le \theta}$, which we denote by $e$ as
well; its tail in $G_{\le \theta}$ is the \emph{$G$-tail} $\tau(e)$ of $e$, a
vertex of $R$ for an entry arc and the tail of $e$ for an internal arc.
\end{definition}

The instance digraph $H(I)$ has no arc into $r$, and its underlying graph
is a minor of that of $G$ (Lemma~\ref{lem:plstruct}\ref{lem:plstruct:minor}), so it lies in
every minor-closed class that contains the underlying graph of $G$.
Separators are defined for a digraph with a spanning out-arborescence:

\begin{definition}\label{def:plsep}
Let $H$ be a digraph, $r$ a vertex of $H$, $A$ an out-arborescence of $H$
rooted at $r$ that contains every vertex of $H$, and $K'\subseteq
V(H)\setminus\{r\}$. For an integer $p\ge1$, a \emph{$p$-root-path
separator} of $(H,A,K')$ is a union $S$ of at least one and at most $p$ root paths of
$A$ such that every weakly connected component of $H-S$ contains at most
$|K'|/2$ vertices of $K'$. Every root path contains $r$, so $r\in S$.
\end{definition}

\subsection{The algorithm}
\label{sec:plalg}

In this subsection we present the algorithm. Fix an integer $p\ge1$ and a
$p$-root-path separable class $\mathcal G$ that contains the underlying graph of
$G$, and call the algorithm of Definition~\ref{def:plclass} for
$\mathcal G$ the \emph{class algorithm}.

\paragraph{Parameters.} Separators consist of at most
$\sigma:=\min\{p,k\}$ root paths, and the algorithm uses the parameters
\[
  \epsilon:=\frac1{2(\bar d+1)},\qquad\delta:=\frac1{100},\qquad
  \beta:=\frac7{4\epsilon}=\frac72(\bar d+1) .
\]
The value of $\beta$ only needs $\beta\ge(e-1)e^\delta/\epsilon$
(Lemma~\ref{lem:plloads}), and $(e-1)e^{1/100}<1.736<7/4$; we take a
rational value so that the weights of step~\ref{step:plweight} of the processing of an
instance have rational exponents. As
$\ln(1/(1-\epsilon))\le\epsilon/(1-\epsilon)$, for every
$d\in\{0,\dots,\bar d\}$,
\begin{equation}\label{eq:plscale}
  (1-\epsilon)^{-d}\le\exp\Bigl(\frac{\epsilon \bar d}{1-\epsilon}\Bigr)
  =\exp\Bigl(\frac{\bar d}{2\bar d+1}\Bigr)\le e^{1/2}.
\end{equation}

\paragraph{Masses.} Let $x$ be the vector computed by
Lemma~\ref{lem:pllp}\ref{lem:pllp:solve}. For an instance $I$ of depth $d$ (the level of the
recursion at which $I$ is processed), the \emph{scaled capacity} of an arc
$e$ of $G_{\le \theta}$ whose head lies in $W_I$ is
$x^I_e:=(1-\epsilon)^{-d}x_e$, and the \emph{entry mass} of a vertex $u$ in
$I$ is
\[
  m_I(u):=\sum_{e\text{ entry arc of }H(I)\text{ with }\tau(e)=u}x^I_e ,
\]
which is $0$ for every vertex $u$ of $G$ outside $R_I$.

\paragraph{Processing an instance.} The function \LevelSep{}$(G,s,K,\theta)$
(Algorithm~\ref{alg:pllevelsep}) first computes $x$. It then processes the
recursion level by level; the set of instances of level $d$ is denoted by
$\mathcal I_d$. Level $0$ consists of the instance
$I_0:=(\{s\},V_{\le \theta}\setminus\{s\},K)$. For $d=0,1,\dots$, as long as
$\mathcal I_d$ is not empty, the function processes the instances of level
$d$ one by one, in any order $I_1,\dots,I_q$, and the children of these
instances form $\mathcal I_{d+1}$. Before the first instance of level $d$
it sets, for every vertex $u$ of $G$, the \emph{load} $\ell(u):=0$ and the
\emph{future mass} $\bar m(u):=\sum_{j=1}^q m_{I_j}(u)$. When it processes $I_j$,
it does the following.
\begin{enumerate}[label=(\arabic*)]
  \item\label{step:plweight} It removes the masses of $I_j$ from the future masses:
    $\bar m(u):=\bar m(u)-m_{I_j}(u)$ for every vertex $u$ of $G$. The \emph{weight}
    of a vertex $u$ of $G$ is then $\omega(u):=\exp(\ell(u)/\sigma+\beta\bar m(u))$. The
    exponent is a rational number, and the function computes, for every
    vertex $u\in R_{I_j}$, a rational number $\tilde\omega(u)$ with
    $\omega(u)\le\tilde\omega(u)\le e^\delta\omega(u)$.
  \item\label{step:plmass} It computes the \emph{weighted entry mass}
    $\Omega_j:=\sum_{u\in R_{I_j}}\tilde\omega(u)m_{I_j}(u)$. An entry arc
    $e$ of $H(I_j)$ is \emph{cheap} if
    $\tilde\omega(\tau(e))\le\Omega_j/\epsilon$.
  \item\label{step:plreach} Let $H'_j$ be the subdigraph of $H(I_j)$ induced by $r$ and the
    vertices that can be reached from $r$ by a directed path with at most
    $\theta$ arcs whose first arc is cheap. The function computes a
    breadth-first out-arborescence $A_j$ from $r$ of the digraph obtained
    from $H'_j$ by deleting its entry arcs that are not cheap. If some
    terminal of $I_j$ is not a vertex of $H'_j$, the function stops with
    failure; Lemma~\ref{lem:plsurvive} shows that this does not happen.
  \item\label{step:plsep} It chooses the \emph{separator} $S_j$ of $I_j$. If $I_j$ has a
    single terminal $t$, then $S_j:=A_j[r,t]$. Otherwise the
    \emph{separator procedure} \Sep{} (in Algorithm~\ref{alg:pllevelsep})
    chooses $S_j$: if $p\le k$, it
    applies the class algorithm to the underlying graph of $H'_j$, the
    vertex $r$, the underlying graph of $A_j$ as the spanning tree, and the
    weights $1$ on $K_{I_j}$ and $0$ elsewhere, and returns
    $S_j:=A_j[r,Z\cup\{r\}]$ for the set $Z$ of vertices of $H'_j$ that
    the class algorithm returns (so $S_j=A_j[r,Z]$ if $Z\neq\emptyset$); if
    $p>k$, it returns the union $S_j$ of the root paths of $A_j$ to the
    terminals of $I_j$. In every case $S_j$ is a union of at least one root
    path of $A_j$, so $r\in S_j$. The \emph{piece} $Q_j$ of $I_j$ is the set
    of arcs of $A_j$ whose heads lie in $S_j$, that is, the arcs of the root
    paths that make up $S_j$; no arc of $A_j$ enters its root $r$.
  \item\label{step:plload} For every vertex $u$ of $G$ it adds to $\ell(u)$ the number $a_u$ of
    arcs of $Q_j$ that leave $r$ and have $G$-tail $u$.
  \item\label{step:plchild} For every weakly connected component $C$ of $H'_j-S_j$ that
    contains a terminal, it creates the \emph{child}
    $I_C:=(R_{I_j}\cup(S_j\setminus\{r\}),C,K\cap C)$ of $I_j$, an
    instance of level $d+1$.
\end{enumerate}

When an instance $I$ is processed as $I_j$, we also write $S_I:=S_j$. A
separator is a union
of root paths, so it contains the parent in $A_j$ of each of its vertices
other than $r$; hence the piece is an out-arborescence rooted at $r$, and
its paths from $r$ are the paths of $A_j$ whose vertex sets make up $S_j$.
When all levels are processed, \LevelSep{}$(G,s,K,\theta)$ returns the
pieces together with their depths. Figure~\ref{fig:plstep} illustrates one
instance.

\begin{algorithm}[tp]
\caption{The separator procedure and the level-by-level separator recursion for a guess $\theta$}
\label{alg:pllevelsep}
\begin{algorithmic}[1]
\Require
  \IOitem{$I_j$: an instance with at least two terminals}
  \IOitem{$H'_j$, $A_j$: the digraph and the arborescence of step~\ref{step:plreach} for $I_j$}
  \IOitem{fixed throughout: $p$, the number $k$ of terminals, and the class algorithm}
\Ensure the separator $S_j$ of step~\ref{step:plsep}, a union of root paths of $A_j$
\Function{\Sep}{$I_j,H'_j,A_j$}
  \State \textbf{if} $p\le k$ \textbf{then return} $A_j[r,Z\cup\{r\}]$ for the set $Z$ that the class algorithm computes as in step~\ref{step:plsep}
  \State \Return the union of the root paths of $A_j$ to the terminals of $I_j$
\EndFunction
\Statex
\Require
  \IOitem{$G$, $s$, $K$: the digraph, the source and the terminals; $\bar d=\lfloor\log_2k\rfloor$}
  \IOitem{$\theta\in\{1,\dots,k-1\}$: the guess for $\OPT$, with $\LP(\theta)$ feasible}
  \IOitem{fixed throughout: $\sigma$ and $\epsilon,\delta,\beta$, the parameters of Appendix~\ref{sec:plalg}}
\Ensure the pieces of all instances, with their depths, or failure
\Function{\LevelSep}{$G,s,K,\theta$}
  \State $x\gets$ a vector as in Lemma~\ref{lem:pllp}\ref{lem:pllp:solve} for $\LP(\theta)$ (Definition~\ref{def:pllp})
  \State $\mathcal I_0\gets\{(\{s\},V_{\le \theta}\setminus\{s\},K)\}$; $d\gets0$
  \While{$\mathcal I_d\neq\emptyset$}
    \State fix an order $I_1,\dots,I_q$ of $\mathcal I_d$; $\mathcal I_{d+1}\gets\emptyset$
    \State $\ell(u)\gets0$ and $\bar m(u)\gets\sum_{j=1}^q m_{I_j}(u)$ for every vertex $u$ of $G$
    \For{$j=1,\dots,q$}
      \State $\bar m(u)\gets \bar m(u)-m_{I_j}(u)$ for every vertex $u$ of $G$
      \State $\tilde\omega(u)\gets$ a rational number in $[\omega(u),e^\delta\omega(u)]$ for every $u\in R_{I_j}$, where $\omega(u)=\exp(\ell(u)/\sigma+\beta\bar m(u))$
      \State $\Omega_j\gets\sum_{u\in R_{I_j}}\tilde\omega(u)\,m_{I_j}(u)$\Comment{an entry arc $e$ of $H(I_j)$ is cheap if $\tilde\omega(\tau(e))\le\Omega_j/\epsilon$}
      \State $H'_j\gets$ the part of $H(I_j)$ that $r$ reaches within $\theta$ arcs through a cheap entry arc (step~\ref{step:plreach})
      \State $A_j\gets$ a breadth-first out-arborescence from $r$ of $H'_j$ without its entry arcs that are not cheap
      \State \textbf{if} some terminal of $I_j$ is not in $H'_j$ \textbf{then return} failure
      \State $S_j\gets A_j[r,t]$ if $I_j$ has a single terminal $t$, and $S_j\gets\Call{\Sep}{I_j,H'_j,A_j}$ otherwise
      \State $Q_j\gets$ the set of arcs of $A_j$ whose heads lie in $S_j$\Comment{the piece of $I_j$}
      \State add to $\ell(u)$ the number $a_u$ of arcs of $Q_j$ that leave $r$ with $G$-tail $u$, for every vertex $u$ of $G$
      \For{every weakly connected component $C$ of $H'_j-S_j$ with $C\cap K\neq\emptyset$}
        \State add $(R_{I_j}\cup(S_j\setminus\{r\}),C,K\cap C)$ to $\mathcal I_{d+1}$
      \EndFor
    \EndFor
    \State $d\gets d+1$
  \EndWhile
  \State \Return the pieces $Q_j$ of all instances, with their depths
\EndFunction
\end{algorithmic}
\end{algorithm}

The schedule is built from the pieces level by level, as follows.

\begin{definition}\label{def:plphase}
For a level $d$ that has at least one instance and a vertex $u$ of $G$, let
$\ell_d(u)$ be the number of arcs of pieces of depth $d$ that leave $r$ and
have $G$-tail $u$ (the value of $\ell(u)$ at the end of level $d$), and let
$\ell^{\max}_d:=\max_{u\in V(G)}\ell_d(u)$. The \emph{phase schedule}
consists of consecutive phases, one for each level $d=0,1,\dots$ that has an
instance, in increasing order of $d$; phase $d$ lasts
$\ell^{\max}_d+\theta+\sigma-2$ rounds. In phase $d$,
\begin{enumerate}[label=(\roman*),ref=(\roman*)]
  \item\label{def:plphase:heads} in each of the first $\ell^{\max}_d$ rounds, every
    vertex $u$ of $G$ calls one of the heads of the arcs that leave $r$ with $G$-tail $u$ in
    the pieces of depth $d$, until it has called all of them, and
  \item\label{def:plphase:inside} every vertex $v\ne r$ of a piece of depth $d$ calls, in every round
    of the phase after the round in which it is informed, one of its
    children in that piece that it has not called yet, as long as one
    exists.
\end{enumerate}
\end{definition}

Algorithm~\ref{alg:pllevelsep} also lists the separator procedure
\Sep{} of step~\ref{step:plsep}, and Algorithm~\ref{alg:planar}
lists the main function \LPSep. The main function first
sets $\bar d$, $\sigma$ and the parameters of Appendix~\ref{sec:plalg} and computes
the fallback schedule $\mathit{best}$ of Lemma~\ref{lem:fallback}, which covers the case
$\OPT\ge k$. Then it tests $\theta=1,\dots,k-1$ with Lemma~\ref{lem:pllp}\ref{lem:pllp:solve}. At the first $\theta$ for
which $\LP(\theta)$ is feasible, which is $\TLP$, it runs
\LevelSep{}$(G,s,K,\theta)$, builds the phase schedule of
Definition~\ref{def:plphase} from the pieces, and returns the shorter of the
two schedules. If \LevelSep{} fails, which
Lemma~\ref{lem:plstruct}\ref{lem:plstruct:halving} excludes, or $\LP(\theta)$ is infeasible for all
these $\theta$, it returns $\mathit{best}$.

\begin{algorithm}[H]
\caption{\LPSep: directed multicast with root-path separators}
\label{alg:planar}
\begin{algorithmic}[1]
\Require
  \IOitem{$G$, $s$, $K$: the digraph, the source and the $k\ge2$ terminals}
  \IOitem{fixed throughout: $p$ and the class algorithm}
\Ensure a schedule that informs every terminal
\Function{\LPSep}{$G,s,K$}
  \State $\bar d\gets\lfloor\log_2k\rfloor$, $\sigma\gets\min\{p,k\}$, and $\epsilon,\delta,\beta$ as in Appendix~\ref{sec:plalg}
  \State $\mathit{best}\gets$ the busy schedule along a shortest-path out-arborescence from $s$ to $K$ (Lemma~\ref{lem:fallback})
  \For{$\theta=1,\dots,k-1$}
    \If{$\LP(\theta)$ is feasible (Lemma~\ref{lem:pllp}\ref{lem:pllp:solve})}
      \State $\mathcal Q\gets\Call{\LevelSep}{G,s,K,\theta}$ (Algorithm~\ref{alg:pllevelsep})
      \State \textbf{if} $\mathcal Q$ is failure \textbf{then return} $\mathit{best}$
      \State $Y\gets$ the phase schedule of the pieces in $\mathcal Q$ (Definition~\ref{def:plphase})
      \State \Return the shorter of $\mathit{best}$ and $Y$
    \EndIf
  \EndFor
  \State \Return $\mathit{best}$
\EndFunction
\end{algorithmic}
\end{algorithm}

\subsection{Analysis}
\label{sec:planalysis}

In this subsection we analyse Algorithms~\ref{alg:pllevelsep}
and~\ref{alg:planar} and prove Theorem~\ref{thm:plclass}. Throughout,
$\theta\in\{1,\dots,k-1\}$ is fixed with $\LP(\theta)$ feasible, $x$ is the vector
computed by \LevelSep{}$(G,s,K,\theta)$, and $f$ is a flow with $(x,f)$ feasible
for $\LP(\theta)$ (Lemma~\ref{lem:pllp}\ref{lem:pllp:solve}). By step~\ref{step:plsep}, every separator $S_j$
is a union of at least one root path of $A_j$. Lemmas~\ref{lem:plsurvive}
and~\ref{lem:plsepinput} and parts~\ref{lem:plstruct:disjoint}, \ref{lem:plstruct:roots}, \ref{lem:plstruct:once}, \ref{lem:plstruct:minor} and the second
sentence of~\ref{lem:plstruct:pieces} of Lemma~\ref{lem:plstruct} use only this, so they hold for
every run of \LevelSep{}$(G,s,K,\theta)$. The other statements,
Lemma~\ref{lem:plstruct}\ref{lem:plstruct:halving}, \ref{lem:plstruct:rootcount} and the first and third sentences of~\ref{lem:plstruct:pieces},
and Lemmas~\ref{lem:plloads}, \ref{lem:plphase} and~\ref{lem:plgeneral}, also assume that
\Sep{} returns a $\sigma$-root-path separator of $(H'_j,A_j,K_{I_j})$
for every instance $I_j$ with at least two terminals; the proof of
Theorem~\ref{thm:plclass} establishes this with Lemma~\ref{lem:plsepinput},
which does not depend on it. Lemma~\ref{lem:plsurvive}, where the cheap
arcs are needed, shows that the scaled capacities keep every terminal
reachable.

\begin{lemma}\label{lem:plsurvive}
Let $I$ be an instance of depth $d$ created by \LevelSep{}$(G,s,K,\theta)$, and
let $t$ be a terminal of $I$. Then $H(I)$ contains a flow from $r$ to $t$
of value at least $1$ along directed paths with at most $\theta$ arcs, which
puts at most $x^I_e$ on every arc $e$ of $H(I)$. Moreover, when $I$ is
processed as $I_j$, the terminal $t$ is a vertex of $H'_j$.
\end{lemma}

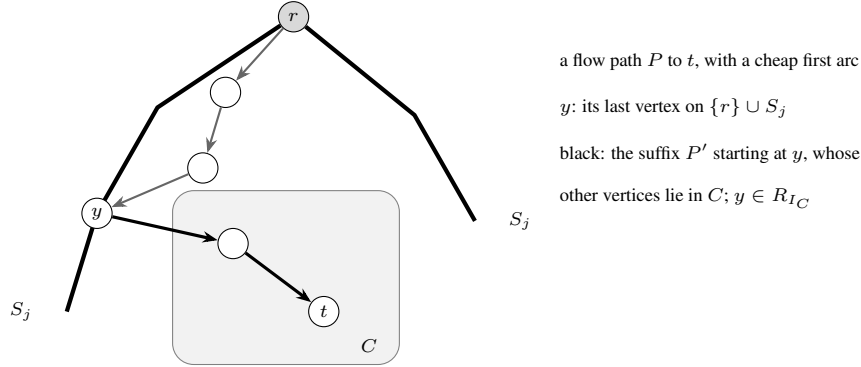
\begin{figure}[htb]
\centering
\begin{tikzpicture}[>={Stealth[length=2mm]},
  v/.style={circle,draw,fill=white,inner sep=0pt,minimum size=4mm,font=\scriptsize},
  lab/.style={font=\scriptsize}]
  \node[v,fill=black!15] (r) at (0,0) {$r$};
  \draw[line width=1.6pt] (r) -- (-1.8,-1.2) -- (-2.6,-2.6) -- (-3.0,-3.9);
  \draw[line width=1.6pt] (r) -- (1.6,-1.3) -- (2.4,-2.7);
  \node[lab] at (-3.6,-3.9) {$S_j$};
  \node[lab] at (3.0,-2.7) {$S_j$};
  \draw[rounded corners=8pt,fill=black!5,draw=black!50] (-1.6,-2.3) rectangle (1.4,-4.6);
  \node[lab] at (1.0,-4.35) {$C$};
  \node[v] (y) at (-2.6,-2.6) {$y$};
  \node[v] (a) at (-0.9,-1.0) {};
  \node[v] (b) at (-1.2,-2.0) {};
  \node[v] (c) at (-0.8,-3.0) {};
  \node[v] (t) at (0.4,-3.9) {$t$};
  \draw[->,thick,black!60] (r) -- (a);
  \draw[->,thick,black!60] (a) -- (b);
  \draw[->,thick,black!60] (b) -- (y);
  \draw[->,very thick] (y) -- (c);
  \draw[->,very thick] (c) -- (t);
  \node[lab,anchor=west,align=left] at (3.4,-0.6)
    {a flow path $P$ to $t$, with a cheap first arc};
  \node[lab,anchor=west,align=left] at (3.4,-1.2)
    {$y$: its last vertex on $\{r\}\cup S_j$};
  \node[lab,anchor=west,align=left] at (3.4,-1.8)
    {black: the suffix $P'$ starting at $y$, whose};
  \node[lab,anchor=west,align=left] at (3.4,-2.4)
    {other vertices lie in $C$; $y\in R_{I_C}$};
\end{tikzpicture}
\caption{The step from an instance to a child in the proof of
Lemma~\ref{lem:plsurvive}. The bold paths are root paths of $A_j$ that form
the separator $S_j$. The suffix $P'$ of a flow path, starting at its last
vertex $y$ on $\{r\}\cup S_j$, is a path of the child's instance digraph,
since $y$ belongs to the child's root set.}
\label{fig:plsurvive}
\end{figure}

\begin{proof}
We prove the first statement of Lemma~\ref{lem:plsurvive} by induction on
the depth $d$ of $I$, and show the second statement for every instance of
depth $d$ on the way.

\emph{Base case, $d=0$.} Then $I=I_0$ and $x^{I_0}=x$. The paths
$P\in\mathcal P_t$ with $f^t_P>0$ are paths of $G_{\le \theta}$ from $s$ to $t$
with at most $\theta$ arcs, and they do not return to $s$, as $G_{\le \theta}$ has
no arcs into $s$. Replacing their first arc, which leaves $s$, by the
corresponding entry arc turns them into paths of $H(I_0)$ from $r$ to $t$,
and $f^t$ becomes a flow of value $1$ that puts at most $x_e$ on every arc
$e$.

\emph{Inductive step} (Figure~\ref{fig:plsurvive}). Assume the first
statement of Lemma~\ref{lem:plsurvive} for the instances of depth $d$, and
let $I=I_j$ be one of them. Let $t$ be a terminal of $I$, and fix a flow as in that statement and a decomposition
of it into paths from $r$ to $t$ with at most $\theta$ arcs, the \emph{flow
paths}. As $H(I)$ has no arc into $r$, each flow path starts with an entry
arc and has no other entry arc. By the definition of $\Omega_j$,
\[
  \Omega_j=\sum_{e\text{ entry arc of }H(I)}\tilde\omega(\tau(e))\,x^I_e ,
\]
and $\tilde\omega\ge\omega\ge1$, so $\Omega_j$ is at least the total
capacity of the entry arcs, which is at least the value of the flow; in
particular $\Omega_j>0$. An entry arc $e$ that is not cheap has
$\tilde\omega(\tau(e))>\Omega_j/\epsilon$. So the flow on flow paths whose
first arc is not cheap is at most
\[
  \sum_{e\text{ not cheap}}x^I_e\;\le\;\frac\epsilon{\Omega_j}\sum_{e\text{
  not cheap}}\tilde\omega(\tau(e))\,x^I_e\;\le\;\frac\epsilon{\Omega_j}\,
  \Omega_j\;=\;\epsilon .
\]
Hence the flow paths whose first arc is cheap carry flow at least
$1-\epsilon>0$. Every vertex of such a path is reached from $r$ by a prefix
of the path, which has at most $\theta$ arcs and a cheap first arc; so all its
vertices lie in $H'_j$, and as $H'_j$ is an induced subdigraph, so do its
arcs. In particular $t$ lies in $H'_j$, which proves the second statement
of Lemma~\ref{lem:plsurvive} for $I$.

For the first statement at depth $d+1$, let $I_C$ be a child of $I$ and let
$t$ be a terminal of $I_C$. Then $t\in C\subseteq W_I$, so $t$ is also a
terminal of $I$, and we use the flow and the flow paths fixed above for
$t$; moreover $t\notin S_j$. For each flow path $P$ with a cheap first arc,
let $y$ be the last vertex of $P$ in $\{r\}\cup S_j$, and let $P'$ be the
subpath of $P$ from $y$ to $t$. As $t\notin\{r\}\cup S_j$, the path $P'$
has at least one arc. Its vertices after $y$ lie in $H'_j$ and not in
$\{r\}\cup S_j$, consecutive ones are joined by arcs, and the last one is
$t\in C$; so they all lie in the component $C$. The first arc of $P'$ goes
from $y$ into $C$. If $y=r$, it is an entry arc of $H(I)$ whose $G$-tail
lies in $R_I$; if $y\neq r$, it is an internal arc of $H(I)$ with tail
$y\in S_j\setminus\{r\}$. In both cases its arc in $G_{\le \theta}$ goes from
$R_{I_C}=R_I\cup(S_j\setminus\{r\})$ into $C$, so it is an entry arc of
$H(I_C)$. The other arcs of $P'$ are arcs of $G_{\le \theta}$ with both ends in
$C$, that is, internal arcs of $H(I_C)$. So $P'$ is a path of $H(I_C)$ from
its root to $t$ with at most $\theta$ arcs. The paths $P'$, with the flow values
of their flow paths $P$, form a flow from the root of $H(I_C)$ to $t$ of
value at least $1-\epsilon$, which puts on every arc $e$ at most the flow
of the flow paths through $e$, hence at most $x^I_e$. Divided by
$1-\epsilon$, it has value at least $1$ and puts at most
$x^I_e/(1-\epsilon)=x^{I_C}_e$ on every arc $e$, as $I_C$ has depth $d+1$.
\end{proof}

Lemma~\ref{lem:plstruct} collects structural facts that the remaining
proofs use; its proof consists of routine inductions.

\begin{lemma}\label{lem:plstruct}
The following hold for \LevelSep{}$(G,s,K,\theta)$.
\begin{enumerate}[label=(\alph*),ref=(\alph*)]
  \item\label{lem:plstruct:disjoint} Every child is an instance. For every level $d$, the regions of the
    instances of level $d$ are pairwise disjoint, so level $d$ has at most
    $k$ instances. Every separator $S_j$ satisfies $S_j\setminus\{r\}
    \subseteq W_{I_j}$.
  \item\label{lem:plstruct:roots} Let $I$ be an instance of depth $d$, and let $J_0,J_1,\dots,J_d$ be
    the instances with $J_0=I_0$, $J_d=I$ and $J_{i+1}$ a child of $J_i$
    for every $i<d$; we call $J_0,\dots,J_{d-1}$ the \emph{ancestors} of
    $I$. Then
    \[
      R_I=\{s\}\cup\bigcup_{i=0}^{d-1}\bigl(S_{J_i}\setminus\{r\}\bigr),
    \]
    that is, the root set of $I$ consists of $s$ and the vertices of $G$
    on the separators of the ancestors of $I$.
  \item\label{lem:plstruct:once} Every vertex of $G$ lies in $S_j\setminus\{r\}$ for at most one
    instance $I_j$.
  \item\label{lem:plstruct:halving} Every instance of level $d$ has at most $k/2^d$ terminals; hence
    every level $d>\bar d$ is empty, and \LevelSep{}$(G,s,K,\theta)$ does not fail.
  \item\label{lem:plstruct:pieces} Every terminal lies in the separator of some instance. Every $A_j$
    contains every vertex of $H'_j$ and has root paths with at most $\theta$
    arcs. Every piece has at most $\sigma$ arcs that leave $r$ and at most
    $\sigma$ leaves.
  \item\label{lem:plstruct:rootcount} For every level $d\ge1$ that has an instance, the union $R_d$ of
    the root sets of the instances of level $d$ satisfies $|R_d|\le1+\sigma \theta+\sigma
    \theta(d-1)k\le\sigma k+\sigma k^3$.
  \item\label{lem:plstruct:minor} For every instance $I$, the underlying graph of $H(I)$ is a minor of
    the underlying graph of $G$.
\end{enumerate}
\end{lemma}

\begin{proof}
\ref{lem:plstruct:disjoint} Let $I_C$ be a child of $I=I_j$. Its root set $R_{I_C}=R_I\cup
(S_j\setminus\{r\})$ and its region $C$ are disjoint, since $C$ avoids
$S_j$ and $C\subseteq W_I$ is disjoint from $R_I$; and $K_{I_C}=K\cap C$.
The separator is a union of root paths of $A_j$, and each of them that has
an arc starts with an entry arc, whose $G$-tail lies in $R_I$, and
continues with internal arcs, which are arcs of $G_{\le \theta}$. So every vertex of
$S_j\setminus\{r\}$ is joined to $R_I$ in $G_{\le \theta}[R_{I_C}]$, and
$G_{\le \theta}[R_{I_C}]$ is weakly connected because $G_{\le \theta}[R_I]$ is. Also
$s\in R_I\subseteq R_{I_C}$, so $I_C$ is an instance. The regions of the
children of $I$ are distinct components of $H'_j-S_j$, so they are
pairwise disjoint subsets of $W_I\setminus S_j$. We show by induction on $d$
that the regions of the instances of level $d$ are pairwise disjoint. For
$d=0$ there is one instance. If this holds for level $d$, two
instances of level $d+1$ are children of one instance of level $d$, and
then their regions are disjoint as just shown, or of two different
instances of level $d$, whose regions are disjoint and contain the regions
of their children. Every instance has a terminal: $I_0$ has the terminal
set $K$, as $K\subseteq V_{\le \theta}$ and $s\notin K$, and a child is created
only for a component with a terminal; so each level has at most $k$
instances. Finally, $S_j\subseteq V(H'_j)\subseteq\{r\}\cup W_I$.

\ref{lem:plstruct:roots} We use induction on the depth of $I$. For depth $0$, $I=I_0$ has the
root set $\{s\}$ and no ancestors. A child $I_C$ of an instance
$I=I_j$ has the root set $R_I\cup(S_j\setminus\{r\})$, and its
ancestors are $I$ and the ancestors of $I$; so~\ref{lem:plstruct:roots} for $I$ gives
\ref{lem:plstruct:roots} for $I_C$.

\ref{lem:plstruct:once} Suppose that $v\in S_I\setminus\{r\}$ and $v\in S_J\setminus\{r\}$ for
instances $I\neq J$. By Lemma~\ref{lem:plstruct}\ref{lem:plstruct:disjoint}, $v\in W_I\cap W_J$. If
$I$ and $J$ have the same depth, this contradicts
Lemma~\ref{lem:plstruct}\ref{lem:plstruct:disjoint}. If $I$ is an ancestor of $J$, then
$W_J\subseteq W_I\setminus S_I$, as regions only shrink from an instance to
its children and avoid the separator of the parent; so $v\notin W_J$, a
contradiction. The case that $J$ is an ancestor of $I$ is
symmetric. Otherwise let, say, $J$ be the deeper one, and let $J'$ be the
ancestor of $J$ at the depth of $I$; then $J'\neq I$, $W_J\subseteq W_{J'}$
and $W_{J'}\cap W_I=\emptyset$ by Lemma~\ref{lem:plstruct}\ref{lem:plstruct:disjoint}, a
contradiction.

\ref{lem:plstruct:halving} We show by induction on $d$ that every instance of level $d$ has at
most $k/2^d$ terminals. For $d=0$, the instance $I_0$ has $k$ terminals.
Assume this for level $d$, and let $I_j$ be an instance of level
$d$. By Lemma~\ref{lem:plsurvive}, all its terminals lie in $H'_j$, so
\LevelSep{} does not fail at $I_j$. If $I_j$ has one terminal, that
terminal lies on $S_j$ and $I_j$ has no child. Otherwise $S_j$ is a
$\sigma$-root-path separator of $(H'_j,A_j,K_{I_j})$, so every component of
$H'_j-S_j$, and hence every child, has at most $|K_{I_j}|/2\le
k/2^{d+1}$ terminals. An instance has at least one terminal, so a level
$d$ with an instance has $2^d\le k$, that is, $d\le \bar d$.

\ref{lem:plstruct:pieces} The second sentence holds for every run. Every vertex $v\neq r$ of
$H'_j$ is reached from $r$ by a path with at most $\theta$ arcs whose first arc
is cheap; as $H(I_j)$ has no arc into $r$, the other arcs of this path are
internal arcs, and all of its vertices and arcs lie in the induced
subdigraph $H'_j$. So $v$ is reached by a path with at most $\theta$ arcs in the
digraph obtained from $H'_j$ by deleting the entry arcs that are not cheap,
and the breadth-first arborescence $A_j$ of this digraph contains $v$ and has
root paths with at most $\theta$ arcs. For the first sentence, follow a terminal
$t$ from $I_0$: whenever $t$ is a terminal of an instance $I_j$, it lies in
$H'_j$ by Lemma~\ref{lem:plsurvive}, and either $t\in S_j$ or $t$ lies in
a component of $H'_j-S_j$ with a terminal, that is, in a child. By
Lemma~\ref{lem:plstruct}\ref{lem:plstruct:halving} this ends, with $t$ in a separator. For the
third, the piece of an instance is the union of the paths in $A_j$ from $r$
to at most $\sigma$ vertices, as its separator is a single root path or a
$\sigma$-root-path separator; each of these paths has at most one arc that
leaves $r$, and every leaf of the piece is the end of one of them, so the
piece has at most $\sigma$ arcs out of $r$ and at most $\sigma$ leaves.

\ref{lem:plstruct:rootcount} By Lemma~\ref{lem:plstruct}\ref{lem:plstruct:roots}, $R_d$ is contained in the set of $s$
and the vertices other than $r$ of the separators of instances of depth
less than $d$. Every separator is a single root path or a
$\sigma$-root-path separator, so it consists of at most $\sigma$ root paths,
with at most $\theta$ vertices other than $r$ each by Lemma~\ref{lem:plstruct}\ref{lem:plstruct:pieces};
so it has at most $\sigma \theta$ vertices other than $r$. There is one instance
at depth $0$, and at each depth $d'$ with $1\le d'\le d-1$ there are at most
$k$ instances by Lemma~\ref{lem:plstruct}\ref{lem:plstruct:disjoint}. This gives $|R_d|\le1+\sigma
\theta+\sigma \theta(d-1)k$. As level $d$ has an instance, $d\le \bar d$ by
Lemma~\ref{lem:plstruct}\ref{lem:plstruct:halving}, and as $\theta\le k-1$ and $d-1<\bar d\le k-1$, we have
$1+\sigma \theta\le\sigma k$ and $\sigma \theta(d-1)k\le\sigma k^3$.

\ref{lem:plstruct:minor} Let $I=(R,W,K_I)$, let $U$ be the underlying graph of $G$, and let $U'$
be its subgraph induced by $R\cup W$. As $G_{\le \theta}[R]$ is weakly connected,
$R$ induces a connected subgraph of $U'$, so $U'/R$ is a minor of $U$; we
identify its new vertex with $r$. The underlying graph of $H(I)$ has the
vertex set $W\cup\{r\}$. Two vertices $u,v\in W$ are adjacent in it only if
an internal arc joins them, which comes from an arc of $G_{\le \theta}$ between
$u$ and $v$, so they are adjacent in $U'/R$; and $r$ is adjacent to $v\in
W$ only if an entry arc enters $v$, which comes from an arc $(u,v)$ of
$G_{\le \theta}$ with $u\in R$, so $v$ is adjacent to $r$ in $U'/R$. Hence the
underlying graph of $H(I)$ is a subgraph of $U'/R$, and so a minor of $U$.
\end{proof}

Lemma~\ref{lem:plsepinput} collects what \Sep{} receives; the
proof of Theorem~\ref{thm:plclass} uses it.

\begin{lemma}\label{lem:plsepinput}
Let $I_j$ be an instance that \LevelSep{}$(G,s,K,\theta)$ processes, and let $H$
be the underlying graph of $H'_j$. Then $H$ is connected and is a minor of
the underlying graph of $G$, the underlying graph of $A_j$ is a spanning
tree of $H$ rooted at $r$ whose root paths are the root paths of $A_j$, and
$K_{I_j}\subseteq V(H)\setminus\{r\}$.
\end{lemma}

\begin{proof}
By Lemma~\ref{lem:plstruct}\ref{lem:plstruct:pieces}, $A_j$ contains every vertex of $H'_j$. Its
underlying graph is connected, as $A_j$ joins $r$ to each of these vertices
by a directed path, and it has at most $|V(H)|-1$ edges, as exactly one arc
of $A_j$ enters every vertex other than $r$ and none enters $r$; so it is a
spanning tree of $H$, and $H$ is connected. The path of this tree from $r$
to a vertex $v$ is the directed path of $A_j$ from $r$ to $v$, so the two
have the same root paths. The underlying graph of $H(I_j)$ is a minor of
that of $G$ by Lemma~\ref{lem:plstruct}\ref{lem:plstruct:minor}, and $H$ is a subgraph of it,
since $H'_j$ is a subdigraph of $H(I_j)$; so $H$ is a minor of the
underlying graph of $G$. Finally, the terminals of $I_j$ lie in its
region, so they differ from $r$, and they are vertices of $H'_j$ by
Lemma~\ref{lem:plsurvive}.
\end{proof}

The new step of the analysis bounds the per-level loads.

\begin{lemma}\label{lem:plloads}
Let $\theta\in\{1,\dots,k-1\}$ with $\LP(\theta)$ feasible, and suppose that
in the recursion for $\theta$ the separator $S_j$ of every instance $I_j$ with
at least two terminals is a $\sigma$-root-path separator
(Definition~\ref{def:plsep}): at least one and at most $\sigma$ root paths of
$A_j$ such that every component of $H'_j-S_j$ contains at most half of the
terminals of $I_j$. Then at level $0$, $\ell_0(s)\le\sigma$ and $\ell_0(u)=0$ for every vertex $u\neq
s$ of $G$. For every level $d\ge1$ that has an instance and every vertex $u$
of $G$,
\[
  \ell_d(u)\;\le\;\sigma\bigl(\beta(1-\epsilon)^{-d}\theta+\ln|R_d|\bigr)\;\le\;
  \sigma\bigl(6(\bar d+1)\theta+\ln(\sigma k+\sigma k^3)\bigr).
\]
\end{lemma}

\begin{proof}
At level $0$ there is one instance, whose root set is $\{s\}$; so every arc
that leaves $r$ has $G$-tail $s$, and the piece has at most $\sigma$ of them
by Lemma~\ref{lem:plstruct}\ref{lem:plstruct:pieces}.

Let $d\ge1$ be a level with an instance, and let $I_1,\dots,I_q$ be the
instances of level $d$ in the order of processing; write $m_j:=m_{I_j}$.
For every vertex $u$ of $G$, the sum of its masses over the level satisfies
\begin{align*}
  \sum_{j=1}^q m_j(u)\;&\le\;(1-\epsilon)^{-d}\sum_{j=1}^q\
  \sum_{e=(u,v)\in E(G_{\le \theta}),\ v\in W_{I_j}}x_e\\
  &\le\;(1-\epsilon)^{-d}
  \sum_{e\in E(G_{\le \theta})\text{ leaving }u}x_e\;\le\;(1-\epsilon)^{-d}\,\theta
  =:\Lambda_d ,
\end{align*}
because $m_j(u)$ sums $x^{I_j}_e$ over some of the arcs $(u,v)$ with $v\in
W_{I_j}$ (over all of them if $u\in R_{I_j}$, and over none otherwise), the
regions $W_{I_j}$ are pairwise disjoint (Lemma~\ref{lem:plstruct}\ref{lem:plstruct:disjoint}), so
every arc is counted at most once, and by the out-degree constraint of
$\LP(\theta)$. The masses depend only on the instances, which are all created
before level $d$ starts, so they are fixed during the level. A vertex
outside $R_d$ has mass $0$ in all instances of the level and is the
$G$-tail of no entry arc, so its load and future mass stay $0$.

For $0\le j\le q$, let $\ell^{(j)}$ and $\bar m^{(j)}$ be the loads and future
masses after the processing of $I_j$ (for $j=0$, before $I_1$), so that
$\bar m^{(j)}(u)=\sum_{j'>j}m_{j'}(u)$, and let $\Psi_j:=\sum_{u\in R_d}
\exp(\ell^{(j)}(u)/\sigma+\beta\bar m^{(j)}(u))$. When $I_j$ is processed, the
weights are $\omega(u)=\exp(\ell^{(j-1)}(u)/\sigma+\beta\bar m^{(j)}(u))$, and $\ell^{(j)}(u)=\ell^{(j-1)}(u)+a_u$
for every vertex $u$ of $G$, where $a_u$ is the number of arcs of the piece
$Q_j$ that leave $r$ and have $G$-tail $u$
(step~\ref{step:plload}). Hence
\[
  \Psi_{j-1}-\Psi_j=\sum_{u\in R_d}\omega(u)\bigl(e^{\beta m_j(u)}-e^{a_u/\sigma}\bigr)\;\ge\;\beta\sum_{u\in R_d}\omega(u)m_j(u)-\frac{e-1}\sigma
  \sum_{u\in R_d}\omega(u)a_u .
\]
Here we used $e^y-1\ge y$ for $y=\beta m_j(u)$, and
$e^{z/\sigma}-1\le(e-1)z/\sigma$ for $0\le z\le\sigma$, as the left side is
convex in $z$ and the two sides agree at $z=0$ and $z=\sigma$; we apply it
with $z=a_u$, which lies in $[0,\sigma]$, as the piece has at most $\sigma$
arcs out of $r$ (Lemma~\ref{lem:plstruct}\ref{lem:plstruct:pieces}). Next, suppose that
$a_u>0$, so that the piece $Q_j$ has an arc $e$ that leaves $r$ and has
$G$-tail $u$. This arc is an arc of $A_j$, and $A_j$ contains no entry arc
that is not cheap (step~\ref{step:plreach}); so $e$ is cheap, that is,
$\tilde\omega(u)=\tilde\omega(\tau(e))\le\Omega_j/\epsilon$
(step~\ref{step:plmass}). As $\omega(u)\le\tilde\omega(u)$
(step~\ref{step:plweight}), every $u$ with $a_u>0$ has
$\omega(u)\le\Omega_j/\epsilon$. Since the piece has at most $\sigma$ arcs
out of $r$ in total, $\sum_ua_u\le\sigma$, and hence
$\sum_{u\in R_d}\omega(u)a_u\le\sigma\Omega_j/\epsilon$. Moreover
$\Omega_j=\sum_{u\in R_{I_j}}\tilde\omega(u)m_j(u)\le e^\delta\sum_{u\in
R_d}\omega(u)m_j(u)$.
Together,
\[
  \Psi_{j-1}-\Psi_j\;\ge\;\Bigl(\beta-\frac{(e-1)e^\delta}\epsilon\Bigr)
  \sum_{u\in R_d}\omega(u)m_j(u)\;\ge\;0 ,
\]
as $\beta=7/(4\epsilon)\ge(e-1)e^\delta/\epsilon$. Hence $\Psi_q\le\Psi_0$.
At the end of the level $\bar m^{(q)}=0$, so for every $u\in R_d$,
\[
  e^{\ell_d(u)/\sigma}\le\Psi_q\le\Psi_0=\sum_{u'\in R_d}e^{\beta
  \bar m^{(0)}(u')}\le|R_d|\,e^{\beta\Lambda_d},
\]
that is, $\ell_d(u)\le\sigma(\beta\Lambda_d+\ln|R_d|)$, the first bound of
Lemma~\ref{lem:plloads}; for $u\notin R_d$ the load is $0$. For the second
bound, $d\le \bar d$ by Lemma~\ref{lem:plstruct}\ref{lem:plstruct:halving}, so $(1-\epsilon)^{-d}\le
e^{1/2}$ by~\eqref{eq:plscale}, and $\beta\Lambda_d\le(7/2)(\bar d+1)e^{1/2}\theta
\le6(\bar d+1)\theta$, as $(7/2)e^{1/2}<5.78$. Finally, $|R_d|\le\sigma k+\sigma
k^3$ by Lemma~\ref{lem:plstruct}\ref{lem:plstruct:rootcount}.
\end{proof}

The phase schedule turns the per-level loads into a schedule. The proof of
Lemma~\ref{lem:plphase} is an induction over the phases that uses
Lemma~\ref{lem:plstruct}\ref{lem:plstruct:roots}, \ref{lem:plstruct:once} and~\ref{lem:plstruct:pieces} and the busy schedule of
Lemma~\ref{lem:busy}.

\begin{lemma}\label{lem:plphase}
The phase schedule of Definition~\ref{def:plphase} is a valid schedule. It
informs every vertex other than $r$ of every piece of depth $d$ by the end
of phase $d$,
so it informs all terminals within $\sum_d(\ell^{\max}_d+\theta+\sigma-2)$
rounds, the sum over the levels that have an instance.
\end{lemma}

\begin{proof}
The levels that have an instance are $0,1,\dots,d_{\max}$ for some
$d_{\max}$, as every instance of a level $d\ge1$ is a child of an instance
of level $d-1$. We show by induction on $d\in\{0,\dots,d_{\max}+1\}$ the
following statement $(P_d)$: the calls of the phases before phase $d$ are
valid (every caller is informed, every call follows an arc of $G$, and no
vertex makes two calls in one round), and when phase $d$ starts (for
$d=d_{\max}+1$: when the last phase ends), every vertex other than $r$ of
every piece of depth less than $d$ is informed.

\emph{Base case, $d=0$.} There are no earlier phases and no pieces of
negative depth.

\emph{Inductive step.} Assume $(P_d)$ for some $d\le d_{\max}$. The
callers of type~\ref{def:plphase:heads} in phase $d$ are $G$-tails of entry arcs of instances
of level $d$, so they lie in root sets, which consist of $s$ and vertices
of separators of smaller depth (Lemma~\ref{lem:plstruct}\ref{lem:plstruct:roots}); by $(P_d)$,
they are informed when the phase starts. A vertex $u$ makes
$\ell_d(u)$ calls of type~\ref{def:plphase:heads}, one per round, along the arcs of $G$ from
which its entry arcs come. A caller of type~\ref{def:plphase:inside} is a vertex other than $r$
of a separator of depth $d$; it is not $s$, as $s$ lies in no region, and
it is not a vertex of a separator of smaller depth, by
Lemma~\ref{lem:plstruct}\ref{lem:plstruct:once}; so it is not a caller of type~\ref{def:plphase:heads} in the same
phase. By Lemma~\ref{lem:plstruct}\ref{lem:plstruct:once}, it lies in only one piece of depth
$d$, where it has one parent, and its calls follow the internal arcs of its
piece, which are arcs of $G$; it calls only after it is informed. So the
calls of phase $d$ are valid.

Every head $v$ of an arc that leaves $r$ in a piece of depth $d$ is
informed within the first $\max_{u\in V(G)}\ell_d(u)$ rounds of the phase. Below
$v$, the piece contains an out-arborescence rooted at $v$ with height at
most $\theta-1$ (the paths of the piece from $r$ have at most $\theta$ arcs, the first
of which is the arc to $v$) and at most $\sigma$ leaves
(Lemma~\ref{lem:plstruct}\ref{lem:plstruct:pieces}). Its vertices call as in the busy schedule of
Lemma~\ref{lem:busy}, which informs it within $(\theta-1)+\sigma-1$ rounds after
$v$ is informed. So every vertex other than $r$ of the piece is informed
within $\ell^{\max}_d+\theta+\sigma-2$ rounds of the start of the phase,
that is, by its end, and $(P_{d+1})$ holds.

By $(P_{d_{\max}+1})$, the schedule is valid and informs all vertices other
than $r$ of all pieces within $\sum_d(\ell^{\max}_d+\theta+\sigma-2)$
rounds. Every terminal lies in a
separator (Lemma~\ref{lem:plstruct}\ref{lem:plstruct:pieces}), hence in a piece, and differs
from $r$.
\end{proof}

We combine Lemmas~\ref{lem:plstruct}, \ref{lem:plloads}
and~\ref{lem:plphase} into a bound that holds whenever \Sep{}
returns $\sigma$-root-path separators:

\begin{lemma}\label{lem:plgeneral}
Suppose that \Sep{} runs in polynomial time and returns, for every
instance $I_j$ with at least two terminals that \LevelSep{} processes,
a $\sigma$-root-path separator of $(H'_j,A_j,K_{I_j})$. Then
Algorithm~\ref{alg:planar} runs in polynomial time, and the schedule that
it returns has the following length.
\begin{enumerate}[label=(\roman*),ref=(\roman*)]
  \item\label{lem:plgeneral:largeopt} If $\OPT\ge k$, it has length at most $2\,\OPT$.
  \item\label{lem:plgeneral:smalllp} If $\TLP\le k-1$, it has length at most
    \begin{align*}
      &2\sigma+\TLP+\bar d\bigl(6\sigma(\bar d+1)\TLP+\sigma\ln(\sigma k+\sigma
      k^3)+\TLP+\sigma\bigr)\\
      &\qquad\le\;7\sigma(\bar d+1)^2\,\OPT+2\sigma+\sigma
      \bar d\bigl(1+\ln(\sigma k+\sigma k^3)\bigr).
    \end{align*}
\end{enumerate}
\end{lemma}

\begin{proof}
\ref{lem:plgeneral:largeopt} If $\OPT\ge k$, the fallback schedule $\mathit{best}$ informs all
terminals within $\OPT+k\le2\,\OPT$ rounds by Lemma~\ref{lem:fallback}, and
the output is never longer than it.

\ref{lem:plgeneral:smalllp} Let $\TLP\le k-1$. As $\TLP$ is the least $\theta\ge1$ for which $\LP(\theta)$ is
feasible, the first $\theta$ at which Algorithm~\ref{alg:planar} finds $\LP(\theta)$
feasible is $\TLP$, and it runs \LevelSep{}$(G,s,K,\TLP)$, which does not
fail and has levels $0,\dots,\bar d$ at most (Lemma~\ref{lem:plstruct}\ref{lem:plstruct:halving}). The
output is never longer than the phase schedule, which informs all
terminals within $\sum_d(\ell^{\max}_d+\TLP+\sigma-2)$ rounds by
Lemma~\ref{lem:plphase}. By Lemma~\ref{lem:plloads}, $\ell^{\max}_0\le\sigma$
and, for $d\ge1$, $\ell^{\max}_d\le\sigma(6(\bar d+1)\TLP+\ln(\sigma
k+\sigma k^3))$; adding $\TLP+\sigma-2$ to these positive bounds and summing
over $d=0,\dots,\bar d$ gives the first bound. For the
second, $\TLP\le\OPT$ (Appendix~\ref{sec:pldefs}), $\bar d(\bar d+1)\le(\bar d+1)^2$ and
$\bar d+1\le\sigma(\bar d+1)^2$ give $6\sigma \bar d(\bar d+1)\TLP\le6\sigma(\bar d+1)^2\OPT$ and
$(\bar d+1)\TLP\le\sigma(\bar d+1)^2\OPT$.

Running time: by Lemma~\ref{lem:pllp}\ref{lem:pllp:solve}, testing $\LP(\theta)$ for
$\theta=1,\dots,k-1$ and computing $x$ take polynomial time, and by
Lemma~\ref{lem:plstruct}\ref{lem:plstruct:disjoint} and~\ref{lem:plstruct:halving}, \LevelSep{} processes at most $\bar d+1$
levels with at most $k$ instances each. The masses are sums of entries of
$x$ times $(1-\epsilon)^{-d}$, and $\epsilon$ is rational, so they are
rationals of polynomial bit size. The exponents $\ell(u)/\sigma+\beta\bar m(u)$ are rational and at most
$k+2\beta k\theta$, which is polynomial in $k$: a load is at most $\sigma k$, as
a level has at most $k$ pieces and each adds at most $\sigma$ to it
(Lemma~\ref{lem:plstruct}\ref{lem:plstruct:pieces}), and $\bar m(u)$ is a sum of at most $k$ masses,
each at most $(1-\epsilon)^{-d}\theta\le2\theta$ by the out-degree constraint of
$\LP(\theta)$ and~\eqref{eq:plscale}. So the numbers $\tilde\omega(u)$ can be
computed in polynomial time, for example from a truncated power series with
a bound on its tail. The digraph $H'_j$, the arborescence $A_j$, the
separator, the components and the phase schedule take polynomial time.
\end{proof}

To prove Theorem~\ref{thm:plclass}, we check that \Sep{}
returns $\sigma$-root-path separators and simplify the bound of
Lemma~\ref{lem:plgeneral}.

\begin{proof}[Proof of Theorem~\ref{thm:plclass}]
The algorithm is \LPSep\ (Algorithm~\ref{alg:planar}) with the class algorithm of
$\mathcal G$. We first check the hypothesis of Lemma~\ref{lem:plgeneral}
for an instance $I_j$ with at least two terminals that \LevelSep{}
processes, using Lemma~\ref{lem:plsepinput}, which holds for every run. If $p>k$, then $\sigma=k$, and \Sep{} returns the union
$S$ of the root paths of $A_j$ to the terminals of $I_j$, which are vertices
of $H'_j$ (Lemma~\ref{lem:plsepinput}): at least two and at most $k$ root
paths, and no component of $H'_j-S$ contains a terminal. If $p\le k$, then
$\sigma=p$. By Lemma~\ref{lem:plsepinput}, the underlying graph $H$ of
$H'_j$ is connected and a minor of that of $G$, so it lies in
$\mathcal G$; the underlying graph of $A_j$ is a spanning tree of $H$
rooted at $r$ with the same root paths as $A_j$; and $K_{I_j}\subseteq
V(H)\setminus\{r\}$. So the class algorithm applies, in time polynomial in
$n$, as $|V(H)|\le n+1$. By Definition~\ref{def:plclass} it returns a nonempty set $Z$ of at
most $p$ vertices such that every component of $H-A_j[r,Z]$, that is, every
weakly connected component of $H'_j-A_j[r,Z]$, contains at most
$|K_{I_j}|/2$ terminals of $I_j$. So \Sep{} returns
$A_j[r,Z\cup\{r\}]=A_j[r,Z]$, a union of at least one and at most $p$ root
paths of $A_j$, hence a $\sigma$-root-path separator of
$(H'_j,A_j,K_{I_j})$.

By Lemma~\ref{lem:plgeneral}, Algorithm~\ref{alg:planar} therefore runs in
polynomial time. If $\OPT\ge k$, its schedule has length at most
$2\,\OPT\le14\sigma(\bar d+1)^2\,\OPT$ by Lemma~\ref{lem:plgeneral}\ref{lem:plgeneral:largeopt}. If
$\OPT<k$, then $\TLP\le\OPT\le k-1$, and Lemma~\ref{lem:plgeneral}\ref{lem:plgeneral:smalllp}
bounds the length by $7\sigma(\bar d+1)^2\,\OPT+2\sigma+\sigma \bar d(1+\ln(\sigma
k+\sigma k^3))$. As $\sigma\le k$, we have $\sigma k+\sigma k^3\le2k^4$; and as $k<2^{\bar d+1}$,
we have $\ln k<(\bar d+1)\ln2$. So $1+\ln(2k^4)<1+\ln2+4(\bar d+1)\ln2\le
(1+5\ln2)(\bar d+1)<4.5(\bar d+1)$, and the last term is at most
$4.5\sigma \bar d(\bar d+1)\le4.5\sigma(\bar d+1)^2\,\OPT$, as $\OPT\ge1$; also
$2\sigma\le2\sigma(\bar d+1)^2\,\OPT$. The length is therefore at most
$(7+2+4.5)\sigma(\bar d+1)^2\,\OPT\le14\sigma(\bar d+1)^2\,\OPT$, and
$\sigma=\min\{p,k\}$.
\end{proof}

\subsection{Planar digraphs and bounded treewidth}
\label{sec:plplanartw}

In this subsection we prove Theorem~\ref{cor:plplanar} and
Corollary~\ref{cor:pltw}: their classes are root-path separable
(Lemmas~\ref{lem:plplanarclass} and~\ref{lem:pltwsep}).

\paragraph{Planar digraphs.} For planar graphs, a lemma of Thorup gives
balanced separators made of three root paths:

\begin{lemma}[{Thorup~\cite[Lemma~2.3]{thorup2004compact}; see also \cite[Theorem~3]{friggstad2023planar}}]\label{fact:plthorup}
Let $H$ be a connected planar graph with a spanning tree rooted at a vertex
$r$ and nonnegative vertex weights. In linear time one can find three
vertices whose paths in the tree to $r$ have a union $S$ such that every
component of $H-S$ has at most half of the total weight.
\end{lemma}

Lemma~\ref{lem:plplanarclass} turns these separators into the property of
Definition~\ref{def:plclass} that Theorem~\ref{cor:plplanar} needs:

\begin{lemma}\label{lem:plplanarclass}
The planar graphs form a $3$-root-path separable class.
\end{lemma}

\begin{proof}
Every minor of a planar graph is planar, since the graphs that can be drawn
without crossings in a fixed surface, here the sphere, are closed under
taking minors~\cite[Section~1]{mohar2001minors}; so the planar graphs form
a minor-closed class. Let $H$ be a connected planar graph, $r$ a vertex of
$H$, $A$ a spanning tree of $H$ rooted at $r$, and $\mu$ vertex weights on
$H$. Lemma~\ref{fact:plthorup} gives three vertices, not necessarily
distinct, whose paths in $A$ to $r$ have a union $S$ such that every
component of $H-S$ has weight at most $\mu(V(H))/2$. Let $Z$ be the set of
these vertices. It is nonempty and has at most three vertices, and
$A[r,Z]=S$, as the path in $A$ between $r$ and a vertex $v$ has the vertex
set $A[r,v]$. So every connected planar graph is $3$-root-path separable.
For rational weights, Lemma~\ref{fact:plthorup} computes $Z$ in
linear time.
\end{proof}

\begin{proof}[Proof of Theorem~\ref{cor:plplanar}]
By Lemma~\ref{lem:plplanarclass}, Theorem~\ref{thm:plclass} applies with
$p=3$, and $14\min\{3,k\}\le42$.
\end{proof}

\paragraph{Bounded treewidth.} A \emph{tree decomposition} of a graph $H$
consists of a tree $\mathcal T$ and a set $B_z\subseteq V(H)$, the
\emph{bag} of $z$, for every node $z$ of $\mathcal T$, such that every
vertex of $H$ lies in some bag, the two ends of every edge lie in a common
bag, and for every vertex $v$ the nodes whose bags contain $v$ induce a
subtree of $\mathcal T$. Its \emph{width} is the
largest size of a bag minus one, and the \emph{treewidth} of $H$ is the
least width of a tree decomposition of $H$. The separators come from
balanced bags:

\begin{lemma}[{\cite[Lemma~7.19 and the remark after its proof]{cygan2015parameterized}}]\label{fact:plbag}
Let $H$ be a graph with a tree decomposition of width at most $w$, and let
$\mu$ be vertex weights on $H$. Then some bag $B$ of the decomposition is
such that every component of $H-B$ has weight at most $\mu(V(H))/2$, and
given the decomposition, such a bag can be found in polynomial time.
\end{lemma}

Lemma~\ref{lem:pltwsep} turns such a bag into a separator made of root
paths, which proves Corollary~\ref{cor:pltw}.

\begin{lemma}\label{lem:pltwsep}
For every fixed integer $w\ge1$, the graphs of treewidth at most $w$ form a
$(w+1)$-root-path separable class.
\end{lemma}

\begin{proof}
Let $\mathcal G$ be this class. It is minor-closed: let $H$ have a tree
decomposition $(\mathcal T,(B_z))$ of width at most $w$. Deleting an edge of
$H$ keeps it a tree decomposition, and deleting a vertex from $H$ and from
all bags gives one of the remaining graph. To contract two adjacent
vertices $u,v$ into a new vertex $y$, replace $u$ and $v$ by $y$ in every
bag: an edge $yy'$ of the new graph comes from an edge $uy'$ or $vy'$, whose
common bag now contains $y$ and $y'$, and the other edges and vertices keep
their bags; the nodes whose bags contain $y$ form
the union of the subtrees of $u$ and of $v$, which share a node whose bag
contains $u$ and $v$, so they induce a subtree; and no bag grows. Hence
every minor of $H$ has treewidth at most $w$.

Let $H\in\mathcal G$ be connected, let $r$, $A$ and $\mu$ be as in
Definition~\ref{def:plclass}, let $B$ be a bag as in
Lemma~\ref{fact:plbag} for a tree decomposition of $H$ of width at most
$w$, and let $Z:=B$ if $B\neq\emptyset$ and $Z:=\{r\}$ otherwise, a nonempty
set of at most $w+1$ vertices. As every $v\in B$ lies on its root path
$A[r,v]$, we have $B\subseteq A[r,Z]$, so every component of $H-A[r,Z]$
lies in a component of $H-B$, whose weight is at most $\mu(V(H))/2$. So every
connected graph in $\mathcal G$ is $(w+1)$-root-path separable. For fixed $w$, Bodlaender's
algorithm~\cite[Theorem~1.1]{bodlaender1996linear} computes a tree decomposition of
width at most $w$ of a graph of treewidth at most $w$ in linear time, and
Lemma~\ref{fact:plbag} then finds $B$ in polynomial time.
\end{proof}

\begin{proof}[Proof of Corollary~\ref{cor:pltw}]
By Lemma~\ref{lem:pltwsep}, Theorem~\ref{thm:plclass} applies with
$p=w+1$, as $\bar d=\lfloor\log_2k\rfloor$.
\end{proof}

\section{Hardness of directed multicast}
\label{sec:hard}

This appendix proves our lower bounds: Theorem~\ref{thm:lowopt}, by which
directed multicast stays $(1-\epsilon)\ln k$-hard even when
$\OPT=O(\log k)$, and Proposition~\ref{prop:largeopt}, by which a lower
bound of the same order holds when $\OPT\ge k^\epsilon$. Both reductions
start from Set Cover, in the form of Theorem~\ref{thm:setcover}, and use the
gadgets of Appendix~\ref{sec:gadgets}. We first explain both reductions.

\paragraph{Set Cover inside multicast.} An instance of Set Cover consists
of a universe of $m$ elements and a family of sets, and it is NP-hard to
distinguish the instances that have a cover with $c$ sets from those in
which every cover has more than $(1-\epsilon)c\ln m$ sets, where $c\ge
m^{1/C}$ for a constant $C$. In our version of Fraigniaud's
gadget~\cite{fraigniaud2001approximation} (he uses binomial trees and many
copies; Appendix~\ref{sec:gadgets}), the source calls set vertices,
one for every set, and each set vertex informs the terminals of its
elements through its \emph{set tree}, an out-arborescence in which every vertex has
at most two children. A cover of size $c$ gives a schedule of length
$c+O(\log m)$, and the set vertices that are informed in time must form a
cover. But as the source makes one call per round, the optimum is at least
the size of a cover, which is polynomial in $m$. We want hardness on
instances with $\OPT=O(\log k)$, on which the bound $O(\log
k)\,\OPT+O(\sqrt k)$ of Elkin and Kortsarz~\cite{elkin2006directed} gives
its largest ratio, $O(\sqrt k/\log k)$.

\begin{figure}[tb]
\centering
\begin{tikzpicture}[>={Stealth[length=2mm]},
  v/.style={circle,draw,inner sep=1.3pt,minimum size=5mm,font=\small},
  t/.style={rectangle,draw,fill=black!10,inner sep=2pt,minimum size=4.5mm,font=\small}]
  \node[v] (s) at (0,0) {$s$};
  \draw (s.south) -- (-2.2,-0.95) -- (2.2,-0.95) -- cycle;
  \foreach \i/\x in {1/-2.2,2/0,3/2.2}
    \node[v] (h\i) at (\x,-1.35) {$h_\i$};
  \foreach \j/\x in {1/-3.3,2/-1.1,3/1.1,4/3.3} {
    \node[v] (x\j) at (\x,-2.6) {$x_\j$};
    \draw (x\j.south) -- (\x-0.55,-3.5) -- (\x+0.55,-3.5) -- cycle;
  }
  \foreach \i in {1,2,3}
    \foreach \j in {1,2,3,4}
      \draw[->,black!45] (h\i) -- (x\j);
  \foreach \e/\x in {1/-3.8,2/-1.9,3/0,4/1.9,5/3.8}
    \node[t] (e\e) at (\x,-4.3) {$e_\e$};
  \draw[->] (-3.6,-3.5) -- (e1); \draw[->] (-3.0,-3.5) -- (e2);
  \draw[->] (-1.4,-3.5) -- (e2); \draw[->] (-1.1,-3.5) -- (e3);
  \draw[->] (-0.8,-3.5) -- (e4);
  \draw[->] (0.8,-3.5) -- (e3); \draw[->] (1.4,-3.5) -- (e5);
  \draw[->] (3.0,-3.5) -- (e4); \draw[->] (3.6,-3.5) -- (e5);
  \node[font=\footnotesize,align=left,anchor=west] at (2.6,-0.6)
    {out-tree from $s$\\ to the $q$ hubs};
  \node[font=\footnotesize,align=left,anchor=west] at (4.2,-1.95)
    {every hub may\\ call every set vertex};
  \node[font=\footnotesize,align=left,anchor=west] at (4.2,-3.1)
    {set trees};
\end{tikzpicture}
\caption{The hub gadget for the sets $S_1=\{e_1,e_2\}$,
$S_2=\{e_2,e_3,e_4\}$, $S_3=\{e_3,e_5\}$ and $S_4=\{e_4,e_5\}$. In $t$
rounds the $q$ hubs make at most $qt$ calls, and the called set vertices
must form a cover. With $q=\lceil c/\ell\rceil$ hubs, a cover of size $c$ is
called within $\ell$ rounds after the hubs are informed.}
\label{fig:hubs}
\end{figure}
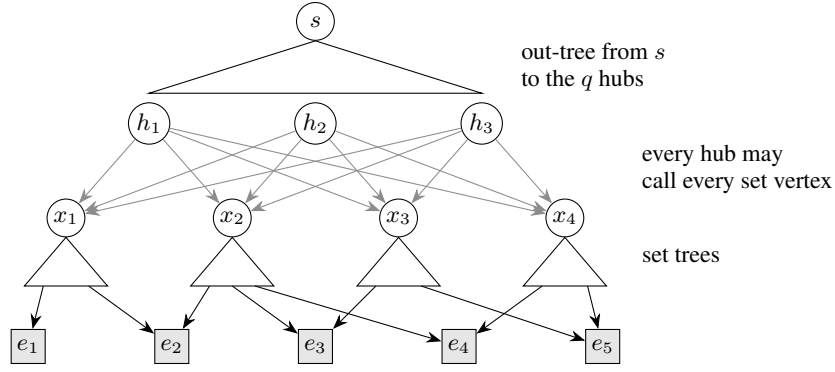

\paragraph{The hub gadget.} Our gadget delegates the calls of the source to
many \emph{hubs} (Figure~\ref{fig:hubs}). For a parameter $\ell\ge1$, the
source first informs, through a binary out-arborescence, $q=\lceil c/\ell\rceil$
hubs, and every hub may call
every set vertex, so a cover of size $c$ is called in $\ell$ rounds after
the hubs are informed. Every set vertex informs the terminals of its
elements through its set tree, of height at most $\lceil\log_2m\rceil+1$,
within $2\lceil\log_2m\rceil+2$ rounds. Conversely, in $t$ rounds the hubs
call at most $qt$ set vertices, and the called set vertices must form a
cover. So if every cover has at least $c'$ sets, then $\OPT\ge c'/q$, while
a cover of size $c$ gives $\OPT\le\ell+O(\log m)$
(Lemma~\ref{lem:hubgadget}).

\paragraph{The choice of parameters.} We take
$\ell=\lambda\lceil\log_2m\rceil$ for a large constant $\lambda$. Then the
additive $O(\log m)$ is a small fraction of $\ell$, so YES instances have
$\OPT\le C_\epsilon\log_2k$ with $k=m$ terminals and a constant
$C_\epsilon$. In NO instances every cover has more than $(1-\epsilon/3)c\ln
m$ sets, so $\OPT$ is at least about $(1-\epsilon/3)\ell\ln m$, which gives
$\OPT\ge(1-\epsilon)\ln k\cdot C_\epsilon\log_2k$ for large $m$. The
instances are acyclic, as Theorem~\ref{thm:lowopt} states.

\paragraph{A large optimum.} Fraigniaud's gadget also gives a lower bound
of order $\ln k$ on instances with a large optimum, which
Corollary~\ref{cor:dtm}\ref{cor:dtm:largeopt} matches up to a constant factor. Elkin and
Kortsarz observed this when $k=n^{\Omega(1)}$ and $\OPT=\Omega(\sqrt k)$
\cite[Section~1.2]{elkin2006directed}, and sketched a proof with many
disjoint copies of a hard instance; we make the exponent of $\OPT\ge
k^\epsilon$ arbitrary and the constant explicit. Take $r$ disjoint copies of the Set Cover
instance and build Fraigniaud's gadget for their union. A smallest cover of
the union consists of a smallest cover of every copy, so it has $r$ times
as many sets, and the gap between YES and NO instances stays about $\ln m$,
up to the additive $O(\log m)$ of the set trees. The union has $k=rm$
elements, and since the source calls one set vertex per round and every
copy needs one of its sets, $\OPT\ge r$. For a fixed $\epsilon\in(0,1)$, the
choice $r=\lceil m^{\epsilon/(1-\epsilon)}\rceil$ gives $r\ge k^\epsilon$,
and $\ln m$ is about $(1-\epsilon)\ln k$. This gives the lower bound
$(1-\delta)(1-\epsilon)\ln k$ of Proposition~\ref{prop:largeopt} on
instances with $\OPT\ge k^\epsilon$, where $\delta/3$ plays the role of the
$\epsilon$ of Theorem~\ref{thm:setcover} (Appendix~\ref{sec:largeopt}).

\subsection{Gadgets}
\label{sec:gadgets}

In this subsection we describe the gadgets from which our reductions are
built, and bound the optimum of the hub gadget
(Lemma~\ref{lem:hubgadget}). A gadget lets the source choose sets, and
every chosen set then informs its elements. All gadgets use out-arborescences in which every vertex has at most
two children. In such a tree, if every vertex calls its children in the
two rounds after it is informed, a vertex at depth $d$ is informed within
$2d$ rounds after the root; a tree with at most $m$ leaves can be chosen
with height $\lceil\log_2m\rceil$.

Both gadgets use \emph{set trees}. For a set $S_j$, its set
tree is an out-arborescence rooted at a vertex $x_j$, whose leaves are the
terminals of the elements of $S_j$, whose other vertices are new, and in
which every vertex has at most two children. Its height is at most
$\lceil\log_2m\rceil+1$, so it is informed within $2\lceil\log_2m\rceil+2$
rounds after $x_j$. Element terminals have no
out-arcs. A terminal can be informed only through the set tree of a set
that contains it.

\begin{figure}[t]
\centering
\begin{tikzpicture}[>={Stealth[length=2mm]},
  v/.style={circle,draw,inner sep=1.3pt,minimum size=5mm,font=\small},
  t/.style={rectangle,draw,fill=black!10,inner sep=2pt,minimum size=4.5mm,font=\small}]
  \node[v] (s) at (0,0) {$s$};
  \foreach \j/\x in {1/-3.6,2/-1.2,3/1.2,4/3.6} {
    \node[v] (x\j) at (\x,-1.4) {$x_\j$};
    \draw[->] (s) -- (x\j);
    \draw (\x,-1.72) -- (\x-0.7,-2.75) -- (\x+0.7,-2.75) -- cycle;
  }
  \foreach \e/\x in {1/-4.0,2/-2.0,3/0,4/2.0,5/4.0}
    \node[t] (e\e) at (\x,-3.9) {$e_\e$};
  \draw[->] (-3.9,-2.75) -- (e1); \draw[->] (-3.3,-2.75) -- (e2);
  \draw[->] (-1.5,-2.75) -- (e2); \draw[->] (-1.2,-2.75) -- (e3);
  \draw[->] (-0.9,-2.75) -- (e4);
  \draw[->] (0.9,-2.75) -- (e3); \draw[->] (1.5,-2.75) -- (e5);
  \draw[->] (3.3,-2.75) -- (e4); \draw[->] (3.9,-2.75) -- (e5);
  \node[font=\footnotesize,align=left,anchor=west] at (5.0,-2.2)
    {set tree of $S_j$};
  \node[font=\footnotesize,align=left,anchor=west] at (5.0,-3.9)
    {terminals (sinks)};
\end{tikzpicture}
\caption{Fraigniaud's gadget for $S_1=\{e_1,e_2\}$, $S_2=\{e_2,e_3,e_4\}$,
$S_3=\{e_3,e_5\}$ and $S_4=\{e_4,e_5\}$. The source calls one set vertex per
round, so the set vertices informed by round $t$ form a cover with at most
$t$ sets. The set trees add only $O(\log m)$ rounds.}
\label{fig:fraigniaud}
\end{figure}
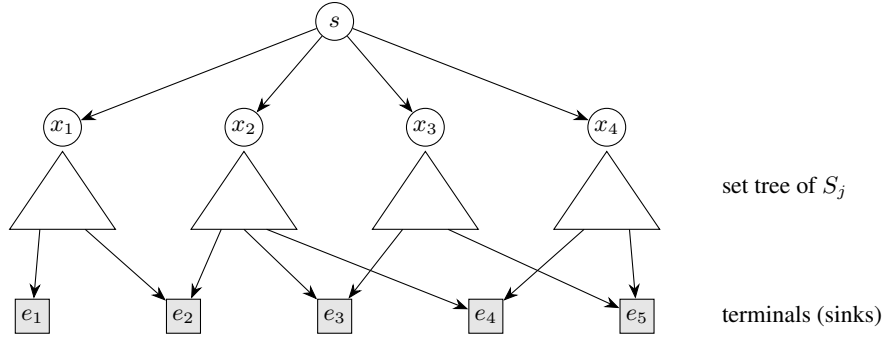

\emph{Fraigniaud's gadget}~\cite{fraigniaud2001approximation}
(Figure~\ref{fig:fraigniaud}) has a source $s$, a vertex $x_j$ with the arc
$(s,x_j)$ for every set, and the set trees. Fraigniaud states his
reduction for another communication model, in which a call may follow a
path and simultaneous calls use vertex-disjoint paths, and under the
assumption $\mathrm{NP}\not\subseteq\mathrm{DTIME}(n^{\log\log n})$. He uses binomial
trees as set trees and attaches about $\log m/\epsilon$ copies of the
digraph to the source, which makes the additive $O(\log m)$ negligible;
one copy suffices here, since $c\ge m^{1/C}$ in
Theorem~\ref{thm:setcover}. His schedules for the YES case use only
single arcs, and every schedule of the telephone model is one of his
model, so his argument applies verbatim.

The source calls one set vertex per round, and if all terminals are
informed by round $t$, the set vertices informed by round $t$ form a cover. Hence $\OPT\le c+2\lceil\log_2m
\rceil+2$ if a cover of size $c$ exists, and $\OPT\ge c'$ if every cover
has at least $c'$ sets. With Theorem~\ref{thm:setcover} this gives
$(1-\epsilon)\ln k$-hardness of \DTM, by combining Fraigniaud's reduction
with the NP-hardness of Set Cover. These instances have a large
optimum: in the NO case $\OPT>(1-\epsilon)c\ln m\ge m^{1/C}$. The \emph{hub
gadget} (Figure~\ref{fig:hubs}) spreads the calls over many vertices and
so brings $\OPT$ down to $O(\log k)$ while keeping the gap.

The hub gadget for a Set Cover instance and parameters $\ell,q$ consists
of the source $s$; an out-arborescence from $s$ of height $\lceil\log_2q\rceil$ in
which every vertex has at most two children and whose leaves are $q$
\emph{hubs} $h_1,\dots,h_q$; a set
vertex $x_j$ for every set, with the arcs $(h_i,x_j)$ for all $i,j$; and
the set trees. It is acyclic and has $k=m$ terminals.

Its optimum is about the size of a smallest cover divided by the number
of hubs:

\begin{lemma}\label{lem:hubgadget}
Let $q=\lceil c/\ell\rceil$. If a cover of size $c$ exists, the hub gadget
has $\OPT\le2\lceil\log_2q\rceil+\ell+2\lceil\log_2m\rceil+2$. If every
cover has at least $c'$ sets, then $\OPT\ge c'/q$.
\end{lemma}
\begin{proof}
The hubs are informed by round
$2\lceil\log_2q\rceil$. Split the cover
into $q$ groups of at most $\ell$ sets each; every hub calls its group in
its next $\ell$ rounds, and the set trees finish within
$2\lceil\log_2m\rceil+2$ further rounds. Conversely, a set vertex can be
informed only by a hub, and in $t$ rounds every hub makes at most $t$
calls, so at most $qt$ set vertices are informed. The informed set
vertices form a cover, so $qt\ge c'$.
\end{proof}

\subsection{Multicast with a small optimum}

In this subsection we show that the logarithmic lower bound for directed
multicast holds already on instances with $\OPT=O(\log k)$. These are the
instances on which the bound $O(\log k)\,\OPT+O(\sqrt k)$ of Elkin and
Kortsarz~\cite{elkin2006directed} gives its largest ratio,
$O(\sqrt k/\log k)$. We restate Theorem~\ref{thm:lowopt}:

\thmlowopt*

\begin{proof}
Let $\delta=\epsilon/3$, and take the instances of
Theorem~\ref{thm:setcover} for $\delta$ in place of $\epsilon$, with their
constant $C$. We may assume that the $\nu$ sets together cover the universe:
this is easy to check, and otherwise no cover exists, so the instance is a
NO instance. We may also assume $c\le \nu\le m^C$: if $c>\nu$, the instance is
a YES instance exactly when all $\nu$ sets together cover the universe. Let
\[
  \lambda=\Bigl\lceil\frac{2C+3}{\delta}\Bigr\rceil,\qquad
  C_\epsilon=\lambda+2C+3,\qquad\ell=\lambda\lceil\log_2m\rceil,
\]
and build the hub gadget with $q=\lceil c/\ell\rceil\le c$ hubs; it is
acyclic, and $k=m$. Its size is polynomial in $m$: there are $q\nu\le
m^{2C}$ arcs from the hubs to the set vertices, and as every vertex of the
tree to the hubs and of the set trees has at most two children, the tree to
the hubs has fewer than $4q$ vertices and each set tree fewer than $8m$, by
their heights $\lceil\log_2q\rceil$ and at most $\lceil\log_2m\rceil+1$. In the bound of the YES case, the hubs contribute
$\lambda\log_2m$ rounds, and the tree to the hubs and the set trees about
$(2C+3)\log_2m$ more. The choice of $\lambda$ makes the second part at most a
$\delta$ fraction of the first: $2C+3\le\delta\lambda$, hence
$C_\epsilon\le(1+\delta)\lambda$, that is,
\begin{equation}\label{eq:lowoptlambda}
  \lambda\;\ge\;\frac{C_\epsilon}{1+\delta}.
\end{equation}

\emph{YES case.} By Lemma~\ref{lem:hubgadget},
$\OPT\le2\lceil\log_2q\rceil+\ell+2\lceil\log_2m\rceil+2$. Here
$\lceil\log_2q\rceil\le C\log_2m+1$, as $q\le c\le m^C$;
$\lceil\log_2m\rceil\le\log_2m+1$; and $\ell\le\lambda\log_2m+\lambda$. So
\[
  \OPT\;\le\;(\lambda+2C+2)\log_2m+\lambda+6\;\le\;C_\epsilon\log_2k
\]
as soon as $\log_2m\ge\lambda+6$.

\emph{NO case.} Every cover has more than $(1-\delta)c\ln m$ sets, so
$\OPT\ge(1-\delta)c\ln m/q$ by Lemma~\ref{lem:hubgadget}. Here
$q\le c/\ell+1=(c/\ell)(1+\ell/c)$, and $\ell\ge\lambda\log_2m\ge
C_\epsilon\log_2m/(1+\delta)$ by~\eqref{eq:lowoptlambda}. Hence
\[
  \OPT\;\ge\;\frac{(1-\delta)\ln m\cdot\ell}{1+\ell/c}\;\ge\;
  \frac{1-\delta}{1+\delta}\cdot\frac{1}{1+\ell/c}\cdot\ln m\cdot
  C_\epsilon\log_2m .
\]
As $c\ge m^{1/C}$ grows polynomially in $m$ and $\ell=O(\log m)$, we have
$\ell/c\le\delta$ for large $m$, and then $1/(1+\ell/c)\ge1-\delta$.
Moreover $(1-\delta)/(1+\delta)\ge(1-\delta)^2$, as
$(1-\delta)(1+\delta)\le1$. So, for large $m$,
\[
  \OPT\;\ge\;(1-\delta)^3\ln m\cdot C_\epsilon\log_2m\;\ge\;
  (1-3\delta)\ln k\cdot C_\epsilon\log_2k\;=\;(1-\epsilon)\ln k\cdot
  C_\epsilon\log_2k .
\]
Instances with fewer elements than these thresholds have constant size and
are solved directly. So an algorithm that distinguishes the two cases of
Theorem~\ref{thm:lowopt} decides the Set Cover instances of
Theorem~\ref{thm:setcover}.
\end{proof}

\subsection{Multicast with a large optimum}
\label{sec:largeopt}

In this subsection we show that the logarithmic lower bound for directed
multicast also holds on instances whose optimum is polynomially large,
namely at least $k^\epsilon$; on these instances
Corollary~\ref{cor:dtm}\ref{cor:dtm:largeopt} has ratio $O(\log k)$, so there the lower bound
is tight up to a constant factor. Elkin and Kortsarz observed this when
$k=n^{\Omega(1)}$ and $\OPT=\Omega(\sqrt k)$
\cite[Section~1.2]{elkin2006directed}, and sketched a proof that takes many
disjoint copies of a hard instance; Proposition~\ref{prop:largeopt} extends
their observation to $\OPT\ge k^\epsilon$ for every fixed $\epsilon<1$,
with an explicit constant. The reduction applies Fraigniaud's gadget to many
disjoint copies of a Set Cover instance. The copies multiply the size of a
smallest cover, and with it the optimum, by the number of copies, so the
gap stays the same, while every schedule needs at least one call of the
source for each copy. We restate Proposition~\ref{prop:largeopt}:

\proplargeopt*

\begin{proof}
Let $\delta'=\delta/3$, and take the instances of
Theorem~\ref{thm:setcover} for $\delta'$ in place of $\epsilon$, with their
constant $C$: a universe of $m$ elements, sets $S_1,\dots,S_\nu$, and an
integer $c\ge m^{1/C}$. We may assume that the sets together cover the
universe: this is easy to check, and otherwise no cover exists, so the
instance is a NO instance. Let
\[
  r=\bigl\lceil m^{\epsilon/(1-\epsilon)}\bigr\rceil .
\]
The \emph{union} consists of $r$ disjoint copies of the instance: its
elements are the pairs $(i,e)$ and its sets the pairs $(i,j)$, for
$i\in\{1,\dots,r\}$, an element $e$ and a set $S_j$, where the set $(i,j)$
contains the elements $(i,e)$ with $e\in S_j$. Every set of the union lies
in one copy, so a family of sets covers the union exactly when, for every
$i$, its sets in copy $i$ cover that copy. Hence a smallest cover of the
union has exactly $r$ times as many sets as a smallest cover of the
instance; we denote its size by $c^*$.

We build Fraigniaud's gadget for the union: the source $s$, a vertex
$x_{i,j}$ with the arc $(s,x_{i,j})$ for every set $(i,j)$, and the set tree
of every set $(i,j)$, whose leaves are the terminals of its elements. Every
terminal lies in some set tree, so it is reachable from $s$. The digraph is
acyclic, it has $k=rm$ terminals, and its size is polynomial in the size of
the instance, as $r\le m^{\epsilon/(1-\epsilon)}+1$ and $\epsilon<1$ is
fixed. Let
\[
  \theta=rc+2\lceil\log_2m\rceil+2 .
\]

\emph{Lower bound on $\OPT$.} The only arc into a set vertex $x_{i,j}$
comes from $s$, which makes one call per round, so at most $\OPT$ set
vertices are informed within $\OPT$ rounds. As stated in
Appendix~\ref{sec:gadgets}, a terminal can be informed only through the set
tree of a set that contains its element, and hence only after the vertex of
that set. So the sets whose vertices are informed within $\OPT$ rounds
cover the union, and $\OPT\ge c^*$.

\emph{Every instance has $\OPT\ge k^\epsilon$.} Every copy needs at least
one set, so $\OPT\ge c^*\ge r$. As $r\ge m^{\epsilon/(1-\epsilon)}$, we have
$r^{1-\epsilon}\ge m^\epsilon$, hence $r=r^\epsilon r^{1-\epsilon}\ge
r^\epsilon m^\epsilon=k^\epsilon$.

\emph{YES case.} If the instance has a cover of size $c$, the union has a
cover with $rc$ sets. The source calls the vertices of these sets in
rounds $1,\dots,rc$. Every set has at most $m$ elements, so its set tree is
informed within $2\lceil\log_2m\rceil+2$ rounds after its root, and every
terminal lies in the set tree of a set of the cover. Hence $\OPT\le \theta$.

\emph{NO case.} Every cover of the instance has more than
$(1-\delta')c\ln m$ sets, so $\OPT\ge c^*>(1-\delta')rc\ln m$. As $rc\ge
c\ge m^{1/C}$, we have $\theta\le(1+\delta')rc$ as soon as
$2\lceil\log_2m\rceil+2\le\delta'm^{1/C}$, which holds for large $m$. Then
\[
  \OPT\;>\;\frac{1-\delta'}{1+\delta'}\,\ln m\cdot \theta\;\ge\;(1-\delta')^2\ln
  m\cdot \theta ,
\]
as $(1-\delta')(1+\delta')\le1$. Moreover
$k=rm\le(m^{\epsilon/(1-\epsilon)}+1)\,m\le2m^{1/(1-\epsilon)}$, so $\ln
k\le\ln2+\ln m/(1-\epsilon)$, that is,
\[
  \ln m\;\ge\;(1-\epsilon)(\ln k-\ln2)\;\ge\;(1-\epsilon)(1-\delta')\ln k
\]
as soon as $\ln k\ge\ln2/\delta'$, which holds for large $m$, as $k\ge m$.
Hence, for large $m$,
\[
  \OPT\;>\;(1-\delta')^3(1-\epsilon)\ln k\cdot \theta\;\ge\;(1-3\delta')
  (1-\epsilon)\ln k\cdot \theta\;=\;(1-\delta)(1-\epsilon)\ln k\cdot \theta .
\]

Instances with fewer elements than these thresholds have constant size and
are solved directly. So an algorithm that distinguishes the two cases of
Proposition~\ref{prop:largeopt} decides the Set Cover instances of
Theorem~\ref{thm:setcover}. Finally, an algorithm with ratio
$(1-\delta)(1-\epsilon)\ln k$ on the instances with $\OPT\ge k^\epsilon$
distinguishes the two cases: on the instances built here, it returns a
schedule of length at most $(1-\delta)(1-\epsilon)\ln k\cdot \theta$ in the YES
case, while in the NO case every schedule is longer.
\end{proof}

\section*{AI methodology statement}
This work grew out of our previous research on telephone broadcasting in undirected graphs~\cite{aminian2025complexity, bringolf2025structured}. That work motivated us to investigate the more general directed telephone multicast problem, where a substantial gap remained between known approximation guarantees and hardness bounds. Our initial investigation led to an unpublished draft presenting a polylogarithmic-type approximation in quasi-polynomial time through connections to directed Steiner trees with degree and height constraints. With technical assistance from Claude (Anthropic), we built on that draft to develop an improved algorithm that, among other guarantees, achieves an $O(k^\epsilon)$ approximation in polynomial time for every fixed $\epsilon>0$.

Our initial idea for this algorithm was to guide a recursive greedy approach using a potential that penalizes high loads at individual vertices. The algorithm scores each candidate partial solution by dividing the increase it would cause in the potential by the number of additional terminals it would cover, and selects a candidate with the smallest score. Through discussions with Claude, we corrected and refined this approach, developed the specific exponential potential and its analysis, and completed the remaining algorithmic and proof details. These interactions materially influenced the general-digraph results in Section~\ref{sec:ovdtm} and Appendix~\ref{sec:dtm}.

We also began investigating multicasting in planar digraphs, aiming to exploit planar separability to obtain stronger approximation guarantees. Our starting point included ideas about selecting short directed separator paths and using potential arguments to control vertex loads. One difficulty was that different regions could require calls from the same previously informed vertex, creating congestion even when the regions themselves were disjoint. Through discussions with Claude, we refined the selection of paths and entry arcs, developed the LP-guided separator recursion, and completed the potential-function analysis. These interactions materially influenced the planar algorithm in Section~\ref{sec:ovplanar} and its proofs in Appendix~\ref{sec:planar}.

We also used Claude for literature searches, reference checks, preparation of figures, and drafting substantial portions of the manuscript. 
We subsequently used ChatGPT (OpenAI) for technical review and to assist with refining technical arguments and revising the exposition and this statement.

The authors take full responsibility for the correctness, originality, attribution, references, and presentation of the paper, including all material developed with AI assistance.

\bibliography{references}

\newcommand{\etalchar}[1]{$^{#1}$}
\providecommand{\hyph}{-}
\begin{thebibliography}{BNGNS00}

\bibitem[AAF{\etalchar{+}}97]{aspnes1997online}
James Aspnes, Yossi Azar, Amos Fiat, Serge Plotkin, and Orli Waarts.
\newblock On-line routing of virtual circuits with applications to load
  balancing and machine scheduling.
\newblock {\em Journal of the ACM}, 44(3):486--504, 1997.
\newblock Preliminary version in STOC 1993.
\newblock \href {https://doi.org/10.1145/258128.258201}
  {\path{doi:10.1145/258128.258201}}.

\bibitem[AG06]{abraham2006object}
Ittai Abraham and Cyril Gavoille.
\newblock Object location using path separators.
\newblock In {\em Proceedings of the 25th Annual ACM Symposium on Principles of
  Distributed Computing (PODC)}, pages 188--197, 2006.
\newblock \href {https://doi.org/10.1145/1146381.1146411}
  {\path{doi:10.1145/1146381.1146411}}.

\bibitem[AKSJS25]{aminian2025complexity}
Aida Aminian, Shahin Kamali, Seyed-Mohammad Seyed-Javadi, and Sumedha.
\newblock On the complexity of telephone broadcasting from cacti to bounded
  pathwidth graphs.
\newblock In {\em Proceedings of the 52nd International Colloquium on Automata,
  Languages, and Programming (ICALP)}, volume 334 of {\em LIPIcs}, pages
  10:1--10:17, 2025.
\newblock \href {https://doi.org/10.4230/LIPIcs.ICALP.2025.10}
  {\path{doi:10.4230/LIPIcs.ICALP.2025.10}}.

\bibitem[BHKSJ25]{bringolf2025structured}
Jeffrey Bringolf, Hovhannes~A. Harutyunyan, Shahin Kamali, and Seyed-Mohammad
  Seyed-Javadi.
\newblock On hardness and approximation of broadcasting in structured graphs.
\newblock arXiv:2510.20026, 2025.

\bibitem[BLR26]{bamas2026randomized}
{\'E}tienne Bamas, Shi Li, and Lars Rohwedder.
\newblock Randomized rounding over dynamic programs.
\newblock In {\em Proceedings of the 58th Annual ACM Symposium on Theory of
  Computing (STOC)}, pages 1857--1868, 2026.
\newblock Full version arXiv:2511.03490.
\newblock \href {https://doi.org/10.1145/3798129.3800892}
  {\path{doi:10.1145/3798129.3800892}}.

\bibitem[BNGNS00]{bar2000message}
Amotz Bar-Noy, Sudipto Guha, Joseph Naor, and Baruch Schieber.
\newblock Message multicasting in heterogeneous networks.
\newblock {\em SIAM Journal on Computing}, 30(2):347--358, 2000.
\newblock Preliminary version in STOC 1998.
\newblock \href {https://doi.org/10.1137/S0097539798347906}
  {\path{doi:10.1137/S0097539798347906}}.

\bibitem[Bod96]{bodlaender1996linear}
Hans~L. Bodlaender.
\newblock A linear-time algorithm for finding tree-decompositions of small
  treewidth.
\newblock {\em SIAM Journal on Computing}, 25(6):1305--1317, 1996.
\newblock Preliminary version in STOC 1993.
\newblock \href {https://doi.org/10.1137/S0097539793251219}
  {\path{doi:10.1137/S0097539793251219}}.

\bibitem[CCC{\etalchar{+}}99]{charikar1999dst}
Moses Charikar, Chandra Chekuri, {To-yat} Cheung, Zuo Dai, Ashish Goel, Sudipto
  Guha, and Ming Li.
\newblock Approximation algorithms for directed {Steiner} problems.
\newblock {\em Journal of Algorithms}, 33(1):73--91, 1999.
\newblock Preliminary version in SODA 1998.
\newblock \href {https://doi.org/10.1006/jagm.1999.1042}
  {\path{doi:10.1006/jagm.1999.1042}}.

\bibitem[CEK06]{chekuri2006greedy}
Chandra Chekuri, Guy Even, and Guy Kortsarz.
\newblock A greedy approximation algorithm for the group {Steiner} problem.
\newblock {\em Discrete Applied Mathematics}, 154(1):15--34, 2006.
\newblock \href {https://doi.org/10.1016/j.dam.2005.07.010}
  {\path{doi:10.1016/j.dam.2005.07.010}}.

\bibitem[CFK{\etalchar{+}}15]{cygan2015parameterized}
Marek Cygan, Fedor~V. Fomin, {\L}ukasz Kowalik, Daniel Lokshtanov, D{\'a}niel
  Marx, Marcin Pilipczuk, Micha{\l} Pilipczuk, and Saket Saurabh.
\newblock {\em Parameterized Algorithms}.
\newblock Springer, 2015.
\newblock \href {https://doi.org/10.1007/978-3-319-21275-3}
  {\path{doi:10.1007/978-3-319-21275-3}}.

\bibitem[CJ26]{chekuri2026length}
Chandra Chekuri and Rhea Jain.
\newblock Length-constrained network design in planar digraphs.
\newblock In {\em Approximation, Randomization, and Combinatorial Optimization.
  Algorithms and Techniques (APPROX/RANDOM)}, volume 392 of {\em LIPIcs}, pages
  15:1--15:24, 2026.
\newblock \href {https://doi.org/10.4230/LIPIcs.APPROX/RANDOM.2026.15}
  {\path{doi:10.4230/LIPIcs.APPROX/RANDOM.2026.15}}.

\bibitem[CJK{\etalchar{+}}24]{chekuri2024polymatroid}
Chandra Chekuri, Rhea Jain, Shubhang Kulkarni, Da~Wei Zheng, and Weihao Zhu.
\newblock From directed {Steiner} tree to directed polymatroid {Steiner} tree
  in planar graphs.
\newblock In {\em Proceedings of the 32nd Annual European Symposium on
  Algorithms (ESA)}, volume 308 of {\em LIPIcs}, pages 42:1--42:19, 2024.
\newblock \href {https://doi.org/10.4230/LIPIcs.ESA.2024.42}
  {\path{doi:10.4230/LIPIcs.ESA.2024.42}}.

\bibitem[DG10]{diot2010path}
Emilie Diot and Cyril Gavoille.
\newblock Path separability of graphs.
\newblock In {\em Proceedings of the 4th International Workshop on Frontiers in
  Algorithmics (FAW)}, volume 6213 of {\em LNCS}, pages 262--273, 2010.
\newblock \href {https://doi.org/10.1007/978-3-642-14553-7_25}
  {\path{doi:10.1007/978-3-642-14553-7_25}}.

\bibitem[DS14]{dinur2014analytical}
Irit Dinur and David Steurer.
\newblock Analytical approach to parallel repetition.
\newblock In {\em Proceedings of the 46th Annual ACM Symposium on Theory of
  Computing (STOC)}, pages 624--633, 2014.
\newblock \href {https://doi.org/10.1145/2591796.2591884}
  {\path{doi:10.1145/2591796.2591884}}.

\bibitem[EGH{\etalchar{+}}25]{egami2025broadcasting}
Yudai Egami, Tatsuya Gima, Tesshu Hanaka, Yasuaki Kobayashi, Michael Lampis,
  Valia Mitsou, Edouard Nemery, Yota Otachi, Manolis Vasilakis, and Daniel Vaz.
\newblock Broadcasting under structural restrictions.
\newblock In {\em Proceedings of the 50th International Symposium on
  Mathematical Foundations of Computer Science (MFCS)}, volume 345 of {\em
  LIPIcs}, pages 42:1--42:18, 2025.
\newblock \href {https://doi.org/10.4230/LIPIcs.MFCS.2025.42}
  {\path{doi:10.4230/LIPIcs.MFCS.2025.42}}.

\bibitem[EK05]{elkin2005combinatorial}
Michael Elkin and Guy Kortsarz.
\newblock A combinatorial logarithmic approximation algorithm for the directed
  telephone broadcast problem.
\newblock {\em SIAM Journal on Computing}, 35(3):672--689, 2005.
\newblock Preliminary version in STOC 2002.
\newblock \href {https://doi.org/10.1137/S0097539704440740}
  {\path{doi:10.1137/S0097539704440740}}.

\bibitem[EK06a]{elkin2006directed}
Michael Elkin and Guy Kortsarz.
\newblock An approximation algorithm for the directed telephone multicast
  problem.
\newblock {\em Algorithmica}, 45(4):569--583, 2006.
\newblock Preliminary version in ICALP 2003.
\newblock \href {https://doi.org/10.1007/s00453-005-1196-4}
  {\path{doi:10.1007/s00453-005-1196-4}}.

\bibitem[EK06b]{elkin2006sublogarithmic}
Michael Elkin and Guy Kortsarz.
\newblock Sublogarithmic approximation for telephone multicast.
\newblock {\em Journal of Computer and System Sciences}, 72(4):648--659, 2006.
\newblock Preliminary version in SODA 2003.
\newblock \href {https://doi.org/10.1016/j.jcss.2005.12.002}
  {\path{doi:10.1016/j.jcss.2005.12.002}}.

\bibitem[Fei98]{feige1998threshold}
Uriel Feige.
\newblock A threshold of $\ln n$ for approximating set cover.
\newblock {\em Journal of the ACM}, 45(4):634--652, 1998.
\newblock Preliminary version in STOC 1996.
\newblock \href {https://doi.org/10.1145/285055.285059}
  {\path{doi:10.1145/285055.285059}}.

\bibitem[FM23]{friggstad2023planar}
Zachary Friggstad and Ramin Mousavi.
\newblock An {$O(\log k)$}-approximation for directed {Steiner} tree in planar
  graphs.
\newblock In {\em Proceedings of the 50th International Colloquium on Automata,
  Languages, and Programming (ICALP)}, volume 261 of {\em LIPIcs}, pages
  63:1--63:14, 2023.
\newblock Journal version in ACM Transactions on Algorithms 21(4), 2025.
\newblock \href {https://doi.org/10.4230/LIPIcs.ICALP.2023.63}
  {\path{doi:10.4230/LIPIcs.ICALP.2023.63}}.

\bibitem[Fra01]{fraigniaud2001approximation}
Pierre Fraigniaud.
\newblock Approximation algorithms for minimum-time broadcast under the
  vertex-disjoint paths mode.
\newblock In {\em Proceedings of the 9th Annual European Symposium on
  Algorithms (ESA)}, volume 2161 of {\em LNCS}, pages 440--451, 2001.
\newblock \href {https://doi.org/10.1007/3-540-44676-1_37}
  {\path{doi:10.1007/3-540-44676-1_37}}.

\bibitem[GKL{\etalchar{+}}22]{guo2022degree}
Xiangyu Guo, Guy Kortsarz, Bundit Laekhanukit, Shi Li, Daniel Vaz, and Jiayi
  Xian.
\newblock On approximating degree-bounded network design problems.
\newblock {\em Algorithmica}, 84(5):1252--1278, 2022.
\newblock Preliminary version in APPROX/RANDOM 2020, LIPIcs 176, 39:1--39:21.
\newblock \href {https://doi.org/10.1007/s00453-022-00924-0}
  {\path{doi:10.1007/s00453-022-00924-0}}.

\bibitem[GKR00]{garg2000polylogarithmic}
Naveen Garg, Goran Konjevod, and R.~Ravi.
\newblock A polylogarithmic approximation algorithm for the group {Steiner}
  tree problem.
\newblock {\em Journal of Algorithms}, 37(1):66--84, 2000.
\newblock Preliminary version in SODA 1998.
\newblock \href {https://doi.org/10.1006/jagm.2000.1096}
  {\path{doi:10.1006/jagm.2000.1096}}.

\bibitem[GLL23]{grandoni2019dst}
Fabrizio Grandoni, Bundit Laekhanukit, and Shi Li.
\newblock {$O(\log^2 k/\log\log k)$}-approximation algorithm for directed
  {S}teiner tree: A tight quasi-polynomial time algorithm.
\newblock {\em SIAM Journal on Computing}, 52(2):STOC19\hyph 298--STOC19\hyph
  322, 2023.
\newblock Preliminary version in STOC 2019.
\newblock \href {https://doi.org/10.1137/20M1312988}
  {\path{doi:10.1137/20M1312988}}.

\bibitem[GN22]{ghuge2022quasi}
Rohan Ghuge and Viswanath Nagarajan.
\newblock Quasi-polynomial algorithms for submodular tree orienteering and
  directed network design problems.
\newblock {\em Mathematics of Operations Research}, 47(2):1612--1630, 2022.
\newblock Preliminary version in SODA 2020.
\newblock \href {https://doi.org/10.1287/moor.2021.1181}
  {\path{doi:10.1287/moor.2021.1181}}.

\bibitem[HHL88]{hedetniemi1988survey}
Sandra~M. Hedetniemi, Stephen~T. Hedetniemi, and Arthur~L. Liestman.
\newblock A survey of gossiping and broadcasting in communication networks.
\newblock {\em Networks}, 18(4):319--349, 1988.
\newblock \href {https://doi.org/10.1002/net.3230180406}
  {\path{doi:10.1002/net.3230180406}}.

\bibitem[HK03]{halperin2003polylogarithmic}
Eran Halperin and Robert Krauthgamer.
\newblock Polylogarithmic inapproximability.
\newblock In {\em Proceedings of the 35th Annual ACM Symposium on Theory of
  Computing (STOC)}, pages 585--594, 2003.
\newblock \href {https://doi.org/10.1145/780542.780628}
  {\path{doi:10.1145/780542.780628}}.

\bibitem[HKR26]{hathcock2026telephone}
Daniel Hathcock, Guy Kortsarz, and R.~Ravi.
\newblock The telephone $k$-multicast problem.
\newblock {\em Algorithmica}, 88(2):Article~25, 2026.
\newblock Preliminary version in APPROX/RANDOM 2024, LIPIcs~317, 21:1--21:16.
\newblock \href {https://doi.org/10.1007/s00453-026-01377-5}
  {\path{doi:10.1007/s00453-026-01377-5}}.

\bibitem[HRZ01]{helvig2001improved}
C.~S. Helvig, Gabriel Robins, and Alexander Zelikovsky.
\newblock An improved approximation scheme for the group {Steiner} problem.
\newblock {\em Networks}, 37(1):8--20, 2001.
\newblock Preliminary version in DATE 1998.
\newblock \href
  {https://doi.org/10.1002/1097-0037(200101)37:1<8::AID-NET2>3.0.CO;2-R}
  {\path{doi:10.1002/1097-0037(200101)37:1<8::AID-NET2>3.0.CO;2-R}}.

\bibitem[IRRS17]{iglesias2017plane}
Jennifer Iglesias, Rajmohan Rajaraman, R.~Ravi, and Ravi Sundaram.
\newblock Plane gossip: Approximating rumor spread in planar graphs.
\newblock arXiv:1612.01492v2, 2017.
\newblock Full version of the LATIN 2018 paper.

\bibitem[IRRS18]{iglesias2018plane}
Jennifer Iglesias, Rajmohan Rajaraman, R.~Ravi, and Ravi Sundaram.
\newblock Plane gossip: Approximating rumor spread in planar graphs.
\newblock In {\em Proceedings of the 13th Latin American Symposium on
  Theoretical Informatics (LATIN)}, volume 10807 of {\em LNCS}, pages 611--624,
  2018.
\newblock \href {https://doi.org/10.1007/978-3-319-77404-6_45}
  {\path{doi:10.1007/978-3-319-77404-6_45}}.

\bibitem[KN22]{kortsarz2022minimum}
Guy Kortsarz and Zeev Nutov.
\newblock The minimum degree group {Steiner} problem.
\newblock {\em Discrete Applied Mathematics}, 309:229--239, 2022.
\newblock Preliminary version in IWOCA 2020.
\newblock \href {https://doi.org/10.1016/j.dam.2021.12.003}
  {\path{doi:10.1016/j.dam.2021.12.003}}.

\bibitem[KP95]{kortsarz1995approximation}
Guy Kortsarz and David Peleg.
\newblock Approximation algorithms for minimum-time broadcast.
\newblock {\em SIAM Journal on Discrete Mathematics}, 8(3):401--427, 1995.
\newblock Preliminary version in ISTCS 1992.
\newblock \href {https://doi.org/10.1137/S0895480193245923}
  {\path{doi:10.1137/S0895480193245923}}.

\bibitem[Moh01]{mohar2001minors}
Bojan Mohar.
\newblock Graph minors and graphs on surfaces.
\newblock In J.~W.~P. Hirschfeld, editor, {\em Surveys in Combinatorics, 2001},
  volume 288 of {\em London Mathematical Society Lecture Note Series}, pages
  145--164. Cambridge University Press, 2001.
\newblock \href {https://doi.org/10.1017/CBO9780511721328.008}
  {\path{doi:10.1017/CBO9780511721328.008}}.

\bibitem[Mos15]{moshkovitz2015projection}
Dana Moshkovitz.
\newblock The projection games conjecture and the {NP}-hardness of $\ln
  n$-approximating set-cover.
\newblock {\em Theory of Computing}, 11(7):221--235, 2015.
\newblock Preliminary version in APPROX 2012.
\newblock \href {https://doi.org/10.4086/toc.2015.v011a007}
  {\path{doi:10.4086/toc.2015.v011a007}}.

\bibitem[MR95]{motwani1995randomized}
Rajeev Motwani and Prabhakar Raghavan.
\newblock {\em Randomized Algorithms}.
\newblock Cambridge University Press, 1995.
\newblock \href {https://doi.org/10.1017/CBO9780511814075}
  {\path{doi:10.1017/CBO9780511814075}}.

\bibitem[Rag88]{raghavan1988probabilistic}
Prabhakar Raghavan.
\newblock Probabilistic construction of deterministic algorithms: Approximating
  packing integer programs.
\newblock {\em Journal of Computer and System Sciences}, 37(2):130--143, 1988.
\newblock Preliminary version in FOCS 1986.
\newblock \href {https://doi.org/10.1016/0022-0000(88)90003-7}
  {\path{doi:10.1016/0022-0000(88)90003-7}}.

\bibitem[Rav94]{ravi1994rapid}
R.~Ravi.
\newblock Rapid rumor ramification: Approximating the minimum broadcast time.
\newblock In {\em Proceedings of the 35th Annual Symposium on Foundations of
  Computer Science (FOCS)}, pages 202--213, 1994.
\newblock \href {https://doi.org/10.1109/SFCS.1994.365693}
  {\path{doi:10.1109/SFCS.1994.365693}}.

\bibitem[SCH81]{slater1981nptree}
Peter~J. Slater, Ernest~J. Cockayne, and Stephen~T. Hedetniemi.
\newblock Information dissemination in trees.
\newblock {\em SIAM Journal on Computing}, 10(4):692--701, 1981.
\newblock \href {https://doi.org/10.1137/0210052} {\path{doi:10.1137/0210052}}.

\bibitem[Tho04]{thorup2004compact}
Mikkel Thorup.
\newblock Compact oracles for reachability and approximate distances in planar
  digraphs.
\newblock {\em Journal of the ACM}, 51(6):993--1024, 2004.
\newblock Preliminary version in FOCS 2001.
\newblock \href {https://doi.org/10.1145/1039488.1039493}
  {\path{doi:10.1145/1039488.1039493}}.

\bibitem[Zel97]{zelikovsky1997series}
Alexander Zelikovsky.
\newblock A series of approximation algorithms for the acyclic directed
  {Steiner} tree problem.
\newblock {\em Algorithmica}, 18(1):99--110, 1997.
\newblock \href {https://doi.org/10.1007/BF02523690}
  {\path{doi:10.1007/BF02523690}}.

\end{thebibliography}
\bibliographystyle{alphaurl}

\end{document}